\documentclass{mn2e}
\usepackage{epsfig}
\usepackage{rotating}
\usepackage{longtable}
\usepackage{graphics}
\usepackage{amssymb}
\usepackage{supertabular}
\usepackage{times} 
\title[Recombination-line vs. forbidden-line abundances in H~{\sc ii} regions]
{Heavy elements in Galactic and Magellanic Cloud H~{\sc ii} regions:
recombination-line versus forbidden-line abundances}
\author[Y. G. Tsamis et al.]
{Yiannis G. Tsamis$^1$, M. J.\,Barlow$^1$, X.-W. Liu$^1$, 
I. J. Danziger$^2$ and P. J. Storey$^1$ \\
$^1$Department of Physics and Astronomy, University College London,
      Gower Street, London WC1E 6BT, UK\\
$^2$Osservatorio Astronomico di Trieste, Via G. B. Tiepolo 11, I-34131 Trieste,
Italy}

\date{Received:}
\newcommand{\eld}{$N_{\rm e}$}
\newcommand{\crd}{$N_{\rm cr}$}
\newcommand{\elt}{$T_{\rm e}$}

\newcommand{\cmt}{cm$^{-3}$}

\newcommand{\cpp}{C$^{2+}$}

\newcommand{\op}{O$^+$}
\newcommand{\opp}{O$^{2+}$}
\newcommand{\oppp}{O$^{3+}$}

\newcommand{\np}{N$^+$}
\newcommand{\npp}{N$^{2+}$}

\newcommand{\Hb}{H$\beta$}
\newcommand{\foiii}{[O~{\sc iii}]}

\newcommand{\foii}{[O~{\sc ii}]}
\newcommand{\fsii}{[S~{\sc ii}]}
\newcommand{\fsiii}{[S~{\sc iii}]}
\newcommand{\fnii}{[N~{\sc ii}]}
\newcommand{\fariv}{[Ar~{\sc iv}]}
\newcommand{\fcliii}{[Cl~{\sc iii}]}
\newcommand{\fneiii}{[Ne~{\sc iii}]}

\newcommand{\oiii}{O~{\sc iii}}

\newcommand{\nii}{N~{\sc ii}}

\newcommand{\oii}{O~{\sc ii}}
\newcommand{\cii}{C~{\sc ii}}

\newcommand{\ariv}{Ar~{\sc iv}}
\newcommand{\hi}{H\,{\sc i}}
\newcommand{\hii}{H~{\sc ii}}
\newcommand{\hei}{He~{\sc i}}
\newcommand{\heii}{He~{\sc ii}}
\newcommand{\tmf}{10$^{-4}$}
\newcommand{\tmfi}{10$^{-5}$}
\newcommand{\hp}{H$^+$}
\newcommand{\hep}{He$^+$}

\begin{document}
\maketitle

\begin{abstract}

\noindent We have obtained deep optical, long-slit spectrophotometry of
the Galactic {\hii} regions M\,17, NGC\,3576 and of the Magellanic Cloud
{\hii} regions 30~Doradus, LMC~N11B and SMC~N66, recording the optical
recombination lines (ORLs) of {\cii}, {\nii} and {\oii}.  A spatial
analysis of 30 Doradus is performed, revealing that the forbidden-line
{\foiii} electron temperature is remarkably constant across the nebula.
The forbidden-line {\opp}/{\hp} abundance mapped by the {\foiii}
$\lambda$4959 collisionally excited line (CEL) is shown to be consistently
lower than the recombination-line abundance mapped by the {\oii} V\,1
multiplet at 4650\,\AA. In addition, the spatial profile of the
{\cpp}/{\opp} ratio derived purely from recombination lines is presented
for the first time for an extragalactic nebula. Temperature-insensitive
ORL {\cpp}/{\opp} and {\npp}/{\opp} ratios are obtained for all nebulae
except SMC~N66. The ORL {\cpp}/{\opp} ratios show remarkable agreement
within each galactic system, while also being in agreement with the
corresponding CEL ratios. The disagreement found between the ORL and CEL
{\npp}/{\opp} ratios for M~17 and NGC~3576 can be attributed to the {\nii}
V~3 and V~5 ORLs that were used being affected by fluorescent excitation
effects.

For all five nebulae, the {\opp}/{\hp} abundance derived from multiple
{\oii} ORLs is found to be higher than the corresponding value derived
from the strong {\foiii} $\lambda\lambda$4959, 5007 CELs, by factors of
1.8--2.7 for four of the nebulae.  The LMC~N11B nebula exhibits
a more extreme discrepancy factor for the {\opp} ion, $\sim$5.  Thus these
{\hii} regions exhibit ORL/CEL abundance discrepancy factors that are
similar to those previously encountered amongst planetary nebulae.

Our optical CEL {\opp}/{\hp} abundances agree to within 20-30 per cent
with published {\opp}/{\hp} abundances that were obtained from
observations of infrared fine-structure lines. Since the low excitation
energies of the latter make them insensitive to variations about typical
nebular temperatures, fluctuations in temperature are ruled out
as the cause of the observed ORL/CEL {\opp} abundance discrepancies. We
present evidence that the observed {\oii} ORLs from these {\hii} regions
originate from gas of very similar density ($<$~3500~{\cmt}) to that 
emitting the observed
heavy-element optical and infrared CELs, ruling out models that employ
high-density ionized inclusions in order to explain the abundance
discrepancy. We consider a scenario whereby much of the heavy-element ORL
emission originates from cold ($\leq$~500~K) metal-rich ionized regions.
These might constitute halos that are being evaporated from much denser
neutral cores. The origin of these metal-rich inclusions is not clear --
they may have been ejected into the nebula by evolved, massive Of and
Wolf-Rayet stars, although the agreement found between heavy element ion
ratios derived from ORLs with the ratios derived from CELs provides no 
evidence for nuclear-processed material in the ORL-emitting regions. \\


\end{abstract}

\section{Introduction}

In recent years deep spectroscopic studies, coupled with new theoretical
results in atomic physics, have cast new light on a well established field of
modern astrophysics: the study of elemental abundances in H\,{\sc ii} regions
(ionized gas clouds marking the birthplaces of stars within the galaxy and in
external galaxies) and planetary nebulae (PNe; ionized ejected envelopes of
evolved low- to intermediate-mass stars). On the one hand, an accurate
knowledge of abundances in galactic and extragalactic H\,{\sc ii} regions is of
paramount importance for constraining galactic chemical evolution models (e.g.
Shields 2002). On the other hand, abundance studies of PNe provide useful
constraints on nucleosynthetic theories and our understanding of the late
stages of stellar evolution (Kingsburgh \& Barlow 1994; Henry, Kwitter \& Bates
2000). Nebular abundances of heavy elements such as C, N, O and Ne, relative to
H, have traditionally been derived from observations of strong, and thus easy
to measure, collisionally-excited ionic lines (CELs); e.g. C~{\sc iii}]
$\lambda\lambda$1906, 1909 {\fnii} $\lambda\lambda$6548, 6584, {\foiii}
$\lambda\lambda$4959, 5007 and {\fneiii} $\lambda\lambda$3869, 3967. However,
the analysis of observations of weak heavy-element optical
recombination lines (ORLs) from PNe, including publications
by Peimbert, Storey \& Torres-Peimbert (1993), Liu et al. 
(1995a; 2000; 2001b), Luo, Liu \& Barlow (2001), Garnett \& Dinerstein
(2001), and the recent work of Tsamis et al. (2002a for
observations and CEL analysis; 2002b for ORL analysis), have
yielded CNONe abundances for PNe that are systematically higher than
those derived using the
standard CEL method. The most promising explanation for these results, at least
for PNe, posits the existence within nebulae of a hitherto unknown,
low-temperature component enhanced in heavy elements and emitting mostly in
ORLs, intermingled with a hot component of more normal composition from which
the CEL emission originates (Liu et al. 2000; Liu 2002a,b; P\'{e}quignot et
al. 2002a, b). The exact nature and origin of the putative ORL-emitting
component is currently a matter of intense debate.

The ORL results for PNe become a matter of concern for nebular abundance
studies, all the more so because nebular CEL abundances have been long plagued
by lingering doubts about their reliability, arising from their exponential
sensitivity to the adopted nebular electron temperature and their dependence on
the adopted nebular electron density (for lines of low critical density,
{\crd}; Rubin 1989). On the other hand, elemental abundances relative to
hydrogen derived from ratios of ORLs [e.g. from the $I$(C~{\sc ii}
$\lambda$4267)/$I$(\Hb) intensity ratio for the derivation of {\cpp}/{\hp}, as
opposed to using the CEL/ORL $I$($\lambda$1908)/$I$(\Hb) ratio] are nearly
independent of both the adopted temperature and density; this means that
abundance determinations employing ORLs should be more accurate than those
using CELs, since in real nebulae much of the emission of CELs can be biased
towards regions of high electron temperature, as well as towards regions having
electron densities less than the critical density of the CEL. As a result, ORL
abundance studies of ionized nebulae are now coming to the fore, thanks also to
rapid progress in detector technology.

To date, deep abundance studies of PNe (summarised by Liu 2002b) have yielded
ORL abundances for C, N, O and Ne which, for the majority (90--95\%) of
nebulae, are typically a factor of 2--3 larger than those obtained from UV,
optical or infrared CELs. For the remaining 5-10\% of PNe, even larger
discrepancy factors (5--80) are found between the heavy element abundances
derived from ORLs and CELs. The situation for {\hii} regions is less clear thus
far than for PNe, mainly on account of the small number of {\hii} regions so
far observed specifically for the purpose of detecting heavy element ORLs and
deriving abundances from them. Peimbert et al. (1993) found ORL
O$^{2+}$/H$^{+}$ abundances for M\,42 and M\,17 that were a factor of two
larger than those found from the [O~{\sc iii}] optical CELs, while Esteban et
al. (1998, 1999) found ORL O$^{2+}$ abundances that were larger than the CEL
values by a factor of 1.5 for M\,42 and a factor of two for M\,8. Clearly, if
{\hii} regions are generally found to yield heavy element abundances from ORLs
that exceed those from CELs by similar factors to those summarised above for
PNe and for M\,8, M\,17 and M\,42, then this could have serious implications
for our understanding of the chemical evolution of galaxies, which to date has
relied to a large extent on CEL abundance analyses of {\hii} regions located in
our own and other galaxies. The current contribution aims to increase the
number of {\hii} regions with ORL abundance analyses, by presenting deep,
medium-resolution, long-slit optical spectrophotometry of the bright Galactic
{\hii} regions M\,17 and NGC\,3576, and the Magellanic Cloud {\hii} regions
30~Doradus, LMC~N11B and SMC~N66. These nebulae were selected to be of
relatively high excitation for H~{\sc ii} regions, with O$^{2+}$
being the dominant ion stage of oxygen.
In Section~2 we describe our optical spectroscopic observations and 
present a thorough list of emission line fluxes and dereddened
intensities. In Section~3 we describe the extinction corrections and the plasma
temperature and density analysis. In Section~4 we present an abundance analysis
using optical collisionally excited lines. In Section~5 we present an
abundance study using optical recombination lines, discussing the relative
intensities of {\oii} ORLs and complications arising from the presence of
bright dust-scattered starlight within the nebulae. In Section~6
high-resolution long-slit spectra of 30~Doradus are used to map the electron
temperature and density, and the ionic abundances across the nebular surface.
Finally, we discuss the implications of the results from this extensive study
in Section~7 and state our conclusions in Section~8.

\setcounter{table}{0}
\begin{table*}
\centering
\begin{minipage}{135mm}
\caption[Journal of H~{\sc  ii} regions observations.]{Journal of
observations.}
\begin{tabular}{lccccrrc}
\noalign{\vskip3pt} \noalign{\hrule} \noalign{\vskip3pt}
H {\sc ii} region  &Date    &$\lambda$-range  &FWHM   &PA &RA   &DEC      &Exp. \\
&(UT)       &(\AA)       &(\AA)   &(deg)   &(2000)    &(2000)  &(sec)\\
\noalign{\vskip3pt} \noalign{\hrule} \noalign{\vskip3pt}
\multicolumn{8}{c}{ESO 1.52-m}\\
\noalign{\vskip3pt}
M\,17             &07/7/96  &3995--4978    &1.5    &$-$21 &18 20 40.0    &$-$16 09 29 &3$\times$1200\\
M\,17                      &13/7/96     &3535--7400         &4.5    &$-$21 &"     &" &60, 300, 600\\
NGC\,3576                  &11/2/97     &3995--4978         &1.5    &$-$50  &11 11 56.9          &$-$61 17 25               &6$\times$1800\\
\noalign{\vskip3pt}
\multicolumn{8}{c}{AAT 3.9-m}\\
\noalign{\vskip3pt}
NGC\,3576   &08/02/95   &3509--3908 &1  &$-$50    &11 12 00.5 &$-$61 18 24  &300, 1800\\
NGC\,3576       &"       &3908--4305    &1      &$-$50       &"       &"       &2$\times$1800\\
NGC\,3576       &"       &3635--7360    &8.5    &$-$50       &"       &"       &120, 300, 600\\
\noalign{\vskip3pt}
\multicolumn{8}{c}{NTT 3.5-m}\\
\noalign{\vskip3pt}
30 Doradus             &15/12/95    &3635--4145         &2 &76 &05 38 45.6          &$-$69 05 24   &3$\times$1200\\
30 Doradus             &"       &4060--4520     &2  &76 &"      &"    &3$\times$1200\\
30 Doradus         &"       &4515--4975     &2  &76 &"      &"    &300, 4$\times$1200\\
30 Doradus         &"       &6507--7828             &3.8    &76 &"          &"    &3$\times$1200\\
30 Doradus         &"           &3800--8400             &11     &76 &"              &"         &60, 300, 600\\
\noalign{\vskip3pt}
LMC N11B                   &16/12/95    &3635--4145 &2      &$-$57 &04 56 47.0          &$-$66 25 11         &600\\
LMC N11B                   &"           &4060--4520             &2      &$-$57 &"          &"         &2$\times$1800\\
LMC N11B                   &"           &4515--4975             &2      &$-$57 &"          &"         &2$\times$1800\\
LMC N11B                   &"           &6507--7828             &3.8    &$-$57 &"          &"         &600\\
LMC N11B                   &"           &3800--8400             &11     &$-$57 &"          &"         &600\\
\noalign{\vskip3pt}
SMC N66                    &16/12/95    &3635--4145 &2      &$-$57 &00 58 55.2          &$-$72 12 32         &600\\
SMC N66                    &"           &4060--4520             &2      &$-$57 &"          &"         &2$\times$1800\\
SMC N66                    &"           &4515--4975             &2      &$-$57 &"          &"         &2$\times$1800\\
SMC N66                    &"           &6507--7828             &3.8    &$-$57 &"          &"         &600\\
SMC N66                    &"           &3800--8400             &11     &$-$57 &"          &"         &300, 600\\

\noalign{\vskip3pt} \noalign{\hrule}
\end{tabular}
\end{minipage}
\end{table*}

\section{Observations and data reduction}

The observational dataset consists of long-slit spectra obtained during runs at
the European Southern Observatory (ESO) using the 1.52-m telescope and the
3.5-m New Technology Telescope (NTT). Additional long-slit spectroscopy for one
target was performed at the 3.9-m Anglo-Australian Telescope (AAT). The journal
of observations is presented in Table~1.

The Galactic H~{\sc ii} regions M\,17 and NGC\,3576 were observed at ESO with
the B\&C spectrograph on the 1.52-m telescope. The detector was a Loral $2048
\times 2048$, 15$\mu$m $\times$ 15$\mu$m CCD in July 1996 and in February 1997.
A 2~arcsec wide, 3.5 arcmin long slit was employed. The CCDs were binned by a
factor of two along the slit direction, in order to reduce the read-out noise.
The spatial sampling was 1.63 arcsec per pixel projected on the sky. Two
wavelength regions of M\,17 were observed in July 1996: a 2400 lines mm$^{-1}$
holographic grating was used in first order to cover the 3995--4978\,\AA~range
at a spectral resolution of 1.5\,\AA~(FWHM); a second grating in first order,
along with a WG345 order sorting filter, was used to cover the
3535--7400\,\AA~range at a resolution of 4.5\,\AA. The shortest integration
time was chosen so as to ensure that strong emission lines like H$\alpha$ and
[O~{\sc iii}] $\lambda\lambda$4959, 5007 would not be saturated. NGC\,3576 was
observed in February 1997, in the 3995--4978~\AA~range only, at a resolution of
1.5\,\AA. Additional spectra of NGC\,3576 were taken at the AAT with the RGO
spectrograph and a TEK $1024 \times 1024$, 24$\mu$m $\times$ 24$\mu$m CCD. A
1200 lines mm$^{-1}$ grating was used in second order with two settings to
cover the 3509--3908, 3908--4305~\AA~ranges, at a resolution of 1\,\AA, while a
250 lines mm$^{-1}$ grating was used to cover the 3655--7360\,\AA~range in
first order at a resolution of 8.5\,\AA. The CCD was again binned by a factor
of two along the slit direction, yielding a plate scale of 1.54 arcsec per
pixel.

The Magellanic Cloud H~{\sc ii} regions 30~Doradus, LMC~N11B and SMC~N66 were
observed with the NTT 3.5-m telescope in December 1995. The ESO Multi Mode
Instrument was used in the following modes: red imaging and low dispersion
grism spectroscopy (RILD) and dichroic medium dispersion spectroscopy (DIMD).
The detector was a TEK $1024 \times 1024$, 24$\mu$m $\times$ 24$\mu$m CCD (no.
31), used for blue-arm DIMD observations, and a TEK $2048 \times 2048$,
24$\mu$m $\times$ 24$\mu$m CCD (no. 36), used for red-arm DIMD and RILD
observations. Both cameras were in use when observing in DIMD mode. In this
case, a dichroic prism was inserted into the beam path so that light was
directed to the blue and red grating units in synchronization, allowing
simultaneous exposures to be obtained in the blue and red part of the optical
spectrum. For all exposures, both CCDs were binned by a factor of two in both
directions. The spatial sampling was thus 0.74 and 0.54 arcsec per pixel
projected on the sky, for CCDs no. 31 and no. 36, respectively. Five wavelength
regions were observed with two different gratings (\#3, \#7) and a grism (\#3),
yielding spectral resolutions of approximately
2\,\AA~($\lambda\lambda$3635--4145, $\lambda\lambda$4060--4520,
$\lambda\lambda$4515--4975), 3.8\,\AA~($\lambda\lambda$6507--7828), and
11\,\AA~FWHM ($\lambda\lambda$3800--8400), respectively. An OG530 filter was
used when observing in DIMD mode. The slits used were 5.6~arcmin long and 1.0
and 1.5~arcsec wide. Relevant exposure times, position angles and target
coordinates are listed in Table~1.

The two-dimensional spectra were reduced with the {\sc midas} software package,
following standard procedures. They were bias-subtracted, flat-fielded via
division by normalized flat field frames, cosmic-ray cleaned, and then
wavelength calibrated using exposures of He-Ar, Th-Ar and Cu-Ar calibration
lamps. During the 1995 and 1997 runs, twilight sky flat-fields were also
obtained, in order to correct the small variations in illumination along the
slit. The ESO 1.52-m spectra were reduced to absolute intensity units using
wide-slit (8\,arcsec) observations of the {\it HST} standard stars Feige\,110
and the nucleus of the planetary nebula NGC\,7293 (Walsh 1993), as well as the
CTIO standards LTT\,4364 and LTT\,6248 (Hamuy et al. 1994). All NTT spectra
were flux-calibrated using wide-slit (5\,arcsec) observations of Feige\,110.
The AAT spectra were flux-calibrated using observations of the standard star
LTT\,3218. In all cases, flux-calibration was done using the {\sc iraf}
software package. Sky-subtraction was not attempted since in no case could any
nebular emission-free windows be extracted from the long-slit CCD frames.


\begin{figure*}
\begin{center} \epsfig{file=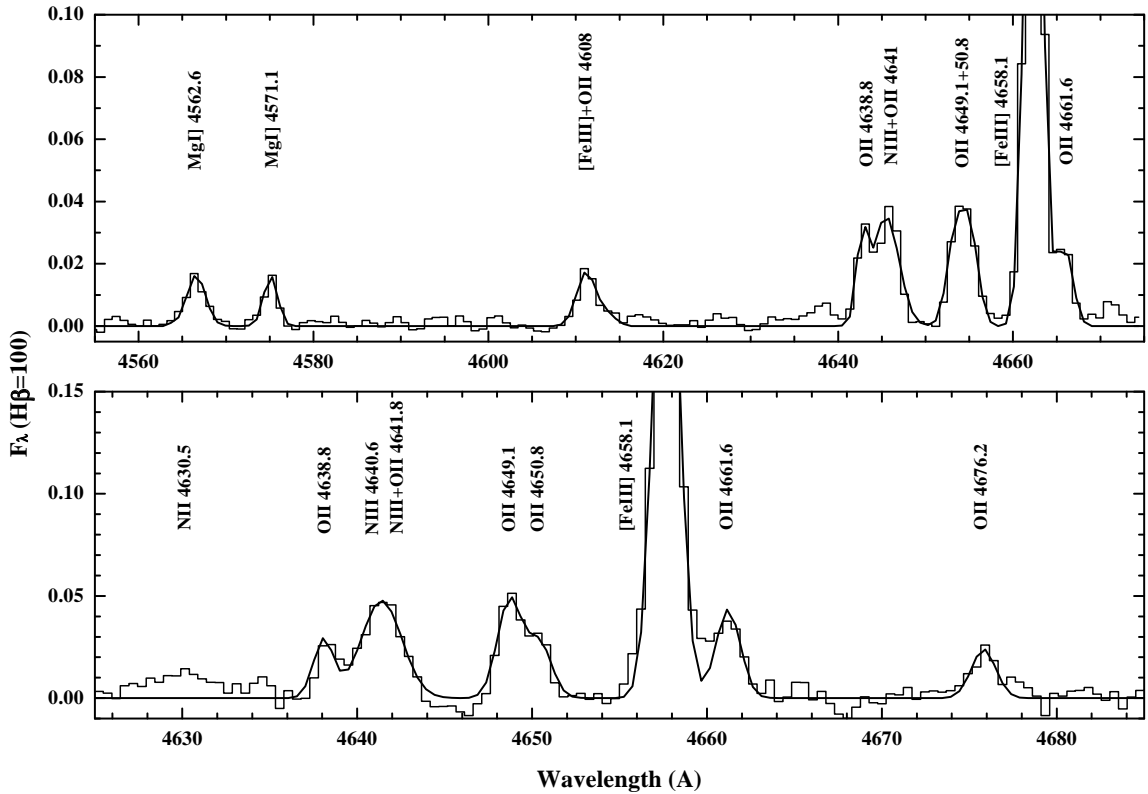, width=15.5 cm, clip=}
\caption {Continuum subtracted spectra of 30~Doradus (\emph{top}; NTT 3.5\,m,
FWHM = 2\,\AA) and NGC\,3576 (\emph{bottom}; ESO 1.52\,m, FWHM = 1.5\,\AA)
showing the {\oii} V\,1 multiplet recombination lines; interstellar extinction
has not been corrected for. The thick overplotted line is the sum of multiple
Gaussian fits to the lines. The intensity is scaled to $F$(H$\beta$) = 100.}
\end{center}
\end{figure*}

Most of the line fluxes -- and certainly all those of heavy element
recombination lines -- were derived using Gaussian line profile fitting
techniques, apart from the strongest ones, for which their fluxes were measured
by simply integrating over the line profiles. In order to deconvolve features
affected by blending, multiple Gaussian fitting was employed. In such cases a
successful estimate of the continuum emission level was deemed to be an
important first step. After subtracting the local continuum, line fluxes were
retrieved by fitting multiple Gaussians of appropriate central wavelength and
usually equal FWHM. The FWHM was taken to be the same as that of nearby
unblended lines of similar strength to the ones fitted. Their relative
wavelength spacings were constrained to be the same as those from laboratory
wavelengths. This procedure assured accurate flux retrieval and aided line
identification in the case of ambiguous features. In Fig.\,~1 we present
high-resolution, continuum-subtracted spectra of 30~Doradus and NGC\,3576
covering the spectral region around the {\oii} V\,1 multiplet at 4650\,\AA,
along with multiple Gaussian fits to the observed emission lines.

A complete list of observed emission lines and their fluxes for  all nebulae
can be found in Table~2. All fluxes are on a scale where $F$(\Hb)~=~100, with
the dereddened flux given by,
\[
I(\lambda) = 10^{c({\rm H}\beta) f(\lambda)}\,F(\lambda).
\]
The amount of interstellar extinction is given by \emph{c}(H$\beta$) which is
the logarithm of the ratio of the dereddened and observed H$\beta$ fluxes,
while \emph{f}($\lambda$) is the adopted extinction curve in each case (see
below) normalized such that \emph{f}(H$\beta$)~=~0. A ratio of total to
selective extinction, $R$~=~$A_{\rm V}/E(B\,-\,V)$~=~3.1 was assumed. All
dereddened line intensities quoted in the remainder of this paper are on a
scale where $I$(\Hb)~=~100.

\setcounter{table}{1}
\begin{table}
\begin{minipage}{75mm}
\centering \caption{Observed and dereddened relative line fluxes
(F($\lambda$) and I($\lambda$), respectively), on a scale where
H$\beta~=~100$.}
\begin{tabular}{l@{\hspace{2.8mm}}l@{\hspace{2.8mm}}l@{\hspace{2.8mm}}c@{\hspace{2.8mm}}c@{\hspace{2.8mm}}c@{\hspace{2.8mm}}}
\noalign{\hrule} \noalign{\vskip3pt}
\multicolumn{6}{c}{M\,17}\\
$\lambda_{\rm obs}$&$F(\lambda)$&$I(\lambda)$&ID&$\lambda_{\rm 0}$&Mult  \\
\noalign{\vskip3pt} \noalign{\hrule} \noalign{\vskip3pt}
       *&         *&         *&H 14     &3721.94&H14    \\
       *&         *&         *&[S III]  &3721.63&F2     \\
 3726.99&3.11($+$1)&9.24($+$1)&[O II]   &3726.03&F1     \\
       *&       *  &         *&[O II]   &3728.82&F1     \\
 3749.86&1.13      &3.32      &H 12     &3750.15&H12    \\
 3770.34&1.31      &3.78      &H 11     &3770.63&H11    \\
 3797.30&1.61      &4.53      &H 10     &3797.90&H10    \\
 3819.01&4.13($-$1)&1.14      &He I     &3819.62&V22    \\
 3834.78&2.41      &6.59      &H 9      &3835.39&H9     \\
 3868.84&6.91      &1.83($+$1)&[Ne III] &3868.75&F1     \\
 3889.14&6.23      &1.62($+$1)&He I     &3888.65&V2     \\
 3967.56&8.24      &2.01($+$1)&[Ne III] &3967.46&F1     \\
 4009.31&1.26($-$1)&2.97($-$1)&He I     &4009.26&V55    \\
 4026.29&9.73($-$1)&2.25      &He I     &4026.21&V18    \\
 4069.00&2.60($-$1)&5.79($-$1)&[S II]   &4068.60&F1     \\
 4076.23&7.05($-$2)&1.56($-$1)&[S II]   &4076.35&F1     \\
 4101.76&1.19($+$1)&2.57($+$1)&H 6      &4101.74&H6     \\
 4119.71&5.44($-$2)&1.15($-$1)&O II     &4119.22&V20    \\
 4121.33&9.00($-$2)&1.91($-$1)&He I     &4120.84&V16    \\
 4143.87&1.35($-$1)&2.80($-$1)&He I     &4143.76&V53    \\
 4267.19&2.61($-$1)&4.81($-$1)&C II     &4267.15&V6     \\
 4317.27&6.36($-$2)&1.12($-$1)&O II     &4317.14&V2     \\
 4319.48&1.89($-$2)&3.30($-$2)&O II     &4319.63&V2      \\
 4340.43&2.79($+$1)&4.77($+$1)&H 5      &4340.47&H5      \\
       *&3.93($-$2)&6.70($-$2)&O II     &4345.56&V2      \\
 4349.01&6.30($-$2)&1.07($-$1)&O II     &4349.43&V2      \\
 4363.22&6.11($-$1)&1.02      &[O III]  &4363.21&F2      \\
 4388.02&4.06($-$1)&6.63($-$1)&He I     &4387.93&V51     \\
 4437.83&5.37($-$2)&8.33($-$2)&He I     &4437.55&V50     \\
 4471.49&3.28      &4.91      &He I     &4471.50&V14     \\
 4607.46&4.56($-$2)&5.95($-$2)&N II     &4607.16&V5      \\
 4630.76&5.22($-$2)&6.02($-$2)&N II     &4630.54&V5      \\
 4634.26&3.42($-$2)&4.34($-$2)&N III    &4634.14&V2      \\
 4638.98&8.36($-$2)&1.06($-$1)&O II     &4638.49&V1      \\
 4640.76&2.77($-$2)&3.48($-$2)&N III    &4640.64&V2      \\
 4641.93&1.28($-$1)&1.52($-$1)&O II     &4641.44&V1      \\
 4649.38&1.10($-$1)&1.38($-$1)&O II     &4649.13&V1      \\
 4651.09&6.99($-$2)&8.71($-$2)&O II     &4650.84&V1      \\
 4658.49&2.25($-$1)&2.79($-$1)&[Fe III] &4658.10&F3      \\
 4662.02&8.47($-$2)&1.04($-$1)&O II     &4661.63&V1      \\
 4676.91&5.19($-$2)&6.30($-$2)&O II     &4676.24&V1      \\
 4711.60&7.01($-$2)&8.20($-$2)&[Ar IV]  &4711.37&F1      \\
 4713.40&4.51($-$1)&5.26($-$1)&He I     &4713.17&V12     \\
 4740.37&6.20($-$2)&7.04($-$2)&[Ar IV]  &4740.17&F1      \\
 4803.20&2.85($-$2)&3.04($-$2)&N II     &4803.29&V20     \\
 4861.55&1.00($+$2)&1.00($+$2)&H 4      &4861.33&H4      \\
 4881.20&1.04($-$1)&1.02($-$1)&[Fe III] &4881.11&F2      \\
 4906.87&5.42($-$2)&5.17($-$2)&O II     &4906.83&V28     \\
 4922.27&1.42      &1.34      &He I     &4921.93&V48     \\
 4931.47&4.49($-$2)&4.18($-$2)&[O III]  &4931.80&F1      \\
 4959.13&1.30($+$2)&1.17($+$2)&[O III]  &4958.91&F1      \\
 5006.43&4.13($+$2)&3.55($+$2)&[O III]  &5006.84&F1      \\
 5198.36&4.04($-$1)&2.84($-$1)&[N I]    &5199.84&F1      \\
 5517.48&8.73($-$1)&4.54($-$1)&[Cl III] &5517.66&F1      \\
 5537.71&7.75($-$1)&3.97($-$1)&[Cl III] &5537.60&F1      \\
 5666.60&1.23($-$1)&5.70($-$2)&N II     &5666.53&V3   \\
 5676.11&3.68($-$2)&1.70($-$2)&N II     &5676.02&V3   \\
 5679.62&1.18($-1$)&5.44($-$2)&N II     &5679.56&V3   \\
 5754.74&7.40($-$1)&3.24($-$1)&[N II]   &5754.60&F3      \\
 5875.44&3.61($+$1)&1.45($+$1)&He I     &5875.66&V11     \\
 6312.09&4.77      &1.43      &[S III]  &6312.10&F3      \\
 6548.71&3.52($+$1)&9.15      &[N II]   &6548.10&F1      \\
 6563.11&1.17($+$3)&3.01($+$2)&H 3      &6562.77&H3       \\
\end{tabular}
\end{minipage}
\end{table}

\setcounter{table}{1}
\begin{table}
\begin{minipage}{75mm}
\centering \caption{{\it --continued}}
\begin{tabular}{l@{\hspace{2.8mm}}l@{\hspace{2.8mm}}l@{\hspace{2.8mm}}c@{\hspace{2.8mm}}c@{\hspace{2.8mm}}c@{\hspace{2.8mm}}}
\noalign{\hrule} \noalign{\vskip3pt}
\multicolumn{6}{c}{M\,17}\\
$\lambda_{\rm obs}$&$F(\lambda)$&$I(\lambda)$&ID&$\lambda_{\rm 0}$&Mult  \\
\noalign{\vskip3pt} \noalign{\hrule} \noalign{\vskip3pt}
 6583.91&1.08($+$2)&2.74($+$1)&[N II]   &6583.50&F1       \\
 6678.59&1.62($+$1)&3.89      &He I     &6678.16&V46    \\
 6717.02&1.72($+$1)&4.04      &[S II]   &6716.44&F2     \\
 6731.39&1.76($+$1)&4.11      &[S II]   &6730.82&F2     \\
 7065.48&1.81($+$1)&3.51      &He I     &7065.25&V10    \\
 7135.90&5.74($+$1)&1.07($+$1)&[Ar III] &7135.80&F1     \\
 7280.02&5.49      &9.50($-$1)&He I     &7281.35&V45    \\
 7319.06&6.61      &1.12      &[O II]   &7318.92&F2     \\
 7329.62&5.17      &8.72($-$1)&[O II]   &7329.67&F2     \\
\noalign{\vskip3pt} \noalign{\hrule} \noalign{\vskip3pt}
\multicolumn{6}{c}{NGC\,3576 (AAT)}\\
$\lambda_{\rm obs}$&$F(\lambda)$&$I(\lambda)$&ID&$\lambda_{\rm 0}$&Mult  \\
\noalign{\vskip3pt} \noalign{\hrule} \noalign{\vskip3pt}
 3531.25 &3.98($-$1)&9.70($-$1)&He I     &3530.50&V36    \\
 3553.94 &1.55($-$1)&3.71($-$1)&He I     &3554.42&V34     \\
 3566.73 &4.25($-$1)&1.00      &?        &3566.95&       \\
 3613.22 &2.17($-$1)&4.92($-$1)&He I     &3613.64&V6     \\
 3634.00 &2.30($-$1)&5.13($-$1)&He I     &3634.25&V28    \\
 3671.17 &2.34($-$1)&5.13($-$1)&H 24     &3671.48&H24    \\
 3673.39 &2.24($-$1)&4.90($-$1)&H 23     &3673.74&H23    \\
 3676.01 &2.80($-$1)&6.11($-$1)&H 22     &3676.36&H22    \\
 3679.01 &2.87($-$1)&6.26($-$1)&H 21     &3679.36&H21     \\
 3682.46 &3.17($-$1)&6.92($-$1)&H 20     &3682.81&H20     \\
 3686.48 &3.70($-$1)&8.04($-$1)&H 19     &3686.83&H19     \\
 3691.21 &4.42($-$1)&9.59($-$1)&H 18     &3691.56&H18     \\
 3696.80 &4.91($-$1)&1.06      &H 17     &3697.15&H17     \\
 3703.50 &6.09($-$1)&1.31      &H 16     &3703.86&H16     \\
 3704.66 &3.01($-$1)&6.49($-$1)&He I     &3705.02&V25     \\
 3711.61 &7.40($-$1)&1.58      &H 15     &3711.97&H15     \\
 3721.49 &1.31      &2.79      &H 14     &3721.94&H14     \\
 3725.66 &3.52($+$1)&7.51($+$1)&[O II]   &3726.03&F1      \\
 3728.45 &2.58($+$1)&5.49($+$1)&[O II]   &3728.82&F1      \\
 3733.99 &1.05      &2.23      &H 13     &3734.37&H13     \\
 3749.77 &1.38      &2.90      &H 12     &3750.15&H12     \\
 3770.25 &1.80      &3.76      &H 11     &3770.63&H11     \\
 3797.52 &2.51      &5.16      &H 10     &3797.90&H10     \\
 3819.21 &5.86($-$1)&1.18      &He I     &3819.62&V22     \\
 3835.00 &3.75      &7.54      &H 9      &3835.39&H9      \\
 3855.69 &1.22($-$1)&2.44($-$1)&Si II    &3856.02&V1      \\
 3862.27 &1.43($-$1)&2.83($-$1)&Si II    &3862.60&V1      \\
 3868.34 &1.04($+$1)&2.06($+$1)&[Ne III] &3868.75&F1      \\
 3871.33 &8.97($-$2)&1.76($-$1)&He I     &3871.82&V60     \\
 3888.64 &8.45      &1.64($+$1)&He I     &3888.65&V2      \\
 3912.83 &3.95($-$2)&7.59($-$2)&?        &3912.83&        \\
 3914.02 &3.39($-$2)&6.52($-$2)&?        &3914.02&        \\
 3918.66 &5.39($-$2)&1.03($-$1)&C II     &3918.98&V4      \\
 3920.37 &4.83($-$2)&9.24($-$2)&C II     &3920.69&V4      \\
 3926.03 &8.01($-$2)&1.52($-$1)&He I     &3926.54&V58     \\
 3964.36 &4.81($-$1)&8.97($-$1)&He I     &3964.73&V5      \\
 3967.04 &3.20      &5.95      &[Ne III] &3967.46&F1      \\
 3969.70 &8.03      &1.49($+$1)&H 7      &3970.07&H7      \\
 4008.84 &1.27($-$1)&2.31($-$1)&He I     &4009.26&V55     \\
 4025.83 &1.11      &2.00      &He I     &4026.21&V18     \\
 4037.19 &1.24($-$2)&2.22($-$2)&?        &4037.19&        \\
 4042.91 &2.56($-$2)&4.54($-$2)&?        &4042.91&        \\
 4059.17 &1.37($-$2)&2.41($-$2)&?        &4059.17&        \\
 4060.28 &9.18($-$3)&1.61($-$2)&[F IV]   &4060.23&F1      \\
 4065.35 &2.37($-$2)&4.14($-$2)&?        &4065.35&        \\
 4068.21 &7.35($-$1)&1.28      &[S II]   &4068.60&F1       \\
 4069.37 &8.08($-$2)&1.41($-$1)&O II     &4069.62&V10      \\
       * &      *      &     * &O II     &4069.89&V10         \\
 4071.77 &6.93($-$2)&1.21($-$1)&O II     &4072.16&V10      \\
 4075.47 &5.81($-$2)&1.01($-$1)&O II     &4075.86&V10     \\
 4075.96 &2.53($-$1)&4.40($-$1)&[S II]   &4076.35&F1      \\
 4080.81 &1.95($-$2)&3.38($-$2)&?        &4080.81&        \\

\end{tabular}
\end{minipage}
\end{table}

\setcounter{table}{1}
\begin{table}
\begin{minipage}{75mm}
\centering \caption{{\it --continued}}
\begin{tabular}{l@{\hspace{2.8mm}}l@{\hspace{2.8mm}}l@{\hspace{2.8mm}}c@{\hspace{2.8mm}}c@{\hspace{2.8mm}}c@{\hspace{2.8mm}}}
\noalign{\hrule} \noalign{\vskip3pt}
\multicolumn{6}{c}{NGC\,3576 (AAT)}\\
$\lambda_{\rm obs}$&$F(\lambda)$&$I(\lambda)$&ID&$\lambda_{\rm 0}$&Mult  \\
\noalign{\vskip3pt} \noalign{\hrule} \noalign{\vskip3pt}
 4084.90 &9.84($-$3)&1.70($-$2)&O II     &4085.11&V10     \\
 4086.93 &2.40($-$2)&4.15($-$2)&O II     &4087.15&V48c    \\
 4089.07 &1.85($-$2)&3.20($-$2)&O II     &4089.29&V48a    \\
 4096.94 &3.79($-$2)&6.50($-$2)&N III    &4097.33&V1       \\
 4101.33 &1.50($+$1)&2.56($+$1)&H 6      &4101.74&H6      \\
 4118.83 &1.61($-$2)&2.71($-$2)&O II     &4119.22&V20      \\
 4120.45 &1.26($-$1)&2.13($-$1)&He I     &4120.84&V16      \\
 4132.64 &3.12($-$2)&5.23($-$2)&O II     &4132.80&V19      \\
 4143.33 &1.85($-$1)&3.07($-$1)&He I     &4143.76&V53      \\
 4152.88 &2.18($-$2)&3.61($-$2)&O II     &4153.30&V19      \\
 4156.11 &1.26($-$2)&2.08($-$2)&O II     &4156.53&V19      \\
 4168.62 &4.03($-$2)&6.58($-$2)&He I     &4168.97&V52      \\
 4170.28 &1.94($-$2)&3.16($-$2)&?        &4170.28&         \\
 4185.44 &2.04($-$2)&3.30($-$2)&O II     &4185.45&V36      \\
 4189.78 &1.95($-$2)&3.14($-$2)&O II     &4189.79&V36      \\
 4253.66 &2.72($-$2)&4.20($-$2)&O II     &4254.00&V101     \\
 4266.73 &2.12($-$1)&3.24($-$1)&C II     &4267.15&V6       \\
 4276.78 &8.81($-$2)&1.34($-$1)&O II     &4275.55&V67a     \\
 4287.00 &7.39($-$2)&1.11($-$1)&[Fe II]  &4287.40&7F       \\
 4340.04 &3.27($+$1)&4.76($+$1)&H 5      &4340.47&H5       \\
 4345.30 &7.48($-$2)&1.08($-$1)&O II     &4345.56&V2       \\
 4349.17 &5.38($-$2)&7.77($-$2)&O II     &4349.43&V2       \\
 4357.14 &5.26($-$2)&7.56($-$2)&O II     &4357.25&V63a     \\
 4358.98 &8.28($-$2)&1.18($-$1)&?        &4358.98&         \\
 4362.80 &1.05      &1.51      &[O III]  &4363.21&F2       \\
 4368.06 &7.49($-$2)&1.06($-$1)&C II?    &4368.14&V45      \\
 4387.49 &3.91($-$1)&5.50($-$1)&He I     &4387.93&V51      \\
 4413.23 &7.58($-$2)&1.04($-$1)&[Fe II]  &4413.78&7F       \\
 4471.07 &3.66      &4.84      &He I     &4471.50&V14      \\
 4603.76 &3.33($-$2)&4.01($-$2)&?        &4603.76&         \\
 4606.98 &4.78($-$2)&5.75($-$2)&N II     &4607.16&V5       \\
 4608.66 &4.45($-$2)&5.34($-$2)&?        &4608.66&         \\
 4610.02 &4.50($-$2)&5.40($-$2)&O II     &4610.20&V92c     \\
 4614.92 &4.04($-$2)&4.83($-$2)&?        &4614.92&         \\
 4620.88 &3.82($-$2)&4.55($-$2)&N II     &4621.39&V5       \\
 4638.46 &5.89($-$2)&6.92($-$2)&O II     &4638.86&V1       \\
 4640.24 &2.19($-$2)&2.57($-$2)&N III    &4640.64&V2        \\
 4641.42 &9.22($-$2)&1.08($-$1)&O II     &4641.81&V1       \\
 4648.60 &1.38($-$1)&1.60($-$1)&O II     &4649.13&V1       \\
 4650.31 &1.31($-$1)&1.52($-$1)&O II     &4650.84&V1       \\
 4661.23 &7.46($-$2)&8.61($-$2)&O II     &4661.63&V1       \\
 4675.84 &5.12($-$2)&5.85($-$2)&O II     &4676.24&V1       \\
 4696.03 &2.80($-$2)&3.16($-$2)&O II     &4696.35&V1        \\
 4701.16 &1.92($-$1)&2.16($-$1)&[Fe III] &4701.62&F3        \\
 4705.72 &3.76($-$2)&4.21($-$2)&O II     &4705.35&V25       \\
 4861.50 &1.00($+$2)&1.00($+$2)&H 4      &4861.33&H4        \\
 4958.86 &1.31($+$2)&1.22($+$2)&[O III]  &4958.91&F1        \\
 5006.82 &4.12($+$2)&3.71($+$2)&[O III]  &5006.84&F1        \\
 5200.35 &8.17($-$1)&6.40($-$1)&[N I]    &5199.84&F1        \\
 5270.91 &4.51($-$1)&3.36($-$1)&[Fe III] &5270.40&1F        \\
 5518.20 &7.41($-$1)&4.70($-$1)&[Cl III] &5517.66&F1        \\
 5538.14 &8.49($-$1)&5.33($-$1)&[Cl III] &5537.60&F1        \\
 5577.88 &4.18      &2.57      &[O I]    &5577.34&F3        \\
5666.60  &6.00($-$2)&3.58($-$2)&N II     &5666.53&V3   \\
5676.11  &2.67($-$1)&1.59($-$2)&N II     &5676.02&V3   \\
5679.60  &1.12($-$1)&6.70($-$2)&N II     &5679.56&V3   \\
 5754.86 &1.05      &5.92($-$1)&[N II]   &5754.60&F3        \\
 5875.76 &2.59($+$1)&1.38($+$1)&He I     &5875.66&V11       \\
 5913.00 &6.56($-$2)&3.42($-$2)&?        &5913.00&          \\
 5931.52 &2.11($-$1)&1.09($-$1)&N II     &5931.52&V28       \\
 5957.35 &2.87($-$1)&1.46($-$1)&Si II    &5957.35&          \\
 5978.71 &1.99($-$1)&1.01($-$1)&Si II    &5978.71&          \\
 6548.11 &4.03($+$1)&1.58($+$1 &[N II]   &6548.10&F1            \\
 6562.97 &7.40($+$2)&2.88($+$2)&H 3      &6562.77&H3        \\
\end{tabular}
\end{minipage}
\end{table}

\setcounter{table}{1}
\begin{table}
\begin{minipage}{75mm}
\centering \caption{{\it --continued}}
\begin{tabular}{l@{\hspace{2.8mm}}l@{\hspace{2.8mm}}l@{\hspace{2.8mm}}c@{\hspace{2.8mm}}c@{\hspace{2.8mm}}c@{\hspace{2.8mm}}}
\noalign{\hrule} \noalign{\vskip3pt}
\multicolumn{6}{c}{NGC\,3576 (AAT)}\\
$\lambda_{\rm obs}$&$F(\lambda)$&$I(\lambda)$&ID&$\lambda_{\rm 0}$&Mult  \\
\noalign{\vskip3pt} \noalign{\hrule} \noalign{\vskip3pt}

 6583.66 &1.36($+$2)&5.25($+$1)&[N II]   &6583.50&F1            \\
 6678.74 &9.91      &3.68      &He I     &6678.16&V46       \\
 6717.02 &1.90($+$1)&6.94      &[S II]   &6716.44&F2        \\
 6731.40 &2.41($+$1)&8.76      &[S II]   &6730.82&F2        \\
 7065.67 &1.70($+$1)&5.45      &He I     &7065.25&V10       \\
 7136.12 &4.87($+$1)&1.51($+$1)&[Ar III] &7135.80&F1        \\
\noalign{\vskip3pt} \noalign{\hrule} \noalign{\vskip3pt}

\multicolumn{6}{c}{NGC\,3576 (ESO)}\\
$\lambda_{\rm obs}$&$F(\lambda)$&$I(\lambda)$&ID&$\lambda_{\rm 0}$&Mult  \\
\noalign{\vskip3pt} \noalign{\hrule} \noalign{\vskip3pt}
 4008.73 &1.41($-$1)&2.52($-$1)&He I     &4009.26&V55        \\
 4025.77 &1.07      &1.89      &He I     &4026.21&V18        \\
 4068.12 &8.95($-$1)&1.54      &[S II]   &4068.60&F1         \\
 4075.73 &3.30($-$1)&5.65($-$1)&[S II]   &4076.35&F1         \\
 4101.16 &1.50($+$1)&2.53($+$1)&H 6      &4101.74&H6         \\
 4120.36 &1.26($-$1)&2.10($-$1)&He I     &4120.84&V16        \\
 4143.17 &1.04($-$1)&1.71($-$1)&He I     &4143.76&V53        \\
 4266.58 &2.06($-$1)&3.12($-$1)&C II     &4267.15&V6         \\
 4276.29 &4.80($-$2)&7.21($-$2)&O II     &4275.99&V67b       \\
 4286.98 &7.73($-$2)&1.15($-$1)&O II     &4285.69&V78b       \\
 4316.94 &1.23($-$1)&1.80($-$1)&O II     &4317.14&V2         \\
 4339.87 &3.25($+$1)&4.68($+$1)&H 5      &4340.47&H5         \\
 4349.02 &3.38($-$2)&4.84($-$2)&O II     &4349.43&V2         \\
 4362.64 &8.61($-$1)&1.22      &[O III]  &4363.21&F2         \\
 4367.45 &8.59($-$2)&1.21($-$1)&O II     &4366.89&V2         \\
 4387.39 &3.18($-$1)&4.44($-$1)&He I     &4387.93&V51        \\
 4412.28 &5.20($-$2)&7.13($-$2)&Ne II    &4413.22&V65        \\
 4413.96 &4.55($-$2)&6.22($-$2)&O II     &4414.90&V5         \\
 4416.03 &5.81($-$2)&7.94($-$2)&O II     &4416.97&V5         \\
 4437.45 &5.83($-$2)&7.84($-$2)&He I     &4437.55&V50        \\
 4449.91 &2.49($-$2)&3.33($-$2)&?        &4449.91&           \\
 4451.61 &4.05($-$2)&5.40($-$2)&O II     &4452.37&V5         \\
 4470.93 &3.24      &4.27      &He I     &4471.50&V14        \\
 4630.03 &2.34($-$2)&3.09($-$2)&N II     &4630.54&V5        \\
 4638.79 &4.34($-$2)&5.09($-$2)&O II     &4638.49&V1         \\
 4640.57 &3.61($-$2)&4.22($-$2)&N III    &4640.64&V2         \\
 4641.74 &7.54($-$2)&8.80($-$2)&O II     &4641.44&V1         \\
 4648.70 &8.00($-$2)&9.29($-$2)&O II     &4649.13&V1         \\
 4650.41 &4.76($-$2)&5.53($-$2)&O II     &4650.84&V1         \\
 4657.68 &5.10($-$1)&5.89($-$1)&[Fe III] &4658.10&F3         \\
 4661.21 &7.95($-$2)&9.16($-$2)&O II     &4661.63&V1         \\
 4675.90 &4.91($-$2)&5.60($-$2)&O II     &4676.24&V1          \\
 4701.14 &1.60($-$1)&1.79($-$1)&[Fe III] &4701.62&F3          \\
 4710.98 &7.82($-$2)&8.70($-$2)&[Ar IV]  &4711.37&F1         \\
 4712.78 &4.43($-$1)&4.91($-$1)&He I     &4713.17&V12         \\
 4733.55 &4.13($-$2)&4.52($-$2)&[Fe III] &4733.93&           \\
 4739.65 &6.92($-$2)&7.54($-$2)&[Ar IV]  &4740.17&F1         \\
 4754.43 &1.01($-$1)&1.09($-$1)&[Fe III] &4754.83&           \\
 4860.93 &1.00($+$2)&1.00($+$2)&H 4      &4861.33&H4         \\
 4880.75 &2.26($-$1)&2.23($-$1)&[Fe III] &4881.11&F2         \\
 4921.62 &1.12      &1.07      &He I     &4921.93&V48        \\
 4924.22 &6.68($-$2)&6.39($-$2)&O II     &4924.53&V28         \\
 4930.97 &8.22($-$2)&7.82($-$2)&[O III]  &4931.80&F1          \\
 4958.53 &1.25($+$2)&1.17($+$2)&[O III]  &4958.91&F1          \\
\noalign{\vskip3pt} \noalign{\hrule} \noalign{\vskip3pt}
\multicolumn{6}{c}{30~Doradus}\\
$\lambda_{\rm obs}$&$F(\lambda)$&$I(\lambda)$&ID&$\lambda_{\rm 0}$&Mult  \\
\noalign{\vskip3pt} \noalign{\hrule} \noalign{\vskip3pt}
 3729.15 &3.80($+$1) &5.02($+$1) &[O II]   &3726.03&F1        \\
 3731.94 &4.21($+$1) &5.54($+$1) &[O II]   &3728.82&F1        \\
 3753.37 &2.69      &3.52      &H 12     &3750.15&H12       \\
 3773.87 &3.29      &4.28      &H 11     &3770.63&H11       \\
 3801.17 &4.35      &5.61      &H 10     &3797.90&H10       \\
 3822.94 &9.27($-$1)&1.18      &He I     &3819.62&V22       \\
 3838.69 &6.48      &8.26      &H 9      &3835.39&H9        \\
 3871.99 &3.07($+$1)&3.87($+$1)&[Ne III] &3868.75&F1        \\
\end{tabular}
\end{minipage}
\end{table}

\setcounter{table}{1}
\begin{table}
\begin{minipage}{75mm}
\centering \caption{{\it --continued}}
\begin{tabular}{l@{\hspace{2.8mm}}l@{\hspace{2.8mm}}l@{\hspace{2.8mm}}c@{\hspace{2.8mm}}c@{\hspace{2.8mm}}c@{\hspace{2.8mm}}}
\noalign{\hrule} \noalign{\vskip3pt}
\multicolumn{6}{c}{30~Doradus}\\
$\lambda_{\rm obs}$&$F(\lambda)$&$I(\lambda)$&ID&$\lambda_{\rm 0}$&Mult  \\
\noalign{\vskip3pt} \noalign{\hrule} \noalign{\vskip3pt}

 3892.31 &1.55($+$1)&1.94($+$1)&He I     &3888.65&V2        \\
 3970.79 &9.37      &1.14($+$1)&[Ne III] &3967.46&F1        \\
 3973.40 &1.40($+$1)&1.71($+$1)&H 7      &3970.07&H7        \\
 4072.50 &4.71($-$1)&5.61($-$1)&[S II]   &4068.60&F1        \\
 4073.66 &1.11($-$1)&1.33($-$1)&O II     &4069.62&V10       \\
 4076.06 &5.73($-$2)&6.83($-$2)&O II     &4072.16&V10       \\
 4080.26 &1.90($-$1)&2.27($-$1)&[S II]   &4076.35&F1        \\
 4082.75 &2.73($-$2)&3.24($-$2)&O II     &4078.84&V10       \\
 4087.98 &9.11($-$3)&1.08($-$2)&O II     &4083.90&V48b      \\
 4089.20 &2.17($-$2)&2.58($-$2)&O II     &4085.11&V10       \\
 4091.24 &1.56($-$2)&1.85($-$2)&O II     &4087.15&V48c      \\
 4093.38 &1.77($-$2)&2.10($-$2)&O II     &4089.29&V48a      \\
 4105.56 &2.20($+$1)&2.60($+$1)&H 6      &4101.74&H6        \\
 4114.32 &2.74($-$2)&3.22($-$2)&O II     &4110.78&V20       \\
 4124.62 &1.77($-$1)&2.08($-$1)&He I     &4120.84&V16       \\
 4136.45 &4.33($-$2)&5.08($-$2)&O II     &4132.80&V19       \\
 4147.48 &2.52($-$1)&2.94($-$1)&He I     &4143.76&V53        \\
 4156.82 &5.22($-$2)&6.09($-$2)&O II     &4153.30&V19       \\
 4160.05 &1.99($-$2)&2.32($-$2)&O II     &4156.53&V19       \\
 4172.75 &5.74($-$2)&6.66($-$2)&O II     &4169.22&V19        \\
 4270.73 &8.11($-$2)&9.19($-$2)&C II     &4267.15&V6         \\
 4279.08 &6.94($-$2)&7.85($-$2)&O II     &4275.55&V67a      \\
 4307.40 &4.62($-$2)&5.19($-$2)&O II     &4303.61&V65a       \\
 4314.99 &2.64($-$2)&2.96($-$2)&O II     &4312.11&V78a      \\
 4315.49 &3.42($-$2)&3.83($-$2)&O II     &4313.44&V78a      \\
 4317.45 &2.46($-$2)&5.33($-$2)&O II     &4315.40&V63c      \\
 4319.19 &6.30($-$2)&7.05($-$2)&O II     &4317.14&V2        \\
 4321.68 &7.83($-$2)&8.76($-$2)&O II     &4319.63&V2        \\
 4344.22 &4.35($+$1)&4.84($+$1)&H 5      &4340.47&H5        \\
 4367.00 &3.03      &3.35      &[O III]  &4363.21&F2         \\
 4391.79 &4.87($-$1)&5.36($-$1)&He I     &4387.93&V51        \\
 4412.55 &2.83($-$2)&3.10($-$2)&Ne II    &4409.30&V55e       \\
 4418.15 &3.03($-$2)&3.32($-$2)&O II     &4414.90&V5         \\
 4420.23 &3.98($-$2)&4.36($-$2)&O II     &4416.97&V5         \\
 4441.41 &5.71($-$2)&6.22($-$2)&He I     &4437.55&V50        \\
 4475.44 &4.03      &4.35      &He I     &4471.50&V14        \\
 4566.56 &4.77($-$2)&5.06($-$2)&Mg I]    &4562.60&           \\
 4575.32 &5.39($-$2)&5.70($-$2)&Mg I]    &4571.10&           \\
 4611.30 &4.82($-$2)&5.06($-$2)&[Fe III] &4607.13&3F         \\
 4613.96 &1.99($-$2)&2.09($-$2)&O II     &4609.44&V92a       \\
 4638.92 &9.71($-$3)&1.01($-$2)&N III    &4634.14&V2         \\
 4643.03 &6.13($-$2)&6.40($-$2)&O II     &4638.49&V1         \\
 4645.99 &8.19($-$2)&8.54($-$2)&O II     &4641.81&V1         \\
 4653.38 &6.16($-$2)&6.42($-$2)&O II     &4649.13&V1         \\
 4655.09 &6.16($-$2)&6.41($-$2)&O II     &4650.84&V1         \\
 4662.27 &6.01($-$1)&6.25($-$1)&[Fe III] &4658.10&3F         \\
 4665.18 &5.44($-$2)&5.65($-$2)&O II     &4661.63&V1         \\
 4671.08 &2.26($-$2)&2.35($-$2)&[Fe III] &4667.00&3F         \\
 4677.91 &1.16($-$2)&1.20($-$2)&O II     &4673.73&V1         \\
 4680.42 &2.15($-$2)&2.24($-$2)&O II     &4676.24&V1         \\
 4689.06 &1.66($-$2)&1.71($-$2)&He II    &4685.68&3.4        \\
 4692.90 &1.09($-$2)&1.12($-$2)&?        &4692.90&           \\
 4696.27 &1.71($-$2)&1.77($-$2)&?        &4696.27&           \\
 4705.74 &1.83($-$1)&1.88($-$1)&[Fe III] &4701.62&3F         \\
 4715.54 &2.04($-$1)&2.10($-$1)&[Ar IV]  &4711.37&F1         \\
 4717.34 &4.67($-$1)&4.80($-$1)&He I     &4713.17&V12        \\
 4738.18 &6.17($-$2)&6.31($-$2)&[Fe III] &4733.93&3F         \\
 4744.40 &1.77($-$1)&1.81($-$1)&[Ar IV]  &4740.17&F1         \\
 4753.57 &1.87($-$2)&1.91($-$2)&?        &      *&           \\
 4758.89 &1.35($-$1)&1.38($-$1)&[Fe III] &4754.83&3F         \\
 4773.70 &6.67($-$2)&6.79($-$2)&[Fe III] &4769.60&3F         \\
 4781.58 &3.16($-$2)&3.21($-$2)&[Fe III] &4777.88&3F          \\
 4800.53 &1.85($-$2)&1.87($-$2)&?        &      *&            \\
 4819.44 &3.05($-$2)&3.07($-$2)&[Fe II]  &4814.55&20F        \\

\end{tabular}
\end{minipage}
\end{table}

\setcounter{table}{1}
\begin{table}
\begin{minipage}{75mm}
\centering \caption{{\it --continued}}
\begin{tabular}{l@{\hspace{2.8mm}}l@{\hspace{2.8mm}}l@{\hspace{2.8mm}}c@{\hspace{2.8mm}}c@{\hspace{2.8mm}}c@{\hspace{2.8mm}}}
\noalign{\hrule} \noalign{\vskip3pt}
\multicolumn{6}{c}{30~Doradus}\\
$\lambda_{\rm obs}$&$F(\lambda)$&$I(\lambda)$&ID&$\lambda_{\rm 0}$&Mult  \\
\noalign{\vskip3pt} \noalign{\hrule} \noalign{\vskip3pt}

 4865.59 &1.00($+$2)&1.00($+$2)&H 4      &4861.33&H4          \\
 4885.14 &1.83($-$1)&1.83($-$1)&[Fe III] &4881.11&2F         \\
 4893.48 &1.72($-$2)&1.71($-$2)&[Fe II]  &4889.63&4F         \\
 4907.18 &3.37($-$2)&3.35($-$2)&[Fe IV]  &4903.50&-F         \\
 4910.38 &3.72($-$2)&3.69($-$2)&O II     &4906.83&V28        \\
 4926.27 &1.16      &1.14      &He I     &4921.93&V48        \\
 4935.46 &7.03($-$2)&6.94($-$2)&[O III]  &4931.80&F1         \\
 4967.44 &1.71($+$2)&1.68($+$2)&[O III]  &4958.91&F1          \\
 5015.56 &5.18($+$2)&5.05($+$2)&[O III]  &5006.84&F1          \\
 5279.65 &2.99($-$1)&2.79($-$1)&[Fe II]  &5273.38&18F         \\
 5315.59 &1.71($-$1)&1.58($-$1)&?        &      *&             \\
 5526.27 &5.28($-$1)&4.76($-$1)&[Cl III] &5517.66&F1            \\
 5546.24 &4.10($-$1)&3.69($-$1)&[Cl III] &5537.60&F1           \\
 5762.29 &2.35($-$1)&2.06($-$1)&[N II]   &5754.60&F3           \\
 5883.20 &1.48($+$1)&1.27($+$1)&He I     &5875.66&V11          \\
 6238.42 &1.75($-$1)&1.45($-$1)&[Ni III] &6231.09&             \\
 6261.66 &1.53($-$1)&1.26($-$1)&[Fe II]  &6254.30&             \\
 6306.34 &1.30      &1.07      &[O I]    &6300.34&F1           \\
 6318.12 &2.37      &1.94      &[S III]  &6312.10&F3            \\
 6562.58 &4.04      &3.22      &[N II]   &6548.10&F1            \\
 6576.15 &3.80($+$2)&3.03($+$2)&H 3      &6562.77&H3            \\
 6595.37 &1.09($+$1)&8.73      &[N II]   &6583.50&F1           \\
 6685.50 &4.64      &3.65      &He I     &6678.16&V46           \\
 6722.67 &7.66      &6.00      &[S II]   &6716.44&F2           \\
 6736.74 &6.94      &5.43      &[S II]   &6730.82&F2           \\
 7074.23 &4.69      &3.55      &He I     &7065.25&V10          \\
 7146.92 &1.80($+$1)&1.35($+$1)&[Ar III] &7135.80&F1           \\
 7166.65 &4.65($-$2)&3.49($-$2)&[Fe II]  &7155.14&14F          \\
 7247.38 &3.26($-$1)&2.43($-$1)&C II     &7231.32&V3           \\
 7252.50 &3.41($-$1)&2.54($-$1)&C II     &7236.42&V3            \\
 7296.23 &9.94($-$1)&7.38($-$1)&He I     &7281.35&V45           \\
 7335.88 &2.80      &2.07      &[O II]   &7318.92&F2            \\
 7346.65 &2.30      &1.70      &[O II]   &7329.67&F2            \\
\noalign{\vskip3pt} \noalign{\hrule} \noalign{\vskip3pt}
\multicolumn{6}{c}{LMC~N11B}\\
$\lambda_{\rm obs}$&$F(\lambda)$&$I(\lambda)$&ID&$\lambda_{\rm 0}$&Mult  \\
\noalign{\vskip3pt} \noalign{\hrule} \noalign{\vskip3pt}
 3729.14 &7.92($+$1) &8.27($+$1) &[O II]  & 3726.03&F1             \\
 3731.94 &1.05($+$2) &1.10($+$2) &[O II]  & 3728.82&F1             \\
 3872.02 &1.97($+$1)&2.05($+$1)&[Ne III]& 3868.75&F1             \\
 3892.03 &2.15($+$1)&2.23($+$1)&He I    & 3888.65&V2             \\
 3970.38 &3.38      &3.50      &[Ne III]& 3967.46&F1             \\
 3973.23 &1.35($+$1)&1.40($+$1)&H 7     & 3970.07&H7             \\
 4072.97 &8.37($-$1)&8.64($-$1)&[S II]  & 4068.60&F1              \\
 4074.13 &3.07($-$1)&3.17($-$1)&O II    & 4069.62&V10             \\
 4076.53 &9.94($-$2)&1.03($-$1)&O II    & 4072.16&V10            \\
 4080.73 &4.89($-$1)&5.05($-$1)&[S II]  & 4076.35&F1              \\
 4083.22 &1.61($-$1)&1.67($-$1)&O II    & 4078.84&V10            \\
 4086.71 &1.36($-$1)&1.41($-$1)&?       & 4086.71&               \\
 4088.71 &1.49($-$1)&1.54($-$1)&O II    & 4083.90&V48b           \\
 4089.92 &1.05($-$1)&1.09($-$1)&O II    & 4085.11&V10            \\
 4091.97 &1.10($-$1)&1.13($-$1)&O II    & 4087.15&V48c           \\
 4094.11 &1.26($-$1)&1.31($-$1)&O II    & 4089.29&V48a           \\
 4106.03 &2.68($+$1)&2.77($+$1)&H 6     & 4101.74&H6              \\
 4114.41 &1.51($-$1)&1.56($-$1)&O II    & 4110.78&V20             \\
 4116.86 &1.44($-$1)&1.49($-$1)&?       & 4116.86&                \\
 4125.03 &2.74($-$1)&2.82($-$1)&He I    & 4120.84&V16              \\
 4135.89 &1.08($-$1)&1.11($-$1)&O II    & 4129.32&V19               \\
 4139.38 &1.32($-$1)&1.36($-$1)&O II    & 4132.80&V19              \\
 4147.96 &3.07($-$1)&3.16($-$1)&He I    & 4143.76&V53              \\
 4157.64 &9.54($-$2)&9.82($-$2)&O II    & 4153.30&V19              \\
 4160.87 &1.24($-$1)&1.28($-$1)&O II    & 4156.53&V19              \\
 4164.34 &2.30($-$1)&2.37($-$1)&?       & 4164.34&                 \\
 4169.09 &2.29($-$1)&2.35($-$1)&?       & 4169.09&                 \\
       * &         * &      *&O II    & 4169.22&V19               \\

\end{tabular}
\end{minipage}
\end{table}

\setcounter{table}{1}
\begin{table}
\begin{minipage}{75mm}
\centering \caption{{\it --continued}}
\begin{tabular}{l@{\hspace{2.8mm}}l@{\hspace{2.8mm}}l@{\hspace{2.8mm}}c@{\hspace{2.8mm}}c@{\hspace{2.8mm}}c@{\hspace{2.8mm}}}
\noalign{\hrule} \noalign{\vskip3pt}
\multicolumn{6}{c}{LMC~N11B}\\
$\lambda_{\rm obs}$&$F(\lambda)$&$I(\lambda)$&ID&$\lambda_{\rm 0}$&Mult  \\
\noalign{\vskip3pt} \noalign{\hrule} \noalign{\vskip3pt}
 4179.89 &8.61($-$2)&8.85($-$2)&N II    & 4176.16&V43a             \\
 4190.33 &8.63($-$2)&8.87($-$2)&C III   & 4186.90&V18               \\
 4193.23 &1.51($-$1)&1.55($-$1)&O II    & 4189.79&V36              \\
 4207.42 &1.29($-$1)&1.32($-$1)&?       & 4207.42&                  \\
 4212.17 &3.90($-$1)&4.00($-$1)&?       & 4212.17&                 \\
 4218.52 &1.15($-$1)&1.18($-$1)&?       & 4218.52&                 \\
 4221.32 &1.75($-$1)&1.79($-$1)&?       & 4221.32&                 \\
 4241.04 &9.68($-$2)&9.93($-$2)&N II    & 4237.05&                 \\
 4245.50 &2.88($-$1)&2.95($-$1)&N II    & 4241.78&V48a             \\
 4252.40 &2.03($-$1)&2.08($-$1)&?       & 4248.15&                 \\
 4257.04 &2.57($-$1)&2.64($-$1)&?       & 4252.73&                  \\
 4263.04 &1.94($-$1)&1.99($-$1)&?       & 4263.04&                  \\
 4265.80 &2.54($-$1)&2.60($-$1)&?       & 4265.80&                  \\
 4268.90 &1.91($-$1)&1.95($-$1)&C II    & 4267.15&V6                \\
 4277.87 &4.16($-$1)&4.26($-$1)&O II    & 4276.28&V67b                \\
 4282.91 &2.61($-$1)&2.67($-$1)&O II    & 4281.32&V53b                \\
 4287.28 &1.07($-$1)&1.09($-$1)&O II    & 4285.69&V78b                \\
 4292.84 &2.73($-$1)&2.79($-$1)&O II    & 4291.25&V55                 \\
 4316.78 &3.79($-$1)&3.87($-$1)&O II    & 4312.11&V78a                \\
 4320.07 &7.82($-$1)&8.00($-$1)&O II    & 4315.40&V63c                \\
 4318.80 &1.31($-$1)&1.34($-$1)&O II    & 4317.14&V2                  \\
 4321.01 &3.34($-$1)&3.41($-$1)&O II    & 4319.63&V2                  \\
 4329.52 &2.86($-$1)&2.92($-$1)&O II    & 4325.76&V2                  \\
 4332.30 &1.89($-$1)&1.93($-$1)&?       & 4329.08&                     \\
 4334.89 &2.82($-$1)&2.88($-$1)&O II    & 4331.13&V65b                 \\
 4344.77 &4.58($+$1)&4.67($+$1)&H 5     & 4340.47&H5                  \\
 4353.52 &1.31($-$1)&8.71($-$2)&O II    & 4349.43&V2                   \\
 4357.68 &1.31($-$1)&1.34($-$1)&O II    & 4353.59&V76c                \\
 4367.56 &1.65      &1.68      &[O III] & 4363.21&F2                  \\
 4392.35 &4.85($-$1)&4.95($-$1)&He I    & 4387.93&V51                 \\
 4412.04 &2.00($-$1)&2.04($-$1)&?       & 4412.00&                    \\
 4419.80 &2.18($-$1)&2.22($-$1)&O II    & 4414.90&V5                  \\
 4475.98 &4.24      &4.31      &He I    & 4471.50&V14                 \\
 4532.10 &4.14($-$2)&4.20($-$2)&?       & 4541.25&                     \\
 4537.53 &8.80($-$2)&8.92($-$2)&?       & 4537.54&                     \\
 4539.73 &5.77($-$2)&5.85($-$2)&N III   & 4534.58&V3                   \\
 4542.95 &7.87($-$2)&7.97($-$2)&?       & 4542.95&                     \\
 4546.93 &7.77($-$2)&7.87($-$2)&?       & 4546.93&                     \\
 4560.15 &1.49($-$1)&1.51($-$1)&[Fe II] & 4555.00&                      \\
 4561.54 &2.59($-$2)&2.63($-$2)&Fe II   & 4556.39&                      \\
 4567.55 &1.92($-$1)&1.95($-$1)&Mg I]   & 4562.60&                      \\
 4576.06 &2.01($-$1)&2.03($-$1)&Mg I]   & 4571.10&                      \\
 4589.35 &5.92($-$2)&5.99($-$2)&?       & 4584.40&                      \\
 4609.38 &4.64($-$2)&4.69($-$2)&?       & 4604.43&                \\
 4643.65 &1.15($-$1)&1.17($-$1)&O II    & 4638.49&V1              \\
 4646.29 &2.73($-$2)&2.75($-$2)&N III   & 4640.64&V2                     \\
 4647.08 &5.26($-$2)&5.30($-$2)&O II     &4641.81&V1         \\
 4654.05 &7.55($-$2)&7.60($-$2)&O II    & 4649.13&V1                     \\
 4655.76 &6.27($-$2)&6.31($-$2)&O II    & 4650.84&V1        \\
 4662.81 &2.69($-$1)&2.72($-$1)&[Fe III]& 4658.10&3F                     \\
 4665.73 &7.05($-$2)&6.40($-$2)&O II    & 4661.63&V1                   \\
 4670.96 &2.54($-$2)&2.56($-$2)&[Fe III]& 4667.00&3F                    \\
 4677.69 &3.05($-$2)&3.08($-$2)&O II    & 4673.73&V1                    \\
 4686.08 &5.00($-$2)&5.04($-$2)&?       & 4681.44&                      \\
 4695.95 &8.14($-$2)&8.20($-$2)&?       & 4691.31&                      \\
 4700.84 &3.43($-$2)&3.46($-$2)&O II    & 4696.35&V1                    \\
 4706.12 &9.02($-$2)&9.08($-$2)&[Fe III]& 4701.62&3F                     \\
 4711.25 &2.60($-$2)&2.61($-$2)&?       & 4706.60&                       \\
 4717.81 &5.06($-$1)&5.09($-$1)&He I    & 4713.17&V12                    \\
 4724.58 &6.19($-$2)&6.23($-$2)&?       & 4719.94&                       \\
 4732.73 &1.83($-$2)&1.84($-$2)&?       & 4728.09&                      \\
 4744.92 &1.12($-$1)&1.13($-$1)&[Ar IV] & 4740.17&F1                    \\
 4749.76 &2.41($-$2)&2.43($-$2)&[Fe II] & 4745.48&                      \\
 4759.60 &8.73($-$2)&8.77($-$2)&[Fe III]& 4754.83&3F                    \\

\end{tabular}
\end{minipage}
\end{table}

\setcounter{table}{1}
\begin{table}
\begin{minipage}{75mm}
\centering \caption{{\it --continued}}
\begin{tabular}{l@{\hspace{2.8mm}}l@{\hspace{2.8mm}}l@{\hspace{2.8mm}}c@{\hspace{2.8mm}}c@{\hspace{2.8mm}}c@{\hspace{2.8mm}}}
\noalign{\hrule} \noalign{\vskip3pt}
\multicolumn{6}{c}{LMC~N11B}\\
$\lambda_{\rm obs}$&$F(\lambda)$&$I(\lambda)$&ID&$\lambda_{\rm 0}$&Mult  \\
\noalign{\vskip3pt} \noalign{\hrule} \noalign{\vskip3pt}
 4774.42 &3.93($-$2)&3.94($-$2)&[Fe III]& 4769.60&3F                    \\
 4779.56 &2.61($-$2)&2.62($-$2)&[Fe II] & 4774.74&20F                   \\
 4795.55 &7.17($-$2)&7.19($-$2)&?       &       *&                      \\
 4803.51 &2.18($-$2)&2.18($-$2)&?       & 4798.87&                      \\
 4815.88 &6.52($-$2)&6.53($-$2)&?       & 4815.88&                      \\
 4819.93 &9.96($-$2)&9.98($-$2)&[Fe II] & 4814.55&20F                    \\
 4865.95 &1.00($+$2)&1.00($+$2)&H 4     & 4861.33&H4                     \\
 4885.52 &5.33($-$2)&5.32($-$2)&[Fe III]& 4881.11&2F                    \\
 4905.47 &2.92($-$2)&2.91($-$2)&[Fe IV]?& 4900.50&-F                     \\
 4926.64 &1.08      &1.07      &He I    & 4921.93&V48                \\
 4929.24 &1.26($-$1)&1.26($-$1)&O II    & 4924.53&V28                   \\
 4935.89 &7.44($-$2)&7.42($-$2)&[O III] & 4931.80&F1                  \\
 4968.01 &1.10($+$2)&1.09($+$2)&[O III] & 4958.91&F1                  \\
 5016.07 &3.32($+$2)&3.30($+$2)&[O III] & 5006.84&F1                    \\
 5052.34 &1.32($-$1)&1.31($-$1)&?       & 5052.34&                      \\
 5060.59 &4.73($-$1)&4.69($-$1)&?       & 5060.59&                       \\
 5526.40 &5.07($-$1)&4.94($-$1)&[Cl III]& 5517.66&F1                     \\
 5546.37 &4.94($-$1)&4.81($-$1)&[Cl III]& 5537.60&F1                     \\
 5752.46 &1.23($-$1)&1.19($-$1)&?       & 5752.46&                        \\
 5763.87 &2.09($-$1)&2.02($-$1)&[N II]  & 5754.60&F3                       \\
 5883.90 &1.52($+$1)&1.47($+$1)&He I    & 5875.66&V11                     \\
 6564.36 &6.11      &5.79      &[N II]  & 6548.10&F1                      \\
 6578.06 &3.34($+$2)&3.16($+$2)&H 3     & 6562.77&H3                      \\
 6597.40 &1.75($+$1)&1.65($+$1)&[N II]  & 6583.50&F1                      \\
 6687.50 &3.89      &3.68      &He I    & 6678.16&V46                     \\
 6724.72 &1.50($+$1)&1.42($+$1)&[S II]  & 6716.44&F2                      \\
 6738.76 &1.11($+$1)&1.05($+$1)&[S II]  & 6730.82&F2                       \\
 7076.35 &2.48      &2.32      &He I    & 7065.25&V10                      \\
 7149.04 &1.23($+$1)&1.15($+$1)&[Ar III]& 7135.80&F1                       \\
 7158.08 &2.67($-$1)&2.49($-$1)&?       &       *&                        \\
 7294.45 &1.83      &1.71      &He I    & 7281.35&V45                      \\
 7338.55 &4.60      &4.28      &[O II]  & 7318.92&F2                      \\
 7350.41 &6.26      &5.84      &[O II]  & 7329.67&F2                      \\
\noalign{\vskip3pt} \noalign{\hrule} \noalign{\vskip3pt}
\multicolumn{6}{c}{SMC~N66}\\
$\lambda_{\rm obs}$&$F(\lambda)$&$I(\lambda)$&ID&$\lambda_{\rm 0}$&Mult  \\
\noalign{\vskip3pt} \noalign{\hrule} \noalign{\vskip3pt}
 3705.43 &2.39      &2.52      &H 16     &3703.86&H16                 \\
 3713.54 &1.83      &1.93      &H 15     &3711.97&H15                \\
 3723.31 &3.55      &3.73      &[S III]  &3721.63&F2                 \\
 3727.50 &4.77($+$1)&5.02($+$1)&[O II]   &3726.03&F1                 \\
 3730.29 &6.80($+$1)&7.16($+$1)&[O II]   &3728.82&F1                 \\
 3736.06 &2.94      &3.09      &H 13     &3734.37&H13                 \\
 3751.70 &2.83      &2.97      &H 12     &3750.15&H12                 \\
 3772.19 &3.86      &4.05      &H 11     &3770.63&H11                 \\
 3799.47 &5.17      &5.43      &H 10     &3797.90&H10                 \\
 3836.98 &7.27      &7.62      &H 9      &3835.39&H9                 \\
 3870.25 &3.72($+$1)&3.89($+$1)&[Ne III] &3868.75&F1                 \\
       * &        * &        * &He I     &3888.65&V2                     \\
 3890.56 &1.89($+$1)&1.98($+$1)&H 8       &3889.05&H8                 \\
 3969.00 &1.03($+$1)&1.07($+$1)&[Ne III] &3967.46&F1                 \\
 3971.61 &1.49($+$1)&1.56($+$1)&H 7      &3970.07&H7                \\
 4027.79 &2.38      &2.47      &He I     &4026.21&V18                \\
 4071.28 &9.37($-$1)&9.73($-$1)&[S II]   &4068.60&F1                \\
 4076.17 &1.43($-$1)&1.49($-$1)&?        &4073.74&                   \\
 4079.03 &2.85($-$1)&2.96($-$1)&[S II]   &4076.35&F1                  \\
 4081.53 &6.41($-$2)&6.65($-$2)&O II     &4078.84&V10             \\
 4104.35 &2.58($+$1)&2.67($+$1)&H 6      &4101.74&H6                 \\
 4114.66 &5.04($-$2)&5.23($-$2)&?        &4112.59&                    \\
 4123.59 &2.51($-$1)&2.60($-$1)&He I     &4120.84&V16                \\
 4128.30 &7.48($-$2)&7.75($-$2)&?        &4125.62&                   \\
 4146.25 &3.07($-$1)&3.18($-$1)&He I     &4143.76&V53                 \\
 4155.79 &1.07($-$1)&1.11($-$1)&O II     &4153.30&V19                \\
 4171.46 &9.24($-$2)&9.56($-$2)&O II     &4169.22&V19                \\
 4207.77 &1.05($-$1)&1.08($-$1)&?        &4204.93&                   \\
\end{tabular}
\end{minipage}
\end{table}

\setcounter{table}{1}
\begin{table}
\begin{minipage}{75mm}
\centering \caption{{\it --continued}}
\begin{tabular}{l@{\hspace{2.8mm}}l@{\hspace{2.8mm}}l@{\hspace{2.8mm}}c@{\hspace{2.8mm}}c@{\hspace{2.8mm}}c@{\hspace{2.8mm}}}
\noalign{\hrule} \noalign{\vskip3pt}
\multicolumn{6}{c}{SMC~N66}\\
$\lambda_{\rm obs}$&$F(\lambda)$&$I(\lambda)$&ID&$\lambda_{\rm 0}$&Mult  \\
\noalign{\vskip3pt} \noalign{\hrule} \noalign{\vskip3pt}

 4342.97 &4.61($+$1)&4.73($+$1)&H 5      &4340.47&H5                  \\
 4349.37 &5.93($-$2)&6.08($-$2)&S II     &4347.20&                    \\
 4358.29 &6.29($-$2)&6.45($-$2)&?        &4355.74&                    \\
 4361.93 &4.87($-$2)&4.99($-$2)&[Fe II]  &4359.34&7F                   \\
 4365.70 &6.11      &6.26      &[O III]  &4363.21&F2                    \\
 4390.47 &4.61($-$1)&4.71($-$1)&He I     &4387.93&V51                  \\
 4408.60 &8.33($-$2)&8.52($-$2)&[Fe II]  &4406.38&                     \\
 4412.07 &1.01($-$1)&1.03($-$1)&[Fe II]  &4409.85&V55e                 \\
 4416.00 &8.33($-$2)&8.52($-$2)&[Fe II]  &4413.78&                     \\
 4420.38 &8.33($-$2)&8.52($-$2)&[Fe II]  &4418.15&                   \\
 4439.95 &1.11($-$1)&1.13($-$1)&He I     &4437.55&V50                  \\
 4474.10 &3.93      &4.01      &He I     &4471.50&V14                   \\
 4485.62 &6.49($-$2)&6.61($-$2)&?        &      *&                      \\
 4538.63 &5.71($-$2)&5.80($-$2)&?        &4535.96&                      \\
 4543.46 &5.05($-$2)&5.13($-$2)&?        &4541.59&       \\
 4546.67 &4.52($-$2)&4.59($-$2)&N III    &4544.80&V12    \\
 4553.44 &4.55($-$2)&4.62($-$2)&?        &4550.76&                     \\
 4565.18 &2.06($-$1)&2.09($-$1)&Mg I]    &4562.60&                     \\
 4573.68 &1.29($-$1)&1.31($-$1)&Mg I]    &4571.10&                     \\
 4589.44 &5.49($-$2)&5.56($-$2)&?        &4586.75&                     \\
 4593.51 &4.55($-$2)&4.61($-$2)&O II     &4590.97&V15                  \\
 4597.60 &3.92($-$2)&3.97($-$2)&?        &4594.90&V15                  \\
 4598.72 &1.99($-$2)&2.02($-$2)&O II     &4596.18&V15                  \\
 4609.66 &4.23($-$2)&4.29($-$2)&[Fe III] &4607.13&3F                   \\
 4633.28 &3.86($-$2)&3.91($-$2)&N II     &4630.54&V5                    \\
 4641.96 &3.07($-$2)&3.11($-$2)&O II     &4638.49&V1                   \\
 4644.60 &6.30($-$2)&6.37($-$2)&N III    &4640.64&V2                    \\
 4652.34 &5.38($-$2)&5.44($-$2)&O II     &4649.13&V1                   \\
 4654.05 &5.06($-$2)&5.12($-$2)&O II     &4650.84&V1                   \\
 4654.68 &7.01($-$3)&7.08($-$3)&?        &4654.68&                     \\
 4658.66 &9.27($-$2)&9.36($-$2)&?        &4655.92&                     \\
 4661.01 &1.63($-$1)&1.65($-$1)&[Fe III] &4658.10&3F                   \\
 4664.28 &7.91($-$2)&8.00($-$2)&O II     &4661.63&V1                  \\
 4685.79 &3.89($-$2)&3.92($-$2)&?        &4683.25&                      \\
 4696.16 &3.84($-$2)&3.87($-$2)&?        &4693.31&                  \\
 4704.42 &4.37($-$2)&4.40($-$2)&[Fe III] &4701.62&3F                   \\
 4714.07 &3.03($-$1)&3.05($-$1)&[Ar IV]  &4711.37&F1                    \\
 4715.87 &4.87($-$1)&4.90($-$1)&He I     &4713.17&V12                  \\
 4724.87 &5.66($-$2)&5.70($-$2)&?        &4722.83&                     \\
 4742.99 &2.05($-$1)&2.06($-$1)&[Ar IV]  &4740.17&F1                   \\
 4744.50 &5.13($-$2)&5.16($-$2)&O II     &4741.71&25                   \\
 4757.83 &5.77($-$2)&5.80($-$2)&[Fe III] &4754.83&3F                   \\
 4760.61 &3.71($-$2)&3.73($-$2)&Fe II?   &4757.66&                     \\
 4815.82 &4.86($-$2)&4.87($-$2)&Si III   &4813.20&                     \\
 4818.07 &5.85($-$2)&5.86($-$2)&S II     &4815.45&V9                   \\
 4820.77 &7.88($-$2)&7.90($-$2)&?        &4818.10&                      \\
 4828.09 &5.94($-$2)&5.95($-$2)&?        &4825.61&                       \\
 4833.25 &6.59($-$2)&6.60($-$2)&?        &4830.55&                        \\
 4837.60 &4.08($-$2)&4.08($-$2)&?        &4834.70&                       \\
 4842.50 &5.94($-$2)&5.94($-$2)&?        &4840.03&                       \\
 4846.88 &5.94($-$2)&5.94($-$2)&?        &4844.15&                       \\
 4864.06 &1.00($+$2)&1.00($+$2)&H 4      &4861.33&H4                         \\
 4891.36 &7.93($-$2)&7.92($-$2)&[Fe II]  &4889.63&4F                     \\
 4896.47 &3.25($-$2)&3.24($-$2)&?        &4893.75&                     \\
 4905.43 &5.59($-$2)&5.57($-$2)&[Fe IV]  &4903.10&                      \\
 4909.17 &4.44($-$2)&4.43($-$2)&O II     &4906.83&V28                     \\
 4914.39 &4.50($-$2)&4.49($-$2)&?        &4914.84&                        \\
 4917.57 &4.50($-$2)&4.49($-$2)&?        &4921.93&                       \\
 4924.75 &1.07      &1.06      &He I     &4921.93&V48              \\
 4927.35 &7.04($-$2)&7.02($-$2)&O II     &4924.53&V28                    \\
 4933.84 &4.82($-$2)&4.81($-$2)&[O III]  &4931.80&F1               \\
 4937.97 &3.70($-$2)&3.68($-$2)&?        &4934.52&                        \\
 4943.12 &7.05($-$2)&7.02($-$2)&O II     &4941.07&V33                     \\
 4965.36 &1.70($+$2)&1.69($+$2)&[O III]  &4958.91&F1 
\\

\end{tabular}
\end{minipage}
\end{table}

\setcounter{table}{1}
\begin{table}
\begin{minipage}{75mm}
\centering \caption{{\it --continued}}
\begin{tabular}{l@{\hspace{2.8mm}}l@{\hspace{2.8mm}}l@{\hspace{2.8mm}}c@{\hspace{2.8mm}}c@{\hspace{2.8mm}}c@{\hspace{2.8mm}}}
\noalign{\hrule} \noalign{\vskip3pt}
\multicolumn{6}{c}{SMC~N66}\\
$\lambda_{\rm obs}$&$F(\lambda)$&$I(\lambda)$&ID&$\lambda_{\rm 0}$&Mult  \\
\noalign{\vskip3pt} \noalign{\hrule} \noalign{\vskip3pt}

 5013.46 &5.12($+$2)&5.08($+$2)&[O III]  &5006.84&F1                      \\
 5525.05 &3.57($-$1)&3.46($-$1)&[Cl III] &5517.66&F1                      \\
 5545.02 &4.41($-$1)&4.28($-$1)&[Cl III] &5537.60&F1                      \\
 5760.62 &1.31($-$1)&1.26($-$1)&[N II]   &5754.60&F3                       \\
 6142.57 &5.97($-$1)&5.67($-$1)&?        &6136.36&                         \\
 6172.14 &5.97($-$1)&5.66($-$1)&[Mn V]   &6166.00&                         \\
 6204.09 &2.57($-$1)&2.44($-$1)&O II     &6197.92&                         \\
 6237.22 &2.16      &2.04      &[Ni III] &6231.09&                        \\
 6260.46 &1.65      &1.56      &[Fe II]  &6254.30&                        \\
 6301.05 &1.56($+$1)&1.48($+$1)&[O I]    &6300.34&F1                      \\
 6312.81 &1.73      &1.64      &[S III]  &6312.10&F3                      \\
 6467.49 &2.05      &1.93      &C II     &6461.95&                        \\
 6499.73 &2.05      &1.93      &?        &6492.97&                        \\
 6529.90 &2.25      &2.11      &?WR      &6516.12&                        \\
 6543.22 &2.04      &1.91      &?        &6529.41&                        \\
 6561.94 &2.95      &2.76      &[N II]   &6548.10&F1                      \\
 6574.48 &3.05($+$2)&2.86($+$2)&H 3      &6562.77&H3                       \\
 6593.61 &4.08      &3.82      &[N II]   &6583.50&F1                   \\
 6683.76 &3.17      &2.97      &He I     &6678.16&V46                     \\
 6720.72 &8.45      &7.89      &[S II]   &6716.44&F2                       \\
 6735.11 &6.19      &5.78      &[S II]   &6730.82&F2                      \\
 7072.20 &2.46      &2.28      &He I     &7065.25&V10                     \\
 7144.83 &9.68      &8.94      &[Ar III] &7135.80&F1                      \\
 7166.84 &1.39      &1.28      &?        &7159.07&                        \\
 7178.54 &3.61($-$1)&3.34($-$1)&[Ar IV]  &7070.62&                        \\
 7186.58 &2.50($-$1)&2.30($-$1)&?        &7186.58&          \\
 7293.51 &5.73      &5.27      &He I     &7281.35&V45                      \\
 7333.56 &1.92      &1.77      &[O II]   &7318.92&F2                       \\
 7344.32 &1.68      &1.55      &[O II]   &7329.67&F2                       \\
\noalign{\vskip3pt} \noalign{\hrule} \noalign{\vskip3pt}
\end{tabular}
\end{minipage}
\end{table}

\section{Nebular analysis}

\subsection{Reddening correction}

In the case of the galactic nebulae M\,17 and NGC\,3576, the galactic reddening
law of Howarth (1983) was used to correct for interstellar extinction, with the
amount of extinction at H$\beta$ determined by comparing the observed Balmer
H$\alpha$/H$\beta$, H$\gamma$/H$\beta$ and H$\delta$/H$\beta$ decrements to
their Case B theoretical values from Storey \& Hummer (1995). For the M\,17 and
NGC\,3576 sightlines, values of \emph{c}(H$\beta$) = 1.84 and 1.25 respectively
were found.

For the Magellanic Cloud nebulae, the contribution from galactic foreground
reddening was estimated from the reddening maps of Burstein \& Heiles (1982),
using the extinction law of Howarth (1983) in all cases. For the 
direction to 30~Doradus, \emph{c}(\Hb) = 0.087 through the Milky Way was
found. The remaining extinction due to the immediate 30~Doradus
environment was found to be \emph{c}(\Hb) = 0.32, from
the foreground-corrected Balmer decrements, using the LMC extinction law of
Howarth (1983). For LMC~N11B and SMC~N66, no further correction was attempted
after foreground Galactic reddenings of \emph{c}(\Hb) = 0.073
and 0.087, respectively, were corrected for, since the resulting Balmer
line ratios were in good agreement with their Case B theoretical values.


\subsection{Electron temperatures and densities}

\setcounter{table}{2}
\begin{table*}
\centering
\begin{minipage}{125mm}
\caption{Plasma diagnostics.}
\begin{tabular}{lccccc}
\noalign{\vskip3pt} \noalign{\hrule} \noalign{\vskip3pt}
Diagnostic ratio &M\,17    &NGC\,3576   &30~Doradus &LMC~N11B   &SMC~N66\\
\noalign{\vskip3pt} \noalign{\hrule}

\noalign{\vskip5pt}
&\multicolumn{5}{c}{$T_{\rm e}$\,(K)}\\
\noalign{\vskip3pt}

\lbrack O~{\sc iii}\rbrack~($\lambda$4959+$\lambda$5007)/$\lambda$4363 & 8200 & 8850 & 10100 & 9400 & 12400 \\
\lbrack N~{\sc ii}\rbrack~($\lambda$6548+$\lambda$6584)/$\lambda$5754  & 9100 & 9000   & 12275 & 9250 & 14825 \\
                                                                       & 7900$^a$ & 8500$^a$  & 11600$^a$  & *    & * \\
\lbrack S~{\sc ii}\rbrack~$\lambda$4068/($\lambda$6717+$\lambda$6730)  & 8850 & 7900 & 7500  & 6950 & 11050 \\
BJ/H\,11                                                               & 7700 & 8070  &*&*&*   \\
$T_0$                                                                  &7840  &8300   &*&*&*   \\
$t^2$                                                                  &0.011 &0.017   &*&*&*   \\
\noalign{\vskip5pt}
&\multicolumn{5}{c}{$N_{\rm e}$\,(cm$^{-3}$)}\\
\noalign{\vskip3pt}

\lbrack Ar~{\sc iv}\rbrack~$\lambda$4740/$\lambda$4711             &1500       &1700      &1800   &*      &{\tiny$\lesssim$}\,100\\
\lbrack Cl~{\sc iii}\rbrack~$\lambda$5537/$\lambda$5517            &1050       &2700      &480    &1700   &3700\\
\lbrack S~{\sc ii}\rbrack~$\lambda$6731/$\lambda$6716              &600        &1350      &390    &80     &60\\
\lbrack O~{\sc ii}\rbrack~$\lambda$3729/$\lambda$3726              &*          &1300      &370    &110    &50\\

\noalign{\vskip3pt} \noalign{\hrule}
\end{tabular}
\begin{description}
\item[$^a$] [N~{\sc ii}] temperatures after correction for recombination excitation contributions.
\end{description}
\end{minipage}
\end{table*}

Nebular electron temperatures {\elt} and densities {\eld} were derived from
several CEL diagnostic ratios by solving the equations of statistical
equilibrium using the multi-level ($\geq$\,5) atomic model {\sc equib} and are
presented in Table~3. The atomic data sets used for this purpose, as well as
for the derivation of abundances, are the same as those used by Liu et al.
(2000) in the study of the planetary nebula NGC\,6153. The procedure was as
follows: a representative initial {\elt} of 9000\,K was assumed in order to
derive {\eld}(Cl~{\sc iii}) and {\eld}(\ariv); the mean electron density
derived from these diagnostics was then used to derive {\elt}(\oiii), and we
then iterated once to get the final values. In a similar manner, {\elt}(\nii)
was derived in conjunction with {\eld}(O~{\sc ii}) and {\eld}(S~{\sc ii}).
The Balmer jump (BJ) electron temperatures of M\,17 and NGC\,3576 were derived
from the ratio of the {\hi} recombination continuum Balmer discontinuity at
3646\,\AA~to the H\,11 $\lambda$3770 line. These temperatures are presented in
Table~3, along with the mean nebular temperature, $T_0$, and temperature
fluctuation parameter, $t^2$ (Peimbert 1967) implied by the BJ and {\foiii}
temperatures.

The electron temperatures deduced from the {\fnii} nebular to auroral line
ratio are higher than those of the corresponding {\foiii} line ratio---except
for LMC~N11B, where the derived values are probably consistent within the
errors. The differences exceed 2000\,K for 30~Doradus and SMC~N66. Using
photoionization models of metal-rich {\hii} regions, Stasi\'{n}ska (1980) has
argued that the electron temperature increases outward as a function of radius,
probably due to a combination of hardening of the radiation field with
increasing optical depth, plus stronger cooling from fine-structure lines of
{\foiii} in the inner parts of the nebulae where {\opp} dominates; thus the
temperature in zones where singly ionized species exist is predicted to be
higher than {\elt}(\foiii).

However, the contribution of recombination to the excitation of the {\fnii}
$\lambda$5754 line (Rubin 1986), coupled with the potential presence of
high-density inclusions in nebulae (Viegas \& Clegg 1994), may cause
deceptively high temperatures to be derived from the {\fnii} nebular to auroral
line ratio, as well as from the corresponding {\foii}
$\lambda$3727/($\lambda$7320 + $\lambda$7330) ratio. We have therefore
estimated corrections to the {\fnii} temperatures of M\,17, NGC\,3576 and
30~Doradus, making use of the derived ORL {\npp}/{\hp} fractions presented
later in Table~6 and Eq.\,~1 of Liu et al. (2000). The revised {\fnii}
temperatures for these three nebulae respectively, are 7900, 8500 and 11600\,K;
improved agreement with the corresponding {\foiii} temperatures of 8200, 8850
and 10100\,K is found in all cases. Given the inherent uncertainties, the
{\elt}(\nii)'s were not used in the abundance analysis that follows.



Regarding the derived electron densities, it is found that in four out of five
nebulae the densities deduced from the {\fsii} $\lambda$6731/$\lambda$6716
ratio are in good agreement with those derived from the {\foii}
$\lambda$3729/$\lambda$3726 ratio (the spectral resolution was not adequate to
allow the determination of the {\foii} doublet ratio in M\,17), but are lower
than the values given by the {\fariv} and {\fcliii} diagnostics.
This can be seen especially in LMC~N11B and SMC~N66 where the {\foii} and
{\fsii} doublet ratios approach the low-density limit, and certainly imply
{\eld}'s an order of magnitude less than from the doubly ionized species
(Table~3). This behaviour is consistent with the presence of strong density
variations in the nebulae, so that diagnostic line ratios from lines with
higher critical densities\footnote{For {\fcliii} $\lambda\lambda$5517, 5537,
{\crd} = 6400 and 34000 {\cmt}; for {\fariv} $\lambda\lambda$4711, 4740, {\crd}
= 14\,000 and 130,000 {\cmt}; for {\foii} $\lambda\lambda$3726, 3729, {\crd} =
4300 and 1300 {\cmt}; and for {\fsii} $\lambda\lambda$6716, 6730, {\crd} = 1200
and 3300 {\cmt} respectively; the quoted values are for an electron temperature
of 10000\,K.} yield higher derived nebular electron densities (see Rubin 1989,
Liu et al. 2001a).

\section{Ionic and total elemental abundances from CEL{\small s}}

We used the statistical equilibrium code {\sc equib} to derive ionic abundances
from nebular CELs. The ionic abundances deduced are presented in Table~4. For
each nebula, the electron temperature derived from the {\foiii} nebular to
auroral line ratio was adopted for all ionic species, following the discussion
detailed in the previous section. Regarding the choice of electron densities,
we adopted the mean of the values deduced from the {\foii} and {\fsii} doublet
ratios in order to derive abundances of singly ionized species, while the mean
electron densities from the {\fariv} and {\fcliii} ratios were used to derive
abundances for doubly and triply ionized species.

\setcounter{table}{3}
\begin{table*}
\centering
\begin{minipage}{125mm}
\footnotesize \caption{Ionic and elemental abundances for helium
relative to hydrogen, derived from ORLs, and those for heavy
elements, derived from CELs.$^a$}
\begin{tabular}{llcccccc}
\noalign{\vskip3pt} \noalign{\hrule} \noalign{\vskip3pt}
     &                          &M\,17     &NGC\,3576  &30~Doradus &LMC~N11B  &SMC~N66                 \\
\noalign{\vskip3pt} \noalign{\hrule} \noalign{\vskip3pt}

4471 &  He$^{+}$/H$^{+}$        &0.0955(1)  &0.0949(1)  &0.0874(1)  &0.0856(1)  &0.0791(1)         \\
5876 &  He$^{+}$/H$^{+}$        &0.0959(4)  &0.0926(4)  &0.0910(3)  &0.1022(4)  &0.1006(4)         \\
6678 &  He$^{+}$/H$^{+}$        &0.0909(1)  &0.0879(1)  &0.0926(1)  &0.0906(1)  &0.0756(1)     \\
Avg. &        He/H              &0.0950     &0.0922     &0.0906     &0.0975     &0.0928      \\
\noalign{\vskip3pt}
3727 &          O$^{+}$/H$^{+}$ & 9.28($-$5)  &1.10($-$4)   & 4.13($-$5)  &  1.12($-$4) &  3.35($-$5)  \\
7320+7330 &     ...             & 1.18($-$4): &2.06($-$4):  & 5.77($-$5): &  3.79($-$4):&  8.77($-$6):   \\
4931 &          O$^{2+}$/H$^{+}$& 2.23($-$4): &3.31($-$4):  &   *       &  2.36($-$4): & 6.30($-$5):  \\
4959 &          ...             & 2.66($-$4)  &2.21($-$4)   & 1.76($-$4)  &  1.47($-$4)  & 9.41($-$5)   \\
     &  \emph{icf}(O)                  & 1.00      & 1.00      & 1.00      & 1.00      & 1.00          \\
     &           O/H            & 3.59($-$4)  & 3.31($-$4)  & 2.17($-$4)  & 2.59($-$4)  & 1.28($-$4)          \\
\noalign{\vskip3pt}
6548+6584 &     N$^{+}$/H$^{+}$ & 8.14($-$6)  & 1.17($-$5)  & 1.48($-$6)  & 3.66($-$6)  & 5.37($-$7) \\
     &  \emph{icf}(N)                  & 3.87      & 3.01      & 5.25      & 2.31      & 3.82                  \\
     &           N/H            & 3.15($-$5)  & 3.52($-$5)  & 7.77($-$6)  & 8.45($-$6)  & 2.05($-$6) \\
\noalign{\vskip3pt}
3868+3967 &   Ne$^{2+}$/H$^{+}$& 7.17($-$5)&  3.46($-$5)  & 3.71($-$5)  & 2.40($-$5)  & 1.72($-$5)          \\
     &  \emph{icf}(Ne)                 &  1.39     &  1.50     & 1.23      & 1.76      & 1.36      \\
     &           Ne/H           &  9.97($-$5) &  5.19($-$5) & 4.56($-$5)  & 4.22($-$5)  & 2.34($-$5) \\
\noalign{\vskip3pt}
7135 &          Ar$^{2+}$/H$^{+}$& 1.48($-$6) & 1.70($-$6)  & 1.11($-$6)  & 1.12($-$6)  & 4.71($-$7)      \\
4711+4740     &  Ar$^{3+}$/H$^{+}$& 2.72($-$8) & 2.15($-$8)  & 3.39($-$8)  &   2.27($-$8)& 3.16($-$8)      \\
     &  \emph{icf}(Ar)                 &  1.35     &  1.50     & 1.24      &  1.76     & 1.35          \\
     &           Ar/H           &  2.03($-$6) &  2.58($-$6) & 1.42($-$6)  &  2.01($-$6) & 6.79($-$7)         \\
\noalign{\vskip3pt}
4069 &          S$^{+}$/H$^{+}$ & 5.79($-$7): & 7.50($-$7): & 2.53($-$7): &  8.82($-$7):&2.60($-$7):\\
6716+6730 &    ...             & 3.60($-$7)  & 6.20($-$7)  & 2.80($-$7)  & 7.32($-$7) & 2.59($-$7)  \\
6312 &          S$^{2+}$/H$^{+}$& 8.05($-$6)  &   *       & 4.28($-$6)  &    *     & 1.66($-$6)  \\
     &  \emph{icf}(S)                  & 1.19      &   *       &   1.29    &    *     &  1.19  \\
     &           S/H            &  1.00($-$5) & 6.57($-$7)   & 5.88($-$6)  & 4.98($-$6) &  2.28($-$6)  \\
\noalign{\vskip3pt}
5517 &          Cl$^{2+}$/{\hp} & 1.07($-$7)  & 9.95($-$8)   &  5.14($-$8) & 7.69($-$8) & 2.63($-$8)\\
     &  \emph{icf}(Cl)          &  1.24     &  1.52      & 1.37      &  1.79    & 1.37  \\
     &          Cl/H            & 1.33($-$7)  &   1.51($-$7) & 7.04($-$8)  &  1.38($-$7)& 3.60($-$8) \\
\noalign{\vskip3pt} \noalign{\hrule} \noalign{\vskip3pt}
\end{tabular}
\begin{description}
\item[$^a$] Numbers followed by `:' have not been used at any point in the analysis. The numbers in
parentheses following the He$^+$/H$^+$ abundances are the weighting factors used for the derivation
of the average He$^+$/H$^+$ abundance.
\end{description}
\end{minipage}
\end{table*}

Total abundances from CELs have been derived adopting the ICF scheme of
Kingsburgh \& Barlow (1994), apart from Cl which was not discussed by those
authors. For that element and in the cases of M\,17, 30~Doradus and SMC~N66,
the prescription of Liu et al. (2000) was used, according to which Cl/H =
(S/S$^{2+}$)\,$\times$\,Cl$^{2+}$/{\hp}, based on the similarities of the
ionization potentials of chlorine ionic stages to those of sulphur ionic
stages. On the other hand, for NGC\,3576 and LMC~N11B, for which the listed S
abundances are lower limits, we adopted Cl/H =
(Ar/Ar$^{2+}$)\,$\times$\,Cl$^{2+}$/{\hp}. The total oxygen abundance was
adopted to be the sum of singly and doubly ionized oxygen abundances. No
significant amounts of {\oppp} are expected, since no {\heii} recombination
lines are seen.

\section{Ionic abundances from ORL{\small s}}

Since ionic abundances relative to {\hp} derived from intensity ratios of
heavy element ORLs relative to a hydrogen recombination line depend only weakly
on the adopted temperature, and are essentially independent of {\eld}, for each
nebula the {\foiii} temperature was adopted for the calculations. The He
abundances derived from {\hei} recombination lines are given in Table~4. 
Case A recombination was assumed for the triplet lines $\lambda$4471,
$\lambda$5876 and Case B for the singlet $\lambda$6678 line. The effective
recombination coefficients were from Brocklehurst (1972). For the
$\lambda$4471 line, the effective coefficient given by Brocklehurst (1972)
differs by only 1.4 per cent from the calculations of Smits (1996).
The differences between the two calculations are even smaller for the
other two lines. Contributions to the observed fluxes by collisional
excitation from the He$^{\rm 0}$ 2s\,$^3$S metastable level by electron
impacts were corrected for using the formulae derived by Kingdon
\& Ferland (1995a).

In the following subsections we present C, N and O ionic abundances
derived from ORLs.

\subsection{{\cpp}/{\hp} and {\npp}/{\hp}}

The case-insensitive {\cii} $\lambda$4267 (4f--3d) line has been detected from
all nebulae except SMC~N66 and used to derive the {\cpp}/{\hp} abundance ratios
presented in Table~5; an upper limit only has been estimated for SMC~N66. The
doublet from multiplet V\,4 ($\lambda$3920) has also been detected from
NGC\,3576, but not used for abundance determinations since its upper levels are
directly connected to the C$^+$ 2p\,$^2P^{\rm o}$ ground term, and the doublet
is therefore potentially affected by optical depth effects.

Regarding the {\npp}/{\hp} ORL abundance ratios, these have been derived for
M\,17 and NGC\,3576 and are presented in Table~6; lines from the {\nii} 3s--3p
V\,3 and V\,5 multiplets have been detected and Case B recombination has been
assumed. It is estimated that under Case A the abundance ratios deduced from
the V\,5 lines would be several times higher, while V\,3 results would be
larger by only about 20\,per cent. Longer exposure times would be required to
detect the weaker case-insensitive lines from the 3d--4f group. {\nii} ORLs
have not been unambiguously detected from any of the Magellanic Cloud {\hii}
regions; the strongest predicted multiplet V\,3 ($\lambda$5680) is probably
present in 30~Doradus, but the resolution of our spectra is too low (FWHM
11\,\AA) at this wavelength to allow us to be conclusive. In that nebula's
spectrum the strongest {\nii} V\,5 multiplet line, at 4630\,\AA, is marginally
detected; an upper limit to the {\npp}/{\hp} fraction of
5.48\,$\times$\,{\tmfi} was derived from it. From Table~6 we adopt {\npp}/{\hp}
fractions of 3.53\,$\times$\,{\tmf} and 2.70\,$\times$\,{\tmf} for M\,17 and
NGC\,3576, respectively.

\setcounter{table}{4}
\begin{table}
\begin{center}
\caption{Ionic carbon abundances from optical recombination lines.}
\begin{tabular}{l@{\hspace{1.1mm}}c@{\hspace{1.1mm}}c@{\hspace{1.1mm}}c@{\hspace{1.1mm}}c@{\hspace{1.1mm}}c@{\hspace{1.1mm}}}
\noalign{\vskip3pt} \noalign{\hrule} \noalign{\vskip3pt}
                                            &M\,17        &NGC\,3576           &30\,Dor          &LMC\,N11B      &SMC\,N66      \\
\noalign{\vskip3pt} \noalign{\hrule} \noalign{\vskip3pt}
$I(\lambda4267)$                            &0.482        &0.312               &0.0919              &0.196          &$\lesssim$\,0.0419       \\
\bf{10$^4$\,$\times$\,{\cpp}/{\hp}}         &\bf{4.35}    &\bf{2.87}           &\bf{0.882}          &\bf{1.82}      &$\lesssim$\,\bf{0.433}       \\
\noalign{\vskip3pt} \noalign{\hrule} \noalign{\vskip3pt}
\end{tabular}
\end{center}
\end{table}

\setcounter{table}{5}
\begin{table}
\footnotesize
\begin{center}
\caption[Recombination-line {\npp}/{\hp} abundances in H~{\sc ii}
regions.]{Recombination-line {\npp}/{\hp} abundances.}
\begin{tabular}{l@{\hspace{1.7mm}}c@{\hspace{1.7mm}}l@{\hspace{1.7mm}}c@{\hspace{1.7mm}}l@{\hspace{1.7mm}}c@{\hspace{1.7mm}}l@{\hspace{1.7mm}}c@{\hspace{1.7mm}}}
 \noalign{\vskip3pt} \noalign{\hrule} \noalign{\vskip3pt}
$\lambda_{\mathrm{0}}$ &Mult.&$I_{\rm obs}$ &$\displaystyle\frac{\rm{~N}^{2+}}{\rm{H}^+}$  &$I_{\rm obs}$ &$\displaystyle\frac{\rm{~N}^{2+}}{\rm{H}^+}$ &$I_{\rm obs}$ &$\displaystyle\frac{\rm{~N}^{2+}}{\rm{H}^+}$\\
\noalign{\vskip2pt}
(\AA)         &     &          &(\tmf)    &    &(\tmf)    &    &(\tmf)\\
\noalign{\vskip3pt} \noalign{\hrule} \noalign{\vskip3pt}
           &     &\multicolumn{2}{c}{\underline{M\,17}}  &\multicolumn{2}{c}{\underline{NGC\,3576}} &\multicolumn{2}{c}{\underline{30 Doradus}} \\
\noalign{\vskip3pt}
\multicolumn{8}{l}{\bf{V\,3 3s\,$^{3}$P$^\mathrm{{o}}$-3p\,$^{3}$D}}\\
5666.63    &V3      &.0569         &4.35     &.0358            &2.80            &* &*\\
5676.02    &V3      &.0170         &2.93     &.0159            &2.72            &* &*\\
5679.56    &V3      &.0544         &2.23     &.0667           &2.75            &* &*\\
\noalign{\vskip3.5pt}
\multicolumn{8}{l}{\bf{V\,5 3s\,$^{3}$P$^\mathrm{o}$-3p\,$^{3}$P}}\\
4630.54    &V5      &.0602         &4.99     &.0309           &2.56            &.0066    &.548\\
\noalign{\vskip3.5pt}
\multicolumn{8}{l}{\bf{V\,20 3p\,$^{3}$D-3d\,$^{3}$D$^\mathrm{{o}}$}}\\
4803.29    &V20     &.0304         &4.52     &*                &*              &*&*\\
\noalign{\vskip3pt}
\multicolumn{1}{l}{\bf{Sum}} & &\bf{.219}   &\bf{3.53}     &\bf{.149}   &\bf{2.70} &*&\bf{.548}\\
\noalign{\vskip3pt} \noalign{\hrule}
\end{tabular}
\end{center}
\end{table}

\subsection{{\opp}/{\hp}}


\setcounter{table}{6}
\begin{table}
\begin{center} \caption{Comparison of the observed and predicted relative
intensities of O~{\sc ii} ORLs.}
\begin{tabular}{c@{\hspace{1.8mm}}c@{\hspace{1.8mm}}c@{\hspace{1.8mm}}c@{\hspace{1.8mm}}l@{\hspace{1.8mm}}c@{\hspace{1.8mm}}c@{\hspace{1.8mm}}} \noalign{\vskip3pt} \noalign{\hrule}
\noalign{\vskip3pt}
$\lambda_{0}$(\AA) &Mult &Term$_{l}$--Term$_{u}$    &g$_{l}$--g$_{u}$ &I$_{pred}$ &I$_{obs}$ &I$_{obs}$/I$_{pred}$\\
\noalign{\vskip3pt} \noalign{\hrule} \noalign{\vskip3pt}
\multicolumn{7}{c}{\underline{M\,17}}\\

\multicolumn{7}{l}{\em (3s--3p)}\\
\noalign{\vskip3pt}
4638.86   & V1  &3s  4P--3p  4D*  &2--4     &0.21 &0.77[.12] &3.7[0.6] \\
4641.81   & V1  &3s  4P--3p  4D*  &4--6     &0.53 &1.11[.16] &2.1[0.3] \\
4649.13   & V1  &3s  4P--3p  4D*  &6--8     &1.00 &1.00[.10] &1.0[0.1]      \\
4650.84   & V1  &3s  4P--3p  4D*  &2--2     &0.21 &0.63[.11] &3.0[0.5] \\
4661.63   & V1  &3s  4P--3p  4D*  &4--4     &0.27 &0.76[.11] &2.8[0.4] \\
4676.24   & V1  &3s  4P--3p  4D*  &6--6     &0.22 &0.46[.10] &2.1[0.5] \\
\noalign{\vskip3pt}
4317.14   & V2  &3s  4P--3p  4P*&  2--4     &0.44 &1.04[.32]&2.6[0.8]   \\
4319.63   & V2  &3s  4P--3p  4P*&  4--6     &0.43 &0.31[.23]&0.7[0.5]   \\
4345.56   & V2  &3s  4P--3p  4P*&  4--2     &0.40 &0.63[.20]&1.5[0.5]   \\
4349.43   & V2  &3s  4P--3p  4P*&  6--6     &1.00 &1.00[.10]&1.0[0.1]       \\
\noalign{\vskip5pt}
\multicolumn{7}{c}{\underline{NGC\,3576}}\\
\multicolumn{7}{l}{\em (3s--3p)}\\
\noalign{\vskip3pt}
4638.86   & V1  &3s  4P--3p  4D*  &2--4      &0.21 & 0.55[.11] & 2.6[0.5]\\
4641.81   & V1  &3s  4P--3p  4D*  &4--6      &0.53 & 0.95[.16] & 1.8[0.3]\\
4649.13   & V1  &3s  4P--3p  4D*  &6--8      &1.00 & 1.00[.10] & 1.0[0.1] \\
4650.84   & V1  &3s  4P--3p  4D*  &2--2      &0.21 & 0.60[.11] & 2.9[0.5]\\
4661.63   & V1  &3s  4P--3p  4D*  &4--4      &0.27 & 0.99[.16] & 3.7[0.6]\\
4676.24   & V1  &3s  4P--3p  4D*  &6--6      &0.22 & 0.60[.16] & 2.7[0.7]\\
\noalign{\vskip3pt}
\multicolumn{7}{l}{\em(3p--3d)}\\
\noalign{\vskip3pt}
4069.89   & V10 &3p  4D*--3d  4F   &4--6     & 0.74 & 1.40[.20] & 1.9[0.3] \\
4072.16   & V10 &3p  4D*--3d  4F   &6--8     & 0.69 & 1.20[.82] & 1.7[1.2] \\
4075.86   & V10 &3p  4D*--3d  4F   &8--10    & 1.00 & 1.00[.10] & 1.0[0.1]      \\
4085.11   & V10 &3p  4D*--3d  4F   &6--6     & 0.13 & 0.17[.10] & 1.3[0.8] \\
\noalign{\vskip3pt}
\multicolumn{7}{l}{\em(3d--4f)}\\
\noalign{\vskip3pt}
4087.15   & V48c&3d  4F--4f  G3* &4--6       & 0.27 & 1.30[.48] & 4.8[1.8] \\
4089.29   & V48a&3d  4F--4f  G5* &10--12     & 1.00 & 1.00[.10] & 1.0[0.1]      \\
4275.55   & V67a&3d  4D--4f  F4* &8--10      & 1.20 & 4.18[1.1] & 3.5[0.9] \\
4357.25   & V63a&3d  4D--4f  D3* &6--8       & 0.05 &2.36[.85]  &47.[17.]  \\
4609.44   &V92a &3d  2D--4f  F4* &6--8       & 0.14 &1.69[.64]  &12.[5.]   \\
\noalign{\vskip5pt}
\multicolumn{7}{c}{\underline{30 Doradus}}\\
\multicolumn{7}{l}{\em (3s--3p)}\\
\noalign{\vskip3pt}
4638.86   & V1  &3s  4P--3p  4D*  &2--4      &  0.21 & 1.00[.15] & 4.8[0.7] \\
4641.81   & V1  &3s  4P--3p  4D*  &4--6      &  0.53 & 1.33[.18] & 2.5[0.3] \\
4649.13   & V1  &3s  4P--3p  4D*  &6--8      &  1.00 & 1.00[.10] & 1.0[0.1] \\
4650.84   & V1  &3s  4P--3p  4D*  &2--2      &  0.21 & 1.00[.19] & 4.8[0.9] \\
4661.63   & V1  &3s  4P--3p  4D*  &4--4      &  0.27 & 0.88[.12] & 3.3[0.4]  \\
4673.73   & V1  &3s  4P--3p  4D*  &4--2      &  0.04 & 0.19[.03] & 4.8[0.8]  \\
4676.24   & V1  &3s  4P--3p  4D*  &6--6      &  0.22 & 0.35[.05] & 1.6[0.2]  \\
\noalign{\vskip3pt}
4317.14   &V2   &3s  4P--3p  4P*&  2--4      &  1.00 & 1.00[.10] & 1.0[0.1] \\
4319.63   &V2   &3s  4P--3p  4P*&  4--6      &  0.98 & 1.06[.12] & 1.1[0.1] \\
\noalign{\vskip3pt}
\multicolumn{7}{l}{\em(3p--3d)}\\
\noalign{\vskip3pt}
4069.89   & V10 &3p  4D*--3d  4F   &4--6     & 1.00  &1.00[.10]  &1.0[0.1]       \\
4072.16   & V10 &3p  4D*--3d  4F   &6--8     & 0.93  &0.51[.09]  &0.6[0.1]   \\
4078.84   & V10 &3p  4D*--3d  4F   &4--4     & 0.14  &0.24[.05]  &1.7[0.4]   \\
4085.11   & V10 &3p  4D*--3d  4F   &6--6     & 0.17  &0.19[.04]  &1.1[0.2]   \\
\noalign{\vskip3pt}
\multicolumn{7}{l}{\em(3d--4f)}\\
\noalign{\vskip3pt}
4083.90   & V48b&3d  4F--4f  G4* &6--8       & 0.29 & 0.51[.31]  &1.8[1.1] \\
4087.15   & V48c&3d  4F--4f  G3* &4--6       & 0.27 & 0.88[.23]  &3.3[0.9]  \\
4089.29   & V48a&3d  4F--4f  G5* &10--12     & 1.00 & 1.00[.13]  &1.0[0.1]  \\
4275.55   & V67a&3d  4D--4f  F4* &8--10      & 1.33 & 3.73[.71]  &2.8[0.5]  \\
4288.82   & V53c&3d  4P--4f  D1* &2--4       & 0.83 & 11.9[2.6]  &14.[3.0] \\
4303.83   & V53a&3d  4P--4f  D3* &6--8       & 0.50 & 2.47[.44] & 4.9[0.9]  \\
4313.44   & V78a&3d  2F--4f  F4* &8--10      & 0.61 & 2.06[.68] & 3.4[1.1]  \\
4315.69   & V63c&3d  4D--4f  D1* &6--4       & 0.11 & 1.82[.67] &17.[6.0]  \\
4609.44   & V92a&3d  2D--4f  F4* &6--8       & 0.57 & 1.00[.24] & 2.0[0.4]  \\
\noalign{\vskip5pt}
\end{tabular}
\end{center}
\end{table}

\setcounter{table}{6}
\begin{table}
\begin{center} \caption{{\it --continued}}
\begin{tabular}{c@{\hspace{1.8mm}}c@{\hspace{1.8mm}}c@{\hspace{1.8mm}}c@{\hspace{1.8mm}}l@{\hspace{1.8mm}}c@{\hspace{1.8mm}}c@{\hspace{1.8mm}}} \noalign{\vskip3pt} \noalign{\hrule}
\noalign{\vskip3pt}
$\lambda_{0}$(\AA) &Mult &Term$_{l}$--Term$_{u}$    &g$_{l}$--g$_{u}$ &I$_{pred}$ &I$_{obs}$ &I$_{obs}$/I$_{pred}$\\
\noalign{\vskip3pt} \noalign{\hrule} \noalign{\vskip3pt}
\multicolumn{7}{c}{\underline{LMC N11B}}\\
\multicolumn{7}{l}{\em (3s--3p)}\\
\noalign{\vskip3pt}
4638.86   & V1  &3s  4P--3p  4D*  &2--4      &  0.21 &1.54[.15] &7.3 [0.7] \\
4641.81   & V1  &3s  4P--3p  4D*  &4--6      &  0.53 &0.70[.18] &1.3 [0.3] \\
4649.13   & V1  &3s  4P--3p  4D*  &6--8      &  1.00 &1.00[.10] &1.0 [0.1] \\
4650.84   & V1  &3s  4P--3p  4D*  &2--2      &  0.21 &0.83[.19] &4.0 [0.9] \\
4661.63   & V1  &3s  4P--3p  4D*  &4--4      &  0.27 &0.84[.12] &3.1 [0.4]  \\
4673.73   & V1  &3s  4P--3p  4D*  &4--2      &  0.04 &0.41[.03] &10.[1.0]  \\
4676.24   & V1  &3s  4P--3p  4D*  &6--6      &  0.22 &0.36[.05] &1.6 [0.2]  \\

\noalign{\vskip5pt}
\multicolumn{7}{c}{\underline{SMC N66}}\\
\multicolumn{7}{l}{\em (3s--3p)}\\
\noalign{\vskip3pt}
4638.86   & V1  &3s  4P--3p  4D*  &2--4      &  0.21 & 0.57[.15] & 2.7[0.7] \\
4649.13   & V1  &3s  4P--3p  4D*  &6--8      &  1.00 & 1.00[.10] & 1.0[0.1] \\
4650.84   & V1  &3s  4P--3p  4D*  &2--2      &  0.21 & 0.94[.19] & 4.5[0.9] \\
4661.63   & V1  &3s  4P--3p  4D*  &4--4      &  0.27 & 1.47[.32] & 5.4[1.2]  \\
\noalign{\hrule}
\end{tabular}
\end{center}
\end{table}

To our knowledge, this is the first time that {\oii} ORLs arising from
Magellanic Cloud H~{\sc ii} regions have been recorded. We have detected
transitions from 3s--3p, as well as 3p--3d and 3d--4f configurations. In
Table~7 we present a comparison between the observed and predicted intensities
of {\oii} lines relative to the strongest expected line within each multiplet.
The numbers in brackets are the formal absolute errors to the intrinsic
intensities derived from the line-fitting method only; they do not include any
possible systematic errors arising e.g. from the flux calibration of our
spectra.
As with our extensive {\oii} optical recombination-line survey of planetary
nebulae (Tsamis et al. 2002b), it is assumed that \emph{LS}-coupling
holds for the 3s--3p
transitions, while intermediate coupling is assumed for those between 3p--3d
and 3d--4f states (Liu et al. 1995a). Table~7 is important both for checking
whether observation agrees with theory and in excluding some lines from further
consideration when blending or misidentification are suspected.

\subsubsection{Relative intensities of {\oii} ORLs}

In contrast to our results from a comparison of PN {\oii} relative line
intensities (Tsamis et al. 2002b), there is clear evidence that several
transitions are
stronger than expected in all five {\hii} regions studied (Table~7). The effect
is especially pronounced amongst the lines of {\oii} multiplet V\,1; those of
multiplet V\,10, however, are in better agreement with theory even though their
formal measurement errors are somewhat larger due to partial blending with the
{\fsii} $\lambda\lambda$4068, 4076 doublet. A similar situation has been
reported by Esteban et al. (1999) from their echelle observations of the Lagoon
Nebula (M\,8). Esteban et al. (1998) however, found quite good agreement
between {\oii} multiplet V\,1 lines from the Orion Nebula, a result supported
by our examination of the relative intensities of the same transitions using
the recent Orion line atlas by Baldwin et al. (2000), which has better spectral
resolution than the data used by the former authors. Our observations show that
for all five {\hii} regions the $\lambda$4638.86 ($J$~=~1/2--3/2) and
$\lambda$4650.84 ($J$~=~1/2--1/2) V\,1 transitions are enhanced
from each nebula: by a factor ranging from $\sim$\,2.8--7.3 for the
five objects under study; for Orion this factor is only $\sim$\,1.4 for
both lines, but less for the other V\,1 transitions. The $\lambda$4661.63
($J$~=~3/2--3/2) line is also abnormally strong compared to theory.

One possible explanation for this behaviour involves the breakdown of thermal
equilibrium among the fine structure levels of the parent $^3$P$_{0,\,1,\,2}$
ground term of recombining {\opp}. Our analysis of the {\oii} 3s--3p
transitions employed term-averaged effective recombination coefficients from
Storey (1994) which were calculated assuming that the {\oii} 3s and 3p levels
are well described by \emph{LS}-coupling. When this is the case, the population
distribution among the O$^{2+}$ $^3$P$_{0,1,2}$ levels has no effect on the
recombination rate to these levels. If, however, there is a breakdown of
\emph{LS}-coupling so that the recombination rate coefficients from the
individual $^3$P$_J$ levels differ, then the total recombination coefficient to
a particular {\oii} 3p state becomes a function of the population distribution
among the $^3$P$_J$ levels. There are no published results for the {\oii}
recombination coefficients that take account of such effects but we have made a
trial calculation of the inverse process of photoionisation from the 3p
$^4$D$_{7/2}$ level, the upper state of the $\lambda$4649.13 transition. This
calculation shows that the direct recombination to this level comes
overwhelmingly from the O$^{2+}$ $^3$P$_2$ level. Recombination from the
$^3$P$_{0,1}$ levels is negligible. Therefore, if the population of the
O$^{2+}$ $^3$P$_2$ level falls below that expected in thermal equilibrium the
true recombination rate to the 3p $^4$D$_{7/2}$ level will be less than that
predicted by the \emph{LS}-coupling results of Storey (1994) and the
$\lambda$4649.13 line will be observed to be weaker than is predicted by the
\emph{LS}-coupling theory. The relative populations of the O$^{2+}$
$^3$P$_{0,1,2}$ levels will depart from thermal equilibrium if the electron
density of the nebula is lower than the critical densities of one or both of
the $^3$P$_{1,2}$ levels [$\sim$\,500 and 3500\,{\cmt}]. The mean electron
density derived from various diagnostics for the objects under study here is in
all cases lower than 2000\,{\cmt}, while for the three Magellanic Cloud {\hii}
regions, which show the greatest deviations from the predicted
\emph{LS}-coupling relative intensities for {\oii} V~1 multiplet
transitions,\footnote{Large deviations are still present even if corrections
are made for {\oii} absorption features in the nebular continuum component,
attributable to dust-scattered starlight (see Section~5.2.2).} it is lower than
1000\,{\cmt} (Table~3); this is in contrast to M\,42, for which Esteban et al.
(1998) find {\eld} $\simeq$ 5000\,{\cmt}. Thus, it seems that within multiplet
V\,1 the strongest expected line, namely $\lambda$4649.13 ($J$~=~5/2--7/2), is
weakened at the expense of transitions between levels of lower $J$. This
argument seems to be supported by our analysis of {\oii} ORLs originating from
planetary nebulae (Liu et al. 1995a, 2000, 2001b; Tsamis
2002; Tsamis et al. 2002b). For
example, the observed relative intensities of V\,1 lines from the dense
planetary nebulae IC\,4191 and NGC\,5315 are in perfect agreement with theory
(Tsamis et al. 2002b); the derived mean electron densities of these
objects are 10700 and
14100\,{\cmt} respectively. The opposite is true in the case of the PN
NGC\,3132 ({\eld}\,$\simeq$\,600\,{\cmt}), where the V\,1 lines display
abnormal ratios (Tsamis et al. 2002b), just as they do for the {\hii}
regions considered
here. We will return to this issue in a future paper (Tsamis et
al. 2002b), by combining our
{\hii} region ORL dataset from this paper with the PN ORL dataset
of Tsamis et al. (2002b) in order to plot the relative intensities of
{\oii} V~1 multiplet components as a function of nebular electron density.

If the above interpretation is correct, the total observed
intensity of the whole multiplet should not be affected and can be used to
derive a reliable ORL {\opp}/{\hp} abundance. A similar effect may
dictate the relative intensities of lines from the 3d--4f group,
weakening the $\lambda$4089 transition (Liu 2002b).

A major point to be deduced from this analysis is that the observed relative
intensities of {\oii} recombination lines point towards their origin in low
density gas ($<$\,3500\,{\cmt}), similar to that emitting the {\foiii} and
other CELs.

\setcounter{table}{7}
\begin{table*}
\centering
\begin{minipage}{150mm}
\caption{Continuum emission and scattered light, expressed as log\,$I^{\rm
c}(\lambda)$/\emph{I}(H$\beta$).}
\begin{tabular}{lcccccccccc}
\noalign{\vskip3pt}\noalign{\hrule}\noalign{\vskip3pt}
               &\multicolumn{2}{c}{M\,17}                 &\multicolumn{2}{c}{NGC\,3576}                    &\multicolumn{2}{c}{30~Doradus}               &\multicolumn{2}{c}{LMC~N11B}              &\multicolumn{2}{c}{SMC~N66}           \\
               &$I^{\rm c}(\lambda)$   &$I^{\rm c}(\lambda)_{\rm d}$   &$I^{\rm c}(\lambda)$   &$I^{\rm c}(\lambda)_{\rm d}$        &$I^{\rm c}(\lambda)$   &$I^{\rm c}(\lambda)_{\rm d}$    &$I^{\rm c}(\lambda)$   &$I^{\rm c}(\lambda)_{\rm d}$   &$I^{\rm c}(\lambda)$   &$I^{\rm c}(\lambda)_{\rm d}$ \\
\noalign{\vskip3pt}\noalign{\hrule}\noalign{\vskip3pt}
$\lambda$4089     &$-$2.402              &$-$2.508                          &$-$2.218                &$-$2.279                                 &$-$2.433                 &$-$2.564                          &$-$2.063                 &$-$2.135                       &$-$2.103                  &$-$2.168          \\
$\lambda$4650     &$-$2.645              &$-$2.806                          &$-$2.435                &$-$2.522                                 &$-$2.624                 &$-$2.797                          &$-$2.278                 &$-$2.357                       &$-$2.214                  &$-$2.283            \\
\noalign{\vskip3pt} \noalign{\hrule}
\end{tabular}
\end{minipage}
\end{table*}

\subsubsection{Continuum observations and scattered light}

An issue that had to be addressed in the course of this analysis is the
potential influence of dust-scattered stellar light on the intensities of weak
nebular emission features such as the {\oii} ORLs of interest. Unlike PNe,
where the size of the emitting region results in relatively small dust columns,
the situation is different in H~{\sc ii} regions whose large volumes contain
much larger dust columns. The dust effectively scatters light from the nebula's
illuminating stars, which then makes up a major fraction of the observed
continuum at UV and optical wavelengths. It is thus possible that emission or
absorption features in the spectra of the exciting stars in an {\hii} region
may contaminate the nebular emission spectrum. Peimbert et al. (1993) dealt
with this problem in an analysis of the M\,42 and M\,17 {\hii} regions. They
employed previously published medium resolution (5--7\,\AA) spectroscopic data
in order to derive ORL {\opp}/{\hp} abundances; their resolution however was
not adequate to allow for the effect of blending of {\oii} ORLs with other
lines e.g. [Fe~{\sc ii}], N~{\sc iii} and C~{\sc iii}, so they resorted to
older photographic line intensities in order to estimate the necessary
corrections. Our CCD spectrophotometry is of significantly higher resolution
(1, 1.5 and 2\,\AA~for the coverage of {\oii} ORLs) and our line intensities
were consistently corrected for blends with other nebular emission lines.

In order to investigate the potential effect of dust-scattered stellar light on
the nebular line fluxes, we have measured the continuum emission at
4089~\AA\ and 4650~\AA , which coincide with the strongest 3d--4f
{\oii} transition at 4089.3\,\AA, and {\oii} multiplet V\,1 ORLs, respectively.
In Table~8 the continuum intensity, $I^{\rm c}(\lambda)$, and the
scattered
light contribution, $I^{\rm c}(\lambda)_{\rm d}$, to the observed continuum are
presented. The observed continuum is mainly due to two components: the
nebular atomic continua and dust-scattered starlight, so that,
\[
I^{\rm c}(\lambda) = I^{\rm c}(\lambda)_{\rm a} + I^{\rm c}(\lambda)_{\rm d},
\]
where $I^{\rm c}(\lambda)_{\rm a}$ represents the sum of the {\hi} and {\hei}
atomic continua calculated using emissivities from Storey \& Hummer (1995) and
Brown \& Mathews (1970), respectively; the {\elt}, {\eld} and {\hep}/{\hp} for
each object were taken into account. We find that for M\,17, dust-scattered
light accounts for 78~percent and 69\,percent of the observed continuum at
4089~\AA\ and 4650~\AA , respectively; for NGC\,3576 these values are
87~percent and 82\,percent; for 30~Doradus the corresponding values are
74~percent and 67\,percent; for LMC~N11B they are 85~percent and 83\,per cent;
finally, for SMC~N66 they are 86~percent and 85\,percent, respectively
(sky-subtraction has not been carried out; the sky contribution was estimated
to be negligible for these dark of Moon conditions).\footnote{A typical
dark-of-Moon sky brightness at $B$ (4400\,\AA) is 22$^m$.8~per arcsec$^2$
corresponding to
4.73\,$\times$\,10$^{-18}$\,erg\,s$^{-1}$\,cm$^{-2}$\,\AA$^{-1}$\,arcsec$^{-2}$
when $B$~=~0 is
6.24\,$\times$\,10$^{-9}$\,erg\,s$^{-1}$\,cm$^{-2}$\,\AA$^{-1}$; thus for e.g.
30~Doradus the sky contribution to the observed continuum at $\lambda$4650 is
$\sim$~0.082\,per cent only.}

In Table~9 we present estimated equivalent widths, $EW_{\rm abs}(neb)$, for
stellar lines in absorption (in m\AA) in the observed nebular continuum, given
by,
\begin{equation}
EW_{\rm abs}(neb) = EW_{\rm abs}(stellar)I^{\rm c}(\lambda)_{\rm d}/I^{\rm
c}(\lambda),
\end{equation}
with the continuum intensity values being those given in Table~8. This analysis
pertains to the stellar content of 30~Doradus in the following way: according
to Walborn \& Blades 1997 (WB97), the visually brightest stars in the
30~Doradus association are B-type supergiants, along with the Of-type star
R\,139 (= Parker 952). Therefore, we will assume that such stellar spectra
dominate the observed dust-scattered light. We have singled out seven stars
which are the brightest in the central $\sim$1\,arcmin$^2$ of the cluster
(excluding the compact core R\,136): R\,137~=~P\,548 [B0.7--1.5\,I,
$V$~=~12.14]; R\,138 [A0\,Ia, $V$~=~11.87]; R\,139\footnote{Whereas Moffat et
al. (1987) and Moffat (1989) classify R\,139 as a WNL/Of-type binary system,
WB97 classify it as a O7\,Iafp-type single star. Our spectrum shown in Fig.\,~2
is of higher resolution than the one of WB97 and supports a classification of
O6.5\,Ib(f) + WNL for R\,139.}~=~P\,952 [WNL + Of, $V$~=~11.94];
R\,140~=~P\,877, 880 [WN + WC, $V$~=~12.22, 12.79]; R\,141~=~P\,1253
[BN0.5\,Ia, $V$~=~12.57]; R\,142~=~P\,987 [B0.5--0.7\,I, $V$~=~11.91]; and
P\,767 [O3\,If*, $V$~=~12.87].

We have high resolution spectrograms (at 0.9\,\AA/pix) of R\,139 (Fig.\,~2) and
R\,140 (Fig.\,~3), since they fell on our slit; the remaining five stars were
positioned on either side of it. It is assumed that the spectra of R\,137 and
R\,142 are similar to that of Parker~3157 ($V$~=~12.47; Parker et al.
1992)\footnote{Parker et al. (1992) classify P\,3157 as spectral type BC1\,Ia.
However, our analysis of its spectrum (Fig.\,~4) does not reveal the presence
of any C~{\sc iii} lines at 4650\,\AA. The absorption features at that
wavelength can be fitted purely by O~{\sc ii} V\,1 multiplet lines. We
therefore suggest a revised spectral type of B1\,Ia for P\,3157.} a supergiant
belonging to the LMC LH\,10 (N11) association, whose spectrogram (at
0.9\,\AA/pix) was extracted from our LMC~N11B frames (Fig.\,~4). High
resolution spectra of R\,141 and P\,767 can be found in WB97; we do not have a
spectrum of R\,138, but we can safely assume that it is featureless at the
wavelengths of interest judging from its spectral type.

\begin{figure*}
\centering \epsfig{file=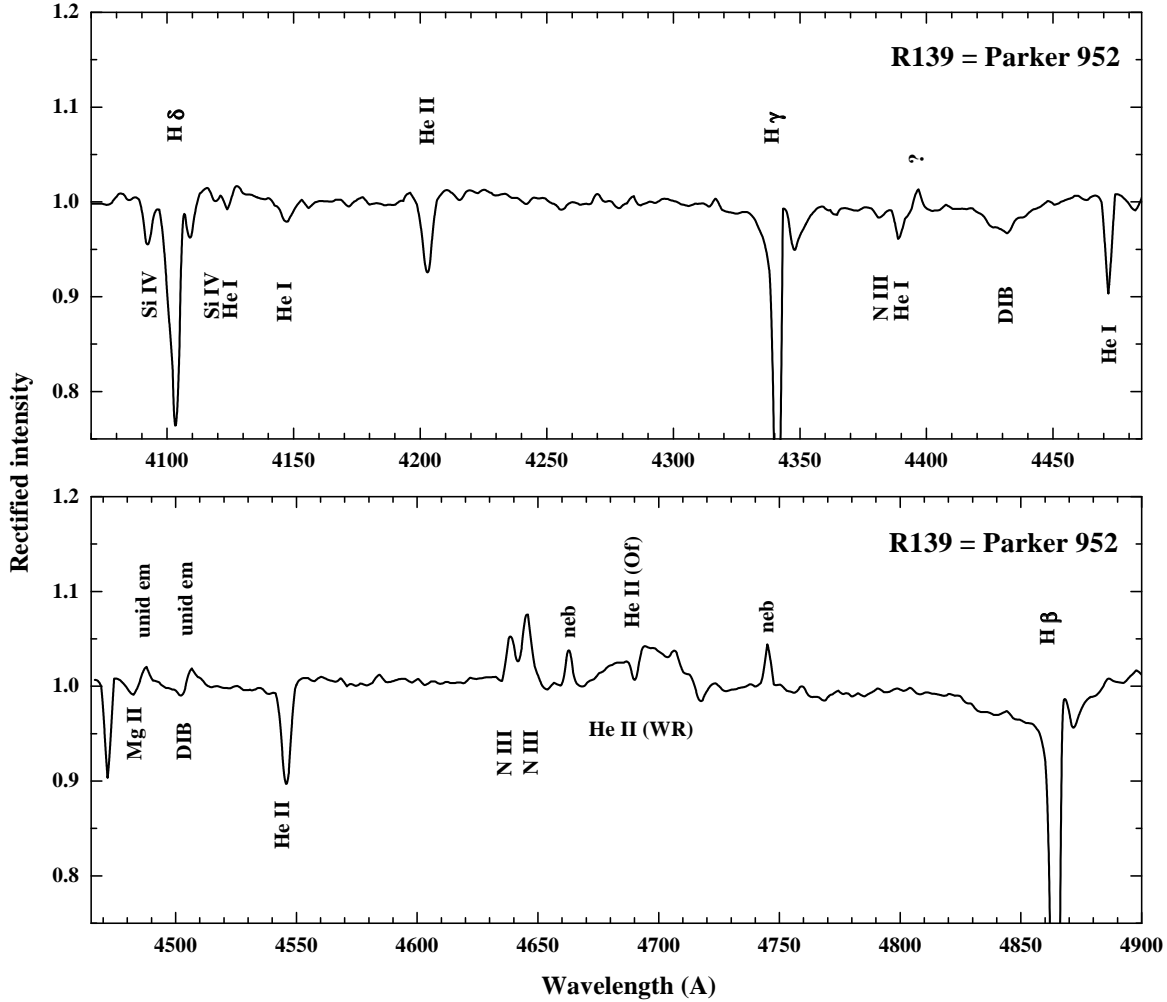, width=16.2 cm, clip=,} \caption {Rectified
blue-violet spectrogram of the O6.5\,Ib(f) + WNL multiple star R\,139 (= Parker
952) in 30~Doradus. The identified features are: H$\delta$ $\lambda$4102,
H$\gamma$ $\lambda$4340, H$\beta$ $\lambda$4861; He~{\sc i}
$\lambda\lambda$4121, 4144, 4387, 4471; He~{\sc ii} $\lambda\lambda$4200, 4541,
4686 (broad WR emission + narrow absorption); N~{\sc iii} $\lambda$4379,
($\lambda\lambda$4638, 4640+4641 emission); Si~{\sc iv} $\lambda\lambda$4089,
4116; Mg~{\sc ii} $\lambda$4481; the diffuse interstellar bands at
$\lambda\lambda$4430, 4502 and the unidentified emission lines at
$\lambda\lambda$4485, 4503. See text for more details.}
\end{figure*}

\begin{figure*}
\centering \epsfig{file=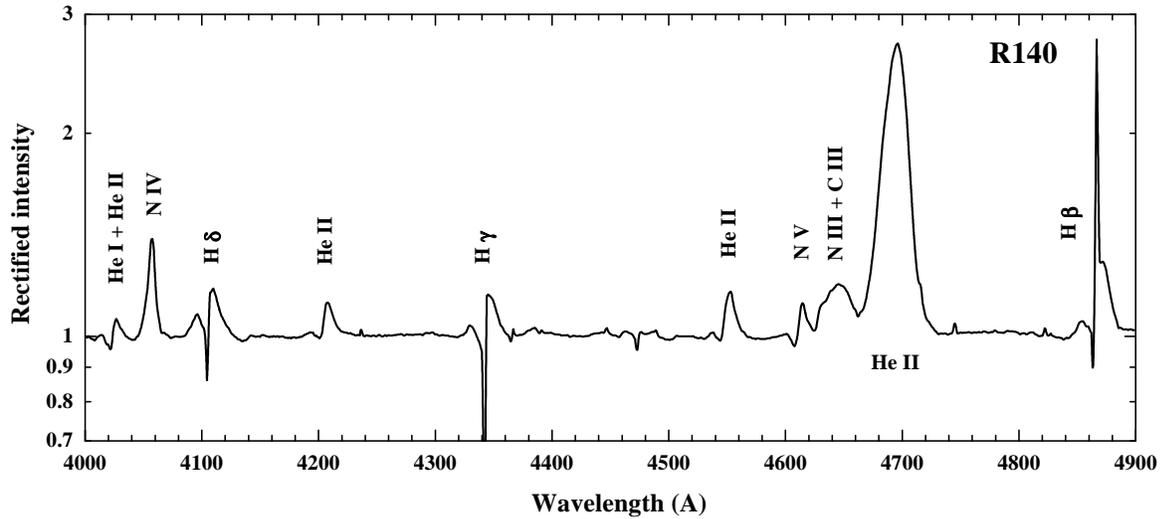, width=16.2 cm, clip=,} \caption[Blue-violet
spectrogram of the WN6(h) component of the multiple star R\,140 in 30 Dor.]
{\label{fig:r140}Rectified log-intensity, blue-violet spectrogram of the WN6(h)
component of the multiple star R\,140 in 30~Doradus. The identified features
are: H$\delta$ $\lambda$4102, H$\gamma$ $\lambda$4340, H$\beta$ $\lambda$4861,
He~{\sc i} $\lambda$4026; He~{\sc ii} $\lambda\lambda$4026, 4200, 4541, 4686;
N~{\sc iii} $\lambda\lambda$4638 + 4640 + 4641; N~{\sc iv} $\lambda$4058;
N~{\sc v} $\lambda\lambda$4603, 4619; see text for more details.}
\end{figure*}

\setcounter{table}{8}
\begin{table}
\begin{center}
\caption{Estimated equivalent widths (in units of m\AA).}
\begin{tabular}{lcccc}
\noalign{\vskip3pt}\noalign{\hrule}\noalign{\vskip3pt}
                &\multicolumn{2}{c}{30~Doradus}   &\multicolumn{2}{c}{LMC~N11B}\\
                &$EW_{\rm em}$  &$EW_{\rm abs}(neb)$          &$EW_{\rm em}$   &$EW_{\rm abs}(neb)$      \\
\noalign{\vskip3pt}\noalign{\hrule}\noalign{\vskip3pt}
$\lambda$4072   &185            &107                          &*              &*                         \\
$\lambda$4089   &57             &87                           &151            &98                       \\
$\lambda$4640   &637            &37                           &286            &167                      \\
$\lambda$4650   &540            &185                          &264            &298                      \\
$\lambda$4661   &238            &40                           &121            &78                       \\
\noalign{\vskip3pt} \noalign{\hrule}
\end{tabular}
\end{center}
\end{table}

The {\oii} spectrum reaches a sharp maximum in absorption in stars of early-B
spectral type (e.g. Walborn \& Fitzpatrick 1990), thus the very existence of
B-type supergiants in nebulae complicates the analysis of nebular {\oii} ORLs.
It is therefore recommended that fields of view relatively clear of such stars
are chosen for future weak-emission line analyses of {\hii} regions. The
stellar absorption and/or emission features that are of relevance to this
analysis are: Si~{\sc iv}\,+\,{\oii} $\lambda$4089, N~{\sc iii}\,+\,{\oii}
$\lambda\lambda$4638, 4640, C~{\sc iii}\,+\,{\oii} $\lambda$4650 and {\oii}
$\lambda$4661.

Values of $EW_{\rm abs}(stellar)$ were estimated at wavelengths of interest on
the stellar spectrograms mentioned above and subsequently averaged, weighted
according to the brightness ratios of the seven illuminating stars in the blue
wavelength region, i.e., 5.6 : 6.3 : 4.3 : 6.8 : 3.2 : 9.8 : 1.0 (for R\,137 :
R~138 : R~139 : R~140 : R~141 : R\,142 : P\,767). We have assumed a
total-to-selective extinction ratio of $R_{B}$~=~$A_{B}/E(B-V)$~=~5 
(Hill et al. 1993) and
adopted apparent magnitudes and ($B$--$V$), $E$($B$--$V$) values from Parker
(1993). We were then able to compute the $EW_{\rm abs}(neb)$ values in Table~9
using Eq.\,~1. To correct the observations, one should add these to the
observed $EW_{\rm em}$ values.

\begin{figure*}
\centering \epsfig{file=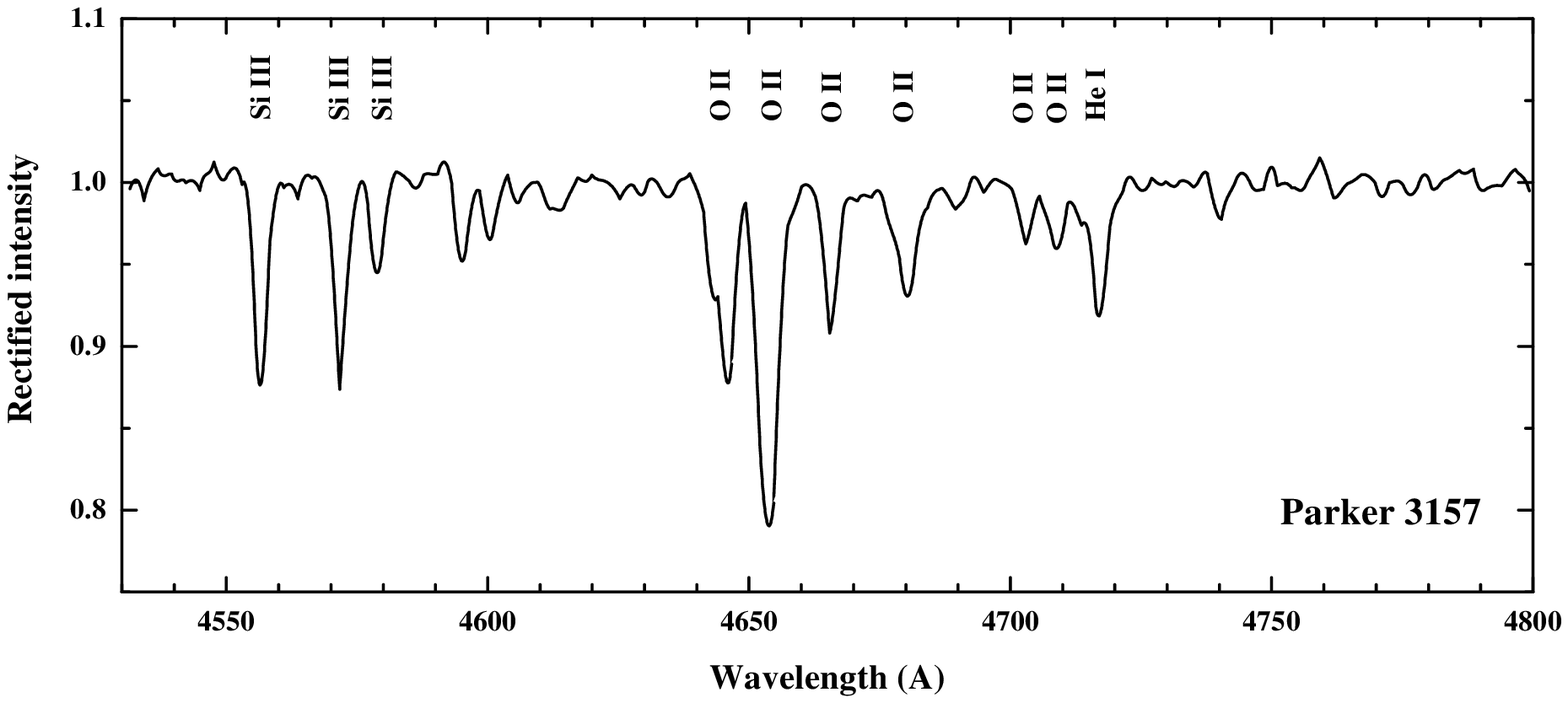, width=16.2 cm, clip=,} \caption {Rectified,
blue spectrogram of the B1\,Ia-type supergiant Parker~3157 in LMC~N11B which
has a full complement of {\oii} V\,1 multiplet absorption lines; see text for
more details.}
\end{figure*}

Considering LMC~N11B, we found the early-B supergiant Parker~3157, with its
extremely rich {\oii} absorption line spectrum, directly in the field of view
of the spectrograph slit (Fig.\,~4). Absorption lines in the scattered light
from this star are responsible for a significant drop in the intensity of the
observed continuum in the region of the $\lambda\lambda$4638, 4640, 4649, 4650
and $\lambda$4661 lines of multiplet V\,1. In contrast to 30~Doradus however,
the dust-scattered light in N11B does not contain any obvious contribution from
Of- and WR-type stars equivalent to e.g. R\,139 or R\,140 (no Wolf-Rayet stars
are listed in the census of N11B by Parker et al. 1992), whose
\emph{emission} lines
in the $\lambda$4650 region can partly compensate absorption lines due to
early-B supergiants, resulting in a fairly level continuum at those wavelengths
for the former nebula. Therefore, together with the supergiant P\,3157, we will
assume that the observed continuum in N11B in dominated by the scattered light
of six other stars, which are visually the brightest amongst those studied by
Parker et al. (1992); they are: P\,3209 [O3\,III(f*), $V$~=~12.66]; 
P\,3252 B2\,II,
$V$~=~11.35]; P\,3271 [B1\,II, $V$~=~12.99]; P\,3070 [O6\,V, $V$~=~12.75];
P\,3223 [O8.5\,IV, $V$~=~12.95]; and P\,3120 [O5.5\,V\,((f*)), $V$~=~12.80].
High resolution rectified spectrograms for these stars were presented by Parker
et al. and were used to estimate mean $EW_{\rm abs}(stellar)$ values. More
accurate measurements were possible for P\,3157 for which we have a digital
spectrogram. A cross-comparison between our digital spectrogram and those
presented by Parker et al. (1992), including P\,3157, shows agreement,
pointing
towards a reliable estimate of the $EW$s.

In Table~10 we present the measured intensities and the derived abundances from
{\oii} ORLs for all five {\hii} regions, {\em before any correction for
scattered light contamination}. Along with values obtained from individual
transitions, abundances derived from total multiplet intensities are also
listed (in bold face) and are discussed in more detail below for each object.
In the context of that discussion revised abundance ratios are then presented,
incorporating corrections for the effects of the scattered stellar continua on
the measured
nebular emission line intensities wherever possible. \\



\setcounter{table}{9}
\begin{table}
\begin{minipage}{75mm}
\caption{Recombination line O$^{2+}$/H$^+$ abundances.$^a$}
\centering
\begin{tabular}{l@{\hspace{5mm}}c@{\hspace{8mm}}l@{\hspace{8mm}}r}
\noalign{\hrule} \noalign{\vskip3pt}
$\lambda_{\mathrm{0}}$ &Mult. &$I_{\rm obs}$ &$\displaystyle\frac{\rm{~O}^{2+}}{\rm{H}^+}$ \\
\noalign{\vskip2pt}
(\AA)         &      &        &(\tmf)\\
\noalign{\vskip3pt} \noalign{\hrule} \noalign{\vskip3pt}

\multicolumn{4}{c}{\underline{M\,17}}\\
\noalign{\vskip3pt}
4638.86   &V1      &.1056          &10.26    \\
4641.81   &V1      &.1524          & 5.87    \\
4649.13   &V1      &.1375          & 2.78    \\
4650.84   &V1      &.0871          & 8.46    \\
4661.63   &V1      &.1043          & 7.93    \\
4676.24   &V1      &.0630          & 5.70    \\   
\noalign{\vskip3pt}
\bf{V\,1 3s$^{4}$P-3p$^{4}$D$^\mathrm{{o}}$}  & &\bf{0.650} &\bf{5.41}  \\
\noalign{\vskip3pt}
4317.14  &V2      & .1115:          & 14.82:   \\    
4319.63  &V2      & .0330           &  4.07    \\
4345.56  &V2      & .0670           &  8.55    \\
4349.43  &V2      & .1070           &  5.67    \\
\noalign{\vskip3pt}
\bf{V\,2 3s$^{4}$P-3p$^{4}$P$^\mathrm{{o}}$}  & &\bf{0.207} &\bf{5.95}  \\
\noalign{\vskip5pt} \noalign{\hrule} \noalign{\vskip3pt}
\bf{\emph{Adopted}}   & &                 & \bf{5.68}   \\
\noalign{\vskip3pt} \noalign{\hrule}

\noalign{\vskip5pt}
\multicolumn{4}{c}{\underline{NGC\,3576}}\\
\noalign{\vskip3pt}
4638.86   &V1    & .0509              & 4.94   \\
4641.81   &V1    & .0880              & 3.39   \\
4649.13   &V1    & .0929              & 1.88   \\
4650.84   &V1    & .0553              & 5.37   \\
4661.63   &V1    & .0916              & 6.96   \\
4676.24   &V1    & .0560              & 5.07    \\  
\noalign{\vskip3pt}
\bf{V\,1 3s$^{4}$P-3p$^{4}$D$^\mathrm{{o}}$}  & &\bf{0.435} &\bf{3.62}  \\
\noalign{\vskip3pt}
4069.62 &V10    &.1409:              &5.45:   \\    
4072.16 &V10    &.1207               &5.01    \\
4075.86 &V10    &.1010               &2.90    \\
4085.11 &V10    &.1700               &3.78     \\
\noalign{\vskip3pt}
\bf{V\,10 3p$^{4}$D$^\mathrm{{o}}$}-3d$^{4}$F  & &\bf{0.239} &\bf{3.77}  \\
\noalign{\vskip3pt}
4087.15  &V48c    & .0415               &    13.49     \\
4089.29  &V48a    & .0320               &     2.82     \\   
4275.55  &V67a    & .1339               &     9.85     \\
\noalign{\vskip3pt}
\bf{3d-4f}  & &\bf{0.207} &\bf{7.40:} \\   
\noalign{\vskip5pt} \noalign{\hrule} \noalign{\vskip3pt}
\bf{\emph{Adopted}}       &&             &\bf{3.70}      \\
\noalign{\vskip3pt} \noalign{\hrule} \noalign{\vskip5pt}

\multicolumn{4}{c}{\underline{30 Doradus}}  \\
\noalign{\vskip3pt}
4638.86   &V1       &  .0640       &   6.11     \\
4641.81   &V1       &  .0854       &   3.23     \\
4649.13   &V1       &  .0642       &   1.28     \\
4650.84   &V1       &  .0641       &   6.12     \\
4661.63   &V1       &  .0565       &   4.23     \\
4673.73   &V1       &  .0120        &  5.80     \\ 
4676.24   &V1       &  .0224        &  1.99     \\
\noalign{\vskip3pt}
\bf{V\,1 3s$^{4}$P-3p$^{4}$D$^\mathrm{{o}}$}  & &\bf{0.369} &\bf{2.97}  \\
\noalign{\vskip3pt}
4317.14   &V2       &  .0705       & 6.96        \\
4319.63   &V2       &  .0876       & 10.91      \\
\noalign{\vskip3pt}
\bf{V\,2 3s$^{4}$P-3p$^{4}$P$^\mathrm{{o}}$}  & &\bf{0.158:} &\bf{8.42:}  \\
\noalign{\vskip3pt}
4414.90   &V5       &.0331:          & 6.27:      \\  
4416.97   &V5       &.0436:          & 14.8:      \\  
\noalign{\vskip3pt}
\bf{V\,5 3s$^{2}$P-3p$^{2}$D$^\mathrm{{o}}$}  & &\bf{0.033:} &\bf{6.27:}  \\
\noalign{\vskip3pt}
4069.89   &V10     & .1328          &  5.14      \\
4072.16   &V10     & .0683          &  2.84      \\
4078.84   &V10     & .0324          &  8.86      \\
4085.11   &V10     & .0258          &  5.74      \\
\noalign{\vskip3pt}
\bf{V\,10 3p$^{4}$D$^\mathrm{{o}}$}-3d$^{4}$F  & &\bf{0.259} &\bf{4.47}  \\ 
\noalign{\vskip3pt}
4132.80   &V19    &.0508            &  9.29   \\  

\end{tabular}

\end{minipage}
\end{table}

\setcounter{table}{9}
\begin{table}
\caption{{\it --continued}}
\centering
\begin{minipage}{75mm}
\begin{tabular}{l@{\hspace{5mm}}c@{\hspace{8mm}}l@{\hspace{8mm}}r}
\noalign{\hrule} \noalign{\vskip3pt}
$\lambda_{\mathrm{0}}$ &Mult. &$I_{\rm obs}$ &$\displaystyle\frac{\rm{~O}^{2+}}{\rm{H}^+}$ \\
\noalign{\vskip2pt}
(\AA)         &      &        &(\tmf)\\
\noalign{\vskip3pt} \noalign{\hrule} \noalign{\vskip3pt}

4153.30   &V19    &.0609            & 7.80    \\
4156.53   &V19    &.0232:           & 18.6:    \\  
\noalign{\vskip3pt}
\bf{V\,19 3p$^{4}$P$^\mathrm{{o}}$}-3d$^{4}$P  & &\bf{0.112} &\bf{8.41:}   \\
\noalign{\vskip3pt}
4110.78   &\bf{V20}     & .0322:         &\bf{13.4:}    \\
\noalign{\vskip3pt}
4906.83   &\bf{V28}     & .0369:          &\bf{14.8:}     \\  
\noalign{\vskip3pt}
4083.90   & V48b   & .0108          &   3.41    \\
4087.15   & V48c   & .0185          &   6.17    \\
4089.29   & V48a   & .0210          &   1.90    \\
4275.55   & V67a   & .0785:          &   5.33:    \\  
4288.82   & V53c   & .1091:          &  11.85:    \\  
4303.83   & V53a   & .0519:          &   9.97:    \\  
4313.44   & V78a   & .0383:          &  28.5:     \\  
4315.69   & V63c   & .0533:          &  43.2:     \\  
4609.44   & V92a   & .0209          &   3.35    \\  
\noalign{\vskip3pt}
\bf{3d-4f}  & &\bf{0.071} &\bf{3.03} \\
\noalign{\vskip5pt} \noalign{\hrule} \noalign{\vskip3pt}
\bf{\emph{Adopted}} &&               &\bf{3.49} \\
\noalign{\vskip3pt} \noalign{\hrule} \noalign{\vskip5pt}

\multicolumn{4}{c}{\underline{LMC N11B}}\\
\noalign{\vskip3pt}
4638.86   &V1     & .1170              &11.24   \\
4641.81   &V1     & .0530               & 2.02   \\
4649.13   &V1     & .0760                & 1.52   \\
4650.84   &V1     & .0631                & 6.06   \\
4661.63   &V1     & .0640               & 4.81   \\
4673.73   &V1     & .0308                &14.93   \\
4676.24   &V1     & .0271                & 2.43   \\
4696.35   &V1     & .0346                & 24.07  \\  
\noalign{\vskip3pt}                                    
\bf{V\,1 3s$^{4}$P-3p$^{4}$D$^\mathrm{{o}}$}  & &\bf{0.466} &\bf{3.14}  \\
\noalign{\vskip3pt}
4349.43   &V2     & .0871               & 4.60   \\
\noalign{\vskip3pt}
\bf{V\,2 3s$^{4}$P-3p$^{4}$P$^\mathrm{{o}}$}  & &\bf{0.087} &\bf{4.60}  \\
\noalign{\vskip3pt}
4072.16   &V10    &.1027               & 4.27   \\
\noalign{\vskip3pt}
\bf{V\,10 3p$^{4}$D$^\mathrm{{o}}$}-3d$^{4}$F  & &\bf{0.103} &\bf{4.27}   \\
\noalign{\vskip3pt}

4153.30   &V19    &.0982                & 12.54  \\
\noalign{\vskip3pt}
\bf{V\,19 3p$^{4}$P$^\mathrm{{o}}$}-3d$^{4}$P  & &\bf{0.098} &\bf{12.54}   \\

\noalign{\vskip3pt}
4089.29   & V48a  &.1305               & 11.41  \\
\noalign{\vskip3pt}
\bf{3d-4f}  & &\bf{0.131}             &\bf{11.41}  \\
\noalign{\vskip5pt} \noalign{\hrule} \noalign{\vskip3pt}
\bf{\emph{Adopted}} & &&\bf{7.19} \\    
\noalign{\vskip3pt} \noalign{\hrule} \noalign{\vskip5pt}

\multicolumn{4}{c}{\underline{SMC N66}}\\
\noalign{\vskip3pt}
4638.49    &V1   &.0311         &2.91       \\
4649.13    &V1   &.0545         &1.06       \\
4650.84    &V1   &.0513         &4.78       \\
4661.63    &V1   &.0800         &5.84       \\
\noalign{\vskip3pt}
\bf{V\,1 3s$^{4}$P-3p$^{4}$D$^\mathrm{{o}}$}  & &\bf{0.217} &\bf{2.15}  \\
\noalign{\vskip3pt} \noalign{\hrule} \noalign{\vskip3pt}
\bf{\emph{Adopted}} & & &\bf{2.15} \\
\noalign{\vskip3pt} \noalign{\hrule}

\end{tabular}
\begin{description}
\item $^a$ Values followed by `:' have not been used when adopting
mean abundance ratios.
\end{description}
\end{minipage}
\end{table}

\noindent{\bf{M\,17}}: The two triplet 3s--3p multiplets V\,1 and V\,2 are
detected, yielding similar results. The mean {\opp}/{\hp} abundance ratio is
5.68\,$\times$\,{\tmf}. Peimbert et al. (1993) have derived a very similar
value of 4.85\,$\times$\,{\tmf} using multiplets V\,1 and V\,5 after correcting
the line intensities for underlying absorption due to dust-scattered stellar
light. We have not taken into account such an effect for this nebula, since
Peimbert et al.'s corrections amount to only 14\,per cent for multiplet V\,1,
the one affected the most. A comparison with the forbidden-line abundance from
Table~3 yields an ORL/CEL abundance ratio of 2.1 for {\opp} ({\S 7.1},
Table~12).
\\

\noindent{\bf{NGC\,3576}}: Multiplets V\,1 and V\,10 yield consistent results;
lines from the 3d--4f group display abnormal intensity ratios. It is likely
that underlying absorption is affecting the $\lambda$4089 transition. We do not
attempt any corrections and adopt for the ORL {\opp}/{\hp} ratio the mean of
the V\,1 and V\,10 results, i.e. 3.70\,$\times$\,{\tmf}. The ORL/CEL ratio for
{\opp} in this case is
1.8 (Table~12).\\

\noindent{\bf{30 Doradus}}: We have detected lines from the 3p--3d multiplets
V\,10, V\,19, V\,20, as well as from the 3s--3p multiplets V\,1, V\,2, V\,5,
and several transitions from the 3d--4f group. Abundances derived from V\,2 and
V\,19 agree very well with each other, but are about a factor of 2.8 higher
than those derived from multiplet V\,1 and 88\,per cent higher than those from
multiplet V\,10. The relative intensities amongst the multiplet V\,19
components are in good agreement with theory (Table~7), apart from
$\lambda$4156.63 ($J$~=~5/2--3/2), which is a factor of 2 stronger than
expected relative to $\lambda$4132.80 ($J$~=~1/2--3/2), probably as a result of
blending.\footnote{This is a ratio of lines originating from the same upper
level and thus invariably fixed by the ratio of their transition probabilities;
the fact that it is observed to differ between several nebulae (e.g. NGC\,3242,
3918, 5882, 6153 and 30~Doradus) by a factor of more than 4 suggests that
blending with an unknown line affects the $\lambda$4156 line.} The V\,1 lines
were discussed in the previous subsection. This abundance discrepancy is quite
striking, since from observations of 10 PNe (Tsamis et al. 2002b) we have
found that on
average multiplet V\,2 yields {\opp} abundances that are 24\,per cent
\emph{lower} than those from V\,1 lines. Furthermore, {\opp} abundances from
multiplets V\,10 and V\,19 generally agree to within 40\,per cent. Case
dependence does not seem to be an issue since multiplets V\,1 and V\,10 are
almost insensitive to optical depth effects, while it is estimated that
multiplets V\,2 and V\,19, which are assumed here to be under Case B, would
yield even higher abundances under Case A.

A single line from V\,20 is detected, $\lambda$4110.78 ($J$~=~3/2--1/2), which
yields a similar abundance to those from the V\,2 and V\,19 lines; the
remaining V\,20 lines coincide with {\hei} $\lambda$4120.84 and cannot be
reliably used to derive an independent abundance estimate. Of the two detected
multiplet V\,5 lines, $\lambda$4416.97 ($J$~=~1/2--3/2) is most probably
blended with [Fe~{\sc ii}] $\lambda$4416.27 and is excluded from the analysis,
while $\lambda$4414.90 ($J$~=~3/2--5/2) yields an abundance ratio rather
consistent with that from the multiplet V\,10 lines.

Multiplet V\,2 is fed from high-lying terms via $^4$P$^{\rm
o}$--$^4$D~\,$\lambda$4119 (V\,20) and less via V\,19 transitions. Multiplet
V\,19 arises from the 3d\,$^4$P term which can be reached via resonance
transitions from the {\op} 2p$^3$\,$^4$S$^{\rm o}$ ground state -- e.g. the
2p$^3$\,$^4$S$^{\rm o}$--3d\,$^4$P~\,$\lambda$430 line -- themselves excited by
resonance fluorescence, either by starlight or by another nebular emission
line. Along with resonance scattering, the excited {\oii} 3d\,$^4$P level will
decay emitting cascade line photons via the multiplets $^4$S$^{\rm
o}$--$^4$P~\,$\lambda$4924 (V\,28), $^4$P$^{\rm o}$--$^4$P~\,$\lambda$4153
(V\,19) and $^4$D$^{\rm o}$--$^4$P~\,$\lambda$3907 (V\,11), with multiplet
strength ratios of about 26.9\,:\,18.5\,:\,1.0. The strongest line of V\,28,
$\lambda$4924.53 ($J$~=~3/2--5/2), coincides in wavelength with [Fe~{\sc iii}]
$\lambda$4924.50 and is blended with the much stronger {\hei} $\lambda$4921.93
line at the resolution of our observations. The second strongest component of
V\,28, $\lambda$4906.83 ($J$~=~3/2--3/2), is detected in our spectra, yielding
an even \emph{higher} abundance ratio than the V\,19 lines, by 70\,per cent.
There is therefore marginal evidence---based on results from V\,28 and
V\,19---that the population of the 3d\,$^4$P term is enhanced by resonance
fluorescence from the ground state; this might explain the overabundances
derived from these lines compared to multiplet V\,1 results; V\,1 itself is on
a cascade route not affected by line fluorescent excitation. However, the
single $\lambda$4110.78 (V\,20) line, which also produces an overabundance
yields results consistent with V\,19, while its upper 3d\,$^4$D term cannot be
reached via permitted transitions from the ground state. It is thus possible
that stellar \emph{continuum}-fluorescence, rather than nebular {\em
line}-fluorescence is instead responsible for this pattern of intensity
enhancements. From a model analysis of the permitted emission line spectrum of
Orion, Grandi (1976) surmised that starlight excitation via the $\lambda$430
resonance line contributes only 20\,per cent as much as recombination to the
$\lambda$4153 (V\,19) line, while he did not discuss the potentially worse
affected $\lambda\lambda$4906, 4924 (V\,28) lines; perhaps in 30~Doradus that
contribution is larger. Due to their possible contamination by line or
continuum fluorescence, V\,19 and V\,28 multiplet lines will be excluded from
our abundance analysis.

Regarding the detected 3d--4f transitions of O~{\sc ii}, several of them are
evidently blended with [Fe~{\sc ii}] and/or Fe~{\sc ii} lines and were omitted
from further consideration. The 3d--4f {\oii} ORLs are insensitive to optical
depth effects and for a significant number of PNe their strengths have
consistently proven to be in excellent agreement with theoretical predictions
under an intermediate coupling scheme (cf. Liu et al. 1995a, 2000, 2001b;
Tsamis et al. 2002b). From this growing body of work it has emerged that
for eleven
thoroughly-analysed PNe the {\opp}/{\hp} abundance ratios derived from the best
detected 3s--3p transitions (those of multiplet V\,1) are \emph{lower} than
those from 3d--4f lines by up to 50\,percent. This discrepancy is removed
however if we assume that these heavy element ORLs arise from ionized regions
of very low electron temperature, of the order of $\sim\,10^3$\,K, much lower
than those indicated by the H\,{\sc i} recombination continuum Balmer
discontinuity, just as one would expect for a dual abundance nebular model (Liu
2002b; P\'{e}quignot et al. 2002b; Tsamis 2002). Transitions of
the 3d--4f group
are close to hydrogenic in nature and the one amongst them with the highest
total angular momentum quantum number ($\lambda$4089, upper state $J$~=~11/2)
is not affected by a change of coupling scheme (Storey 1994; Liu et al. 1995a).
Furthermore, their upper terms cannot be reached by permitted resonance lines
and are thus unaffected by fluorescence effects. Overall, the 3d--4f lines are
the best ORL indicators of {\opp}/{\hp} abundances. In 30~Doradus the strongest
expected 3d--4f line at 4089.13\,\AA~is abnormally weak compared to other
3d--4f lines (Table~7); as a consequence, the abundance ratio derived from this
line alone is lower than that from the V\,1 lines, by 36\,per cent. This is in
stark contrast with the standard observed behaviour of these two sets of lines
as outlined above.

An explanation for this result may involve the effect of dust-scattered stellar
light on the nebular line intensities. Based on observations of the scattered
light continuum described previously, we have estimated upward corrections to
the observed emission line intensities for 30~Doradus (Table~9); these amount
to 58\,per cent ($\lambda$4072); a factor of 2.5 ($\lambda$4089); 6\,per cent
($\lambda\lambda$4638, 4640); 34\,per cent ($\lambda\lambda$4649, 4650); and
17\,per cent ($\lambda$4661), respectively. The resulting corrected
{\opp}/{\hp} values from the various multiplets (Table~11) then display the
more regular pattern already established from the PN analysis; multiplet V\,1
results are now 25\,per cent lower than those from the co-added 3d--4f
transitions and 20\,per cent lower than those from the $\lambda$4072 V\,10
line. Corrected line intensities and resulting {\opp/\hp} abundance ratios are
presented in Table~11. Before any corrections, the {\opp}/{\hp} ORL abundance
is 3.49\,$\times$\,{\tmf}, while after incorporating the corrections it becomes
4.67\,$\times$\,{\tmf}.

For 30~Doradus, the derived ORL/CEL abundance ratio for {\opp} is 2.0 before
correction for underlying stellar absorption lines; after the corrections the
ratio is 2.7 (Table~12).
\\

\noindent{\bf{LMC~N11B}}: In this nebula too the emission line spectrum is
contaminated by the scattered continuum of illuminating stars. Regarding the
nebular {\oii} ORLs, lines of multiplet V\,1 are affected the most. According
to our calculations presented in Tables~8 and 9, the following upward
corrections to the observed nebular intensities have been estimated: 65\,per
cent ($\lambda$4089); 58\,per cent ($\lambda\lambda$4638, 4640); a factor of
2.1 ($\lambda\lambda$4649, 4650); and 64\,per cent ($\lambda$4661). The
corrected line intensities and resulting {\opp}/{\hp} abundance ratios for this
nebula are listed in Table~11. The {\opp}/{\hp} ORL abundance before any
correction is 7.19\,$\times$\,{\tmf}, while that after is
12.0\,$\times$\,{\tmf}.

Therefore, for this nebula the ORL/CEL abundance ratio for {\opp} is 4.9 before
the corrections for absorption, while it rises to 8.2 afterwards (Table~12).
\\

\noindent{\bf{SMC~N66}}: Only the V\,1 multiplet lines of {\oii} are reliably
detected; no corrections for underlying absorption have been estimated.
Co-adding the intensities of the $\lambda$4638, $\lambda\lambda$4649, 4650 and
$\lambda$4661 transitions, we find an {\opp}/{\hp} ORL abundance ratio of
2.15\,$\times$\,{\tmf}. The derived {\opp} ORL/CEL ratio is 2.3
(Table~12).\\

\setcounter{table}{10}
\begin{table}
\begin{center}
\caption{Corrected {\oii} intensities and resulting ORL O$^{2+}$/H$^+$
abundances.}
\begin{tabular}{l@{\hspace{2.5mm}}c@{\hspace{2.5mm}}l@{\hspace{2.5mm}}c@{\hspace{2.5mm}}l@{\hspace{2.5mm}}c@{\hspace{2.5mm}}}
\noalign{\vskip3pt}\noalign{\hrule}\noalign{\vskip3pt}
$\lambda_{\mathrm{0}}$ &Mult.&$I_{\rm cor}$ &$\displaystyle\frac{\rm{~O}^{2+}}{\rm{H}^+}$  &$I_{\rm cor}$ &$\displaystyle\frac{\rm{~O}^{2+}}{\rm{H}^+}$\\
\noalign{\vskip2pt}
(\AA)         &     &          &(\tmf)    &    &(\tmf)    \\
\noalign{\vskip3pt} \noalign{\hrule} \noalign{\vskip3pt}
           &     &\multicolumn{2}{c}{30 Doradus}  &\multicolumn{2}{c}{LMC N11B} \\
\noalign{\vskip2pt}

4072.16   &V10     & .1077         &   4.47            &  *        & *           \\

4638-40   &V1       &  .1601       &   4.34            & .2389     & 6.51        \\
4649-50   &V1       &  .1723       &   2.84            & .2964     & 4.91        \\
4661      &V1       &  .0661       &   4.94            & .1050     & 7.89        \\

\noalign{\vskip3pt}
\bf{V\,1 3s$^{4}$P-3p$^{4}$D$^\mathrm{{o}}$}  & &\bf{0.399} &\bf{3.59}  &\bf{0.640} &\bf{5.26}\\
\noalign{\vskip3pt}

4083.90   &V48b    & .0108          &  3.41               &*  &*                    \\
4087.15   &V48c    & .0185          &  6.17               &*  &*                    \\
4089.29   &V48a    & .0531          &  4.79               &.2154   &18.8            \\
4275.55   &V53c    & .0785          &  5.34               &*  &*                    \\
4609.44   &V92c    & .0209          &  3.35               &*  &*                    \\
\noalign{\vskip3pt}
\bf{3d-4f}  & &\bf{0.182} &\bf{4.76}   &\bf{0.215} &\bf{18.8}\\
\noalign{\vskip3pt} \noalign{\hrule} \noalign{\vskip3pt}
\bf{\emph{Adopted}} & &              &\bf{4.67}      &  &\bf{12.0}   \\
\noalign{\vskip3pt} \noalign{\hrule}
\end{tabular}
\end{center}
\end{table}

\subsubsection{Uncertainties}

Clearly, even though the dust-scattered component to the observed continuum is
found to be very high ($\sim$\,80\,per cent typically) and the various
absorption/emission EWs were carefully measured, not \emph{all} stars that
contribute to the continuum were located. For instance, 30~Doradus contains the
luminous core R\,136 whose effect was neglected. The missing stars most likely
belong to hotter O3\,If$^*$/WN and WR types (Crowther \& Dessart 1998) that
have weaker {\oii} absorption lines. Thus for both 30~Doradus and LMC~N11B the
underlying stellar absorption-line equivalent widths may be weaker than assumed
here. It is therefore likely that our corrections to the observed ORL
intensities and the final ORL abundances, in Table~11, are \emph{upper} limits
only in the cases of 30~Doradus and LMC~N11B.

\section{Spatial variation of nebular properties in 30~Doradus}

The high S/N ratio of our observations of {\oii} ORLs from 30~Doradus (see
Fig.\,~1) prompted us to investigate the abundance discrepancy problem in that
nebula by studying the variations of electron temperature, electron density and
of the ionic recombination-line {\cpp}/{\hp} and {\opp}/{\hp} abundances and
forbidden-line {\opp}/{\hp} abundances along the spectrograph slit. The results
are presented in this section.

\begin{figure}
\begin{center} \epsfig{file=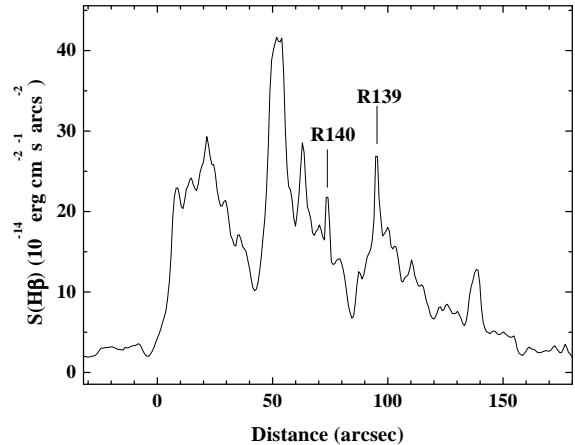, width=10. cm, clip=}
\caption{The {\Hb} surface brightness distribution of 30~Doradus along the
slit; interstellar extinction has not been corrected for. The positions of the
R\,139 and R\,140 systems are marked.}
\end{center}
\end{figure}

The 30~Doradus slit passed across the stars R\,139 and R\,140 and the nebular
analysis presented thus far is based on a spectrum integrated over a spatial
extent of 2\,arcmin approximately. This excludes a region of 40\,arcsec
containing R\,139 and R\,140, in order to minimize the contribution of
dust-scattered starlight to the observed spectrum. In Fig.\,~5 the {\Hb}
surface brightness distribution of the nebula is plotted against the spatial
dimension. Steep variations of the nebular surface brightness are seen. The
positions corresponding to R\,140 and R\,139 are at +73 and +95\,arcsec
respectively.

For the blue spectra, each pixel along the slit corresponded to 0.74\,arcsec on
the sky, while the seeing was about 1\,arcsec FWHM. For the spatial analysis we
used spectra secured with the two overlapping grating set-ups, centred at
4290\,\AA , covering the {\cii} $\lambda$4267 recombination line, H$\gamma$ and
the {\foiii} $\lambda$4363 CEL, and at 4745\,\AA , covering the {\oii} V\,1
recombination multiplet at 4650\,\AA, the {\fariv} $\lambda\lambda$4711, 4740
density-sensitive doublet, H$\beta$ and the {\foiii} $\lambda$4959 forbidden
line. For each grating set-up, spectra were extracted at intervals along the
slit, averaged over about 5 pixels each time, so that a substantial S/N was
achieved. We wanted to examine the nebular properties as close as possible to
stars like R\,139 and R\,140, so spectra were also extracted in the region
between the two, which was not included previously in the integrated spectrum.
The individual extractions were brought to scale via their overlapping portions
and normalized to flux units such that $F$(H$\beta$)~=~100. We then subtracted
the local continuum and measured the fluxes of {\cii} $\lambda$4267 and {\oii}
V\,1 multiplet recombination lines, as well as the {\fariv} doublet and
{\foiii} $\lambda\lambda$4363, 4959 CELs, dereddening them using the extinction
constant derived from the integrated spectrum (Section~3.1).

The electron temperature derived from the {\foiii} $\lambda$4959/$\lambda$4363
ratio, as a function of position along the slit, is shown in Fig.\,~6. The
{\foiii} $\lambda$4959 line was saturated on the four 20-min exposures centred
at 4745\,\AA, so its flux was measured on the 5-min high-resolution spectrum.
The {\foiii} temperature is found to remain practically constant over the
region examined here, with a mean value of 9990\,K and a standard deviation of
200\,K. This is marginally lower than the temperature of 10100\,$\pm$\,250\,K,
derived using the line fluxes from the integrated spectrum (Table~3). No
evidence for substantial temperature fluctuations is seen.

\begin{figure}
\begin{center} \epsfig{file=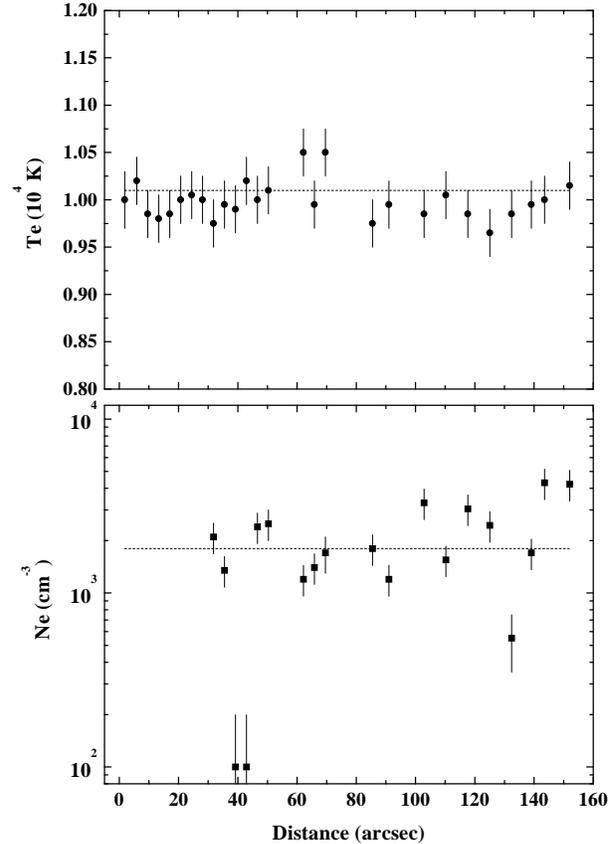, width=10. cm, clip=}
\caption{Variations of (\emph{top}): the electron temperature in 30~Doradus
derived from {\foiii} $\lambda$4959/$\lambda$4363. The dashed line denotes the
temperature derived from lines fluxes obtained from integrating the spectrum
along the slit; (\emph{bottom}): The electron density derived from {\fariv}
$\lambda$4740/$\lambda$4711, with the dashed line having the same meaning as
previously.}
\end{center}
\end{figure}

\begin{figure}
\begin{center} \epsfig{file=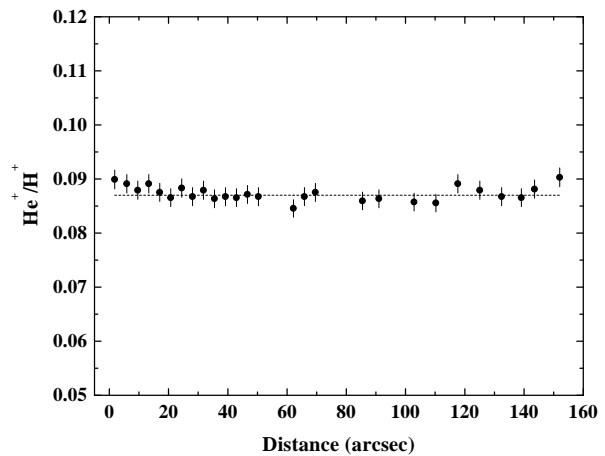, width=10. cm, clip=}
\caption{Variation of the He$^+$/H$^+$ abundance ratio along the slit for
30~Doradus, derived from the {\hei} $\lambda$4471 line. The dashed line marks
the helium abundance derived from integrating the $\lambda$4471 flux along the
slit.}
\end{center}
\end{figure}

\begin{figure}
\begin{center} \epsfig{file=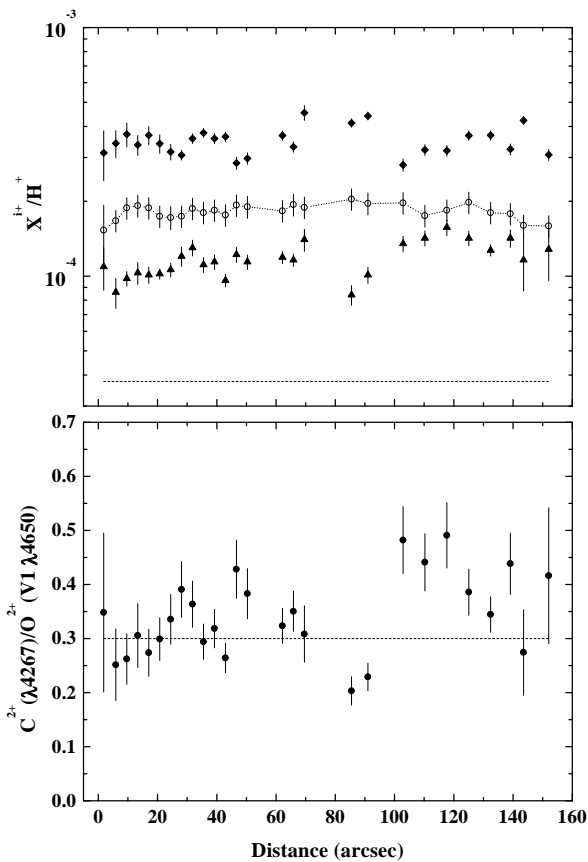, width=10. cm, clip=}
\caption{(\emph{top}): Variations across 30~Doradus of i) the ionic
\emph{recombination}-line abundance of {\opp}/{\hp} (solid diamonds) derived
from the {\oii} V\,1 multiplet at 4650\,\AA; ii) the ionic
\emph{forbidden}-line abundance of {\opp}/{\hp} (open circles, dotted line)
derived from the {\foiii} $\lambda$4959 line; iii) the ionic recombination-line
abundance of {\cpp}/{\hp} (solid triangles) derived from the {\cii}
$\lambda$4267 line; iv) the horizontal dashed line at the bottom denotes the
\emph{forbidden}-line abundance of {\cpp}/{\hp} derived from the C~{\sc iii}]
$\lambda$1909 doublet using the line intensity quoted by Dufour et al. (1982).
(\emph{bottom}): The {\cpp}/{\opp} ratio derived solely from
\emph{recombination}-lines. The dashed line denotes the ratio obtained from the
integrated spectrum.}
\end{center}
\end{figure}

The electron density variation across the nebula has been mapped using the
{\fariv} doublet ratio and is plotted in Fig.\,~6 as well. The density has a
mean value of 1960\,{\cmt} and a standard deviation of 1220\,{\cmt}. This is in
good agreement with the {\fariv} density of 1800\,{\cmt} derived from the
integrated spectrum (Table~3). There is evidence for a density minimum of
{\eld} = 100\,{\cmt} at +40~arcsec in Fig.\,~6, corresponding to a local
minimum in the {\Hb} surface brightness at about +40\,arcsec in Fig.\,~5. Owing
to the high critical densities of the $\lambda$4711, 4740 transitions (14000
and 130\,000\,{\cmt} respectively), the {\fariv} lines are good tracers of
high-density ionized gas, while still sensitive down to
$N_e~\sim~10^3$~cm$^{-3}$. We see no evidence however of any high-density
material at our spatial resolution.

In Fig.\,~7 the variation of the He$^+$/H$^+$ abundance ratio across the
30~Doradus nebula is presented, derived from the {\hei} $\lambda$4471
recombination line. The adopted electron temperature was 9990\,K. The
He$^+$/H$^+$ ratio remains constant over the range plotted here. The helium
abundance remains very close to the mean value even in the nebular region
between R\,139 and R\,140.

Fig.\,~8 shows the spatial variations of the {\cpp}/{\hp} and {\opp}/{\hp} ORL
abundances and of the {\opp}/{\hp} forbidden-line abundance, derived
respectively from the {\cii} $\lambda$4267 line, the {\oii} V\,1 multiplet at
4650\,\AA~and the {\foiii} $\lambda$4959 forbidden line. In should be noted
that in order to circumvent non-LTE effects that affect abundances derived from
individual {\oii} V\,1 components (Section~5.2.1), the total observed intensity
of the $\lambda$4650 multiplet was used to obtain the ORL {\opp}/{\hp} ratio.
An electron temperature of 9990\,K was adopted for the calculation of the ORL
abundances, while the point-by-point {\foiii} $\lambda$4959/$\lambda$4363
temperature measurements (Fig.\,~6) were used for the derivation of the
forbidden-line {\opp}/{\hp} values.

From Fig.\,~8 (\emph{top}) no significant variation in the
\emph{forbidden}-line {\opp}/{\hp} abundance across the nebula can be
discerned. Over the whole range plotted {\opp}/{\hp} has an average value of
1.83\,$\times$\,{\tmf} and a standard deviation of 1.10\,$\times$\,{\tmfi}. The
forbidden-line abundance remains consistently lower than the ORL {\opp}/{\hp}
abundance across the plotted range, by a factor of 2. The {\cpp}/{\hp}
abundance derived from the collisionally-excited C~{\sc iii}] $\lambda$1908
intercombination doublet (Dufour et al. 1982) is also consistently lower than
the mapped ORL abundances of doubly ionized carbon. Could temperature
fluctuations account for these discrepancies? At the spatial resolution of our
mapping, there is no evidence of {\foiii} temperature fluctuations of a
sufficient magnitude to cause this. An electron temperature of 8200\,K would be
needed in order to force the forbidden-line {\opp}/{\hp} abundances to become
equal to the recombination-line abundances. This would necessitate the
existence of fluctuations from the mean temperature (9990\,K) of about 18\,per
cent or -- in the terminology of Peimbert (1967) -- a temperature fluctuation
parameter $t^2 \sim 0.03$, which is however not deduced from our spatial
analysis. We note, however, that while the above observations place
constraints upon temperature fluctuations in the plane of the sky, along
our slit, fluctuations along the line of sight though the nebula are not
constrained by this method. Rubin et al. (2002) discuss this point in
more detail, in the context of HST observations of the planetary nebula
NGC~7009.

Fig.\,~8 also presents, for the first time to our knowledge for an
extragalactic {\hii} region, the spatial variation of the {\cpp}/{\opp} ratio
derived purely from ORLs. The {\cpp}/{\opp} ratio across the nebula has a mean
value of 0.33 and a 1\,$\sigma$ error of 0.02. This is in excellent agreement
with the value of 0.30 derived from the integrated spectrum. This latter value
was calculated adopting as the {\opp}/{\hp} abundance the value derived from
the V\,1 multiplet (Table~10), while the mean {\cpp}/{\hp} abundance is taken
from Table~5.  The {\opp} V\,1 abundances that were used for Fig.\,~8 are those
{\em before} any correction is made for underlying absorption at 4650\,\AA~in
the scattered-light continuum. Such a correction is estimated to raise the
{\opp} recombination-line abundance by a factor of 1.34 (Section~5.2.2).  An
interesting feature of our spatial analysis of heavy-element ORLs is that there
is tentative evidence for a local minimum in the {\cpp}/{\opp} ratio, derived
purely from ORLs, in the nebular region between the massive stars R\,139 and
R\,140. The relevance of this result is discussed in following section.

\section{Discussion}

\subsection{ORL/CEL abundance discrepancies and elemental ratios}

It is clear from the above analysis and Table~12 that substantial ORL versus
CEL abundance discrepancies \emph{are present}, not only in planetary nebulae,
but in {\hii} regions too.

Discrepancies were reported for M\,42 and M\,17 by Peimbert et al. (1993), for
both of which they found an ORL/CEL factor of $\sim$\,1.7 for the {\opp} ion.
Esteban et al. (1998) found ORL/CEL discrepancy factors of 1.4--1.6 and 2.2,
respectively, for {\opp} and {\cpp} in the Orion nebula [adopting
{\cpp}($\lambda$1908)/{\hp} CEL abundances from Walter et al. 1992]. Our
unpublished long-slit spectra, obtained at the AAT 3.9-m and ESO 1.52-m
telescopes, show a factor of only 1.3 discrepancy for {\opp} in that nebula,
but imply an ORL/CEL {\cpp} discrepancy of 2.4, when compared with Walter et
al.'s \emph{IUE} results for C~{\sc iii}] $\lambda$1908. For the {\hii} region
M\,8, Esteban et al. (1999) reported a factor of two discrepancy between the
recombination-line {\opp} abundance and its forbidden-line value. A CEL
abundance study of the Gum 38a complex, which contains NGC\,3576, has been made
by Girardi et al. (1997), but no detection of {\cii} $\lambda$4267 or any other
heavy element optical recombination line was reported. Owing to its high
reddening, NGC\,3576 was never observed by the \emph{IUE} either, so no
comparison can be made with the ORL {\cpp}/{\hp} abundance ratio reported here.

Table~12 shows that the ORL/CEL {\opp} discrepancy factors, $\Re$, for our
{\hii} regions are: 2.1, 1.8, 2.0--2.7, 4.9--8.2 and 2.3, for M\,17, NGC\,3576,
30 Doradus, LMC~N11B and SMC~N66 respectively. Even if we adopt the lower ORL
abundances for 30~Doradus and N11B (i.e. neglect corrections for any underlying
absorption in the dust-scattered continuum), the mean discrepancy factor is
2.6. LMC~N11B displays the largest discrepancy documented so far for an {\hii}
region (=~4.9), in the neighborhood of the mean value (=~5.1) that we have
found for a sample of eighteen PNe (Tsamis et al. 2002b). How is one to
explain such an extreme discrepancy for an {\hii} region?

\setcounter{table}{11}
\begin{table*}
\centering
\begin{minipage}{150mm}
\caption{ORL/CEL discrepancy factors, $\Re$, and comparison of elemental
abundance ratios derived purely from ORLs and CELs.$^a$}
\begin{tabular}{lcccccccc}
\noalign{\vskip3pt} \noalign{\hrule} \noalign{\vskip3pt}
              &                &Sun$^b$     &M\,42$^c$    &M\,17        &NGC\,3576    &30~Doradus         &LMC~N11B       &SMC~N66 \\
\noalign{\vskip3pt} \noalign{\hrule} \noalign{\vskip3pt}
10$^4\times$O/H &({\sc cel}s)  &4.90        &3.32        &3.59         &3.31         &2.17               &2.59           &1.28 \\
$\Re$(\opp)   &({\sc orl/cel}) &*           &1.3         &2.1          &1.8          &2.0--2.7           &4.9--8.2       &2.3             \\
$\Re$(\cpp)   &({\sc orl/cel}) &*           &2.4         &*            &*            &2.6$^d$            &*              &*          \\
{\cpp}/{\opp} &({\sc orl}s)$^e$ &0.50       &0.77        &0.77         &0.78         &0.25               &0.25           &$\la0.20$ \\
{\npp}/{\opp} &({\sc orl}s)$^e$ &0.19       &0.15        &0.62         &0.71         &$\la0.16$          & *             & * \\
{\npp}/{\opp} &(IR {\sc cel}s)$^f$ &0.19    &0.19        &0.19         &0.16         &0.060              &0.075          & * \\
{\np}/{\op}   &({\sc cel}s)    &0.19        &0.15        &0.088        &0.11         &0.036              &0.033          &0.016 \\
Ne/O          &({\sc cel}s)    &0.25        &0.18        &0.28         &0.16         &0.21               &0.16           &0.18        \\
S/O           &({\sc cel}s)    &0.033       &0.028       &0.027        & *           &0.027              &0.019:         &0.018 \\
Ar/O          &({\sc cel}s)    &.0074       &.011        &.0057        &.0078        &.0065              &.0078          &.0053  \\
10$^4\times${\opp}/H$^+$ &(Opt {\sc cel}s) &*     &2.26        &2.66         &2.21         &1.76               &1.47           &0.94 \\
10$^4\times${\opp}/H$^+$ &(IR {\sc cel}s)$^g$ &*  &1.66        &1.80         &2.68         &2.21               &*              &* \\

\noalign{\vskip3pt} \noalign{\hrule} \noalign{\vskip3pt}
\end{tabular}
\begin{description}
\item[$^a$] All nebular data are from this paper, apart from those in rows 6 and 12, and the data in rows 7--10 for M\,42.
\item[$^b$] Solar elemental ratios adopting log\,(X/H) + 12.0  = 8.69 and 8.39 for X = O and C (Allende Prieto et al. 2001,
            2002), together with  N, Ne, S, Ar abundances from Grevesse et al. (1996).
\item[$^c$] Entries 1--5 and entry 11 from our unpublished data; entries 7--10 from Esteban et al. (1998).
\item[$^d$] Adopting a CEL {\cpp}/{\hp} abundance of 3.44$\times$10$^{-5}$ (Dufour et al. 1982).
\item[$^e$] No corrections for scattered light absorption features have been implemented in obtaining the listed nebular ORL {\cpp}/{\opp} and
            {\npp}/{\opp} ratios.
\item[$^f$] From Simpson et al. (1995) for M\,17, NGC\,3576 and 30~Dor; from Rubin et al. (1988) for M\,42; value for LMC~N11B
            derived from the [O~{\sc iii}] 52- and 88-$\mu$m and [N~{\sc iii}] 57-$\mu$m fluxes in ISO LWS spectrum TDT60901217.
            Note that the positions at which these IR data were obtained were not the same as for our optical long-slit spectra.
\item[$^g$] From Simpson et al. (1995) for M\,17, NGC\,3576 and 30~Dor; from Rubin et al. (1991) for M~42.
\end{description}
\end{minipage}
\end{table*}

Before commenting on this issue, we consider the derived elemental
abundance ratios presented in Table~12 for this sample of H~{\sc ii}
regions, where they are compared with solar values. The {\cpp}/{\opp} and
{\npp}/{\opp} ratios are purely from ORL lines and should be almost
completely unaffected by temperature variations. They should give a good
measure of the C/O and N/O ratios -- especially the latter since for these
relatively high-excitation H~{\sc ii} regions, the N$^{2+}$ and O$^{2+}$
zones occupy almost the entire H$^+$ zone (Shaver et al. 1983; Simpson et
al. 1995). From Table~12, we see that the agreement between the
{\cpp}/{\opp} ratios for the three galactic nebulae is excellent, with the
mean value of 0.77 being 50\,per cent higher than the solar value of
Allende Prieto et al. (2001, 2002).

The ORL {\npp}/{\opp} and CEL {\npp}/{\opp} and {\np}/{\op} ratios listed for
M~42 in Table~12 are in good agreement. Our {\npp}/{\opp} ORL results for M\,17
and NGC\,3576 are much \emph{higher} than the respective CEL {\npp}/{\opp}
ratios, by factors of 3--4.5, while they are larger than the {\np}/{\op} CEL
ratios as well. There is the strong possibility however that the {\nii} ORLs
used for these two nebulae yield unreliable abundances because they are
affected by fluorescence effects. Grandi (1976) showed that for Orion the
{\nii} multiplets V\,3, V\,5 and V\,30 are excited by fluorescence via the
{\hei} $\lambda$508.6 resonance line. In the current study, {\npp}/{\hp}
abundances for M\,17 and NGC\,3576 have been derived from {\nii} multiplets
V\,3 and V\,5 (Table~6). On the other hand, Liu et al. (2001b) showed that the
observed N~{\sc ii} ORL intensity ratios for the planetary nebulae M\,1-42 and
M\,2-36 are inconsistent with fluorescent \emph{line}-excitation of the N~{\sc
ii} V\,3 and V\,5 lines and proposed that continuum fluorescence by starlight
is a more plausible cause. In any case, it seems probable that fluorescence of
some sort contributes to the excitation of the N~{\sc ii} V\,3 and V\,5 triplet
lines observed from M\,17 and NGC\,3576. Unfortunately, the {\nii} 3d--4f
high-lying transitions, which are generally not biased by this effect, have not
been detected from either M\,17 or NGC\,3576; their measurement would offer a
means of checking the fluorescence hypothesis. However, the ORL
{\npp}/{\opp} ratio for M\,42 listed in Table~12 was based on our observations
of N~{\sc ii} 3d--4f and singlet lines, and so should be unaffected by
fluorescence effects, consistent with the agreement found between the ORL
{\npp}/{\opp} and CEL {\npp}/{\opp} and {\np}/{\op} ratios for this nebula.

For the two LMC nebulae, the {\cpp}/{\opp} ORL ratios are in excellent
agreement, their value of 0.25 being only 11\,per cent smaller than the mean
CEL C/O ratio of 0.28 found for LMC {\hii} regions by Kurt \& Dufour (1998). In
addition, their {\np}/{\op} ratios are in excellent agreement with the mean CEL
N/O ratio of 0.035 for the LMC found by the same authors, though smaller than
the {\npp}/{\opp} ratios found for the same two nebulae from far-IR CELs.
Finally, for SMC~N66 our {\cpp}/{\opp} ORL upper limit is consistent with the
SMC mean CEL C/O value of 0.15 (Kurt \& Dufour), although our CEL {\np}/{\op}
ratio of 0.016 is only 46\,per cent of Kurt \& Dufour's mean CEL N/O ratio of
0.035 for the SMC.

From the above comparisons an important conclusion can be drawn: it seems that
the C/O abundance ratio derived from ORLs is equal within the uncertainties to
the same ratio derived from CELs, both for the galactic nebulae and, as quoted
by Kurt \& Dufour, for the Magellanic Cloud nebulae as well (a similar
conclusion appears to apply to N/O ratios too, provided care is taken in the
choice of the N~{\sc ii} ORLs used to derive {\npp} abundances). Similar
results for C/O, N/O and Ne/O have been reported for planetary nebulae (Liu et
al. 1995a, 2000, 2001b; Mathis, Torres-Peimbert \& Peimbert 1998; Luo, Liu 
\& Barlow 2001). Apart from the obvious observation
that this strengthens the reliability of C/O or N/O ratios derived from either
pure CEL or pure ORL abundance analyses, it also raises important questions
regarding the chemical history of the ORL-emitting medium with respect to that
of the `normal' CEL-emitting nebular component, given that they appear to have
similar C/O, N/O and Ne/O ratios.

\subsection{Potential causes of the ORL/CEL abundance discrepancies}

\subsubsection{The case against temperature fluctuations}

The existence of temperature fluctuations within the nebular volumes has been
long thought to offer an obvious solution to the conundrum of discrepant ORL
versus CEL abundances, both for PNe and {\hii} regions. Temperature
fluctuations within a nebula can lead to enhanced emission of the
temperature-sensitive, high-excitation energy [O~{\sc iii}] $\lambda$4363
transition from the hotter nebular regions. As a result, the electron
temperature deduced from the net [O~{\sc iii}] ($\lambda$4959 +
$\lambda$5007)/$\lambda$4363 intensity ratio would be biased towards the hotter
regions and would lead to too low an {\opp} abundance being derived from the
net nebular spectrum. The Peimbert (1967) temperature fluctuation parameter,
$t^2$, derived from a comparison between the hydrogen BJ and {\foiii}
forbidden-line temperatures is 0.011 and 0.017 for M\,17 and NGC\,3576,
respectively (Table~3). On the other hand the ORL/CEL {\opp} discrepancy
factors for these two nebulae of 2.1 and 1.7, respectively, would require
values of $t^2$\,$\sim$\,0.038, i.e. considerably larger.\footnote{On the other
hand, Liu et al. (1995b) found very similar values of T$_{\rm e}$(O~{\sc iii})
and T$_{\rm e}$(BJ) for M\,42, the nebula in Table~12 with the smallest {\opp}
ORL/CEL discrepancy factor (=~1.3).} For the three Magellanic Cloud nebulae the
$t^2$ factors implied by the abundance discrepancies are even greater, reaching
the uncomfortably high value of $\sim$\,0.1 in the case of LMC~N11B, very
similar to the value needed to reconcile ORL and CEL {\opp} abundances for the
rather extreme planetary nebula NGC\,7009 (Liu et al. 1995a). This is a factor
of 10 larger than typical values predicted by photoionization models of
chemically and density homogeneous nebulae, which yield $t^2$\,$\sim$\,0.01
(Garnett 1992; Gruenwald \& Viegas 1995; Kingdon \& Ferland 1995b). As in the
case of planetary nebulae, it seems unlikely that temperature fluctuations are
to blame for the discrepancies observed here, especially for extreme objects
like LMC~N11B.

If temperature fluctuations were indeed responsible for the low heavy element
ion abundances deduced from optical CELs compared to those derived from ORLs
(see rows 2 and 3 of Table~12), one would then expect infrared CELs to yield
abundances similar to the high values obtained from ORLs, due to the low
excitation energies of IR fine structure transitions and their consequent lack
of sensitivity to temperature fluctuations at typical nebular temperatures. The
last two rows of Table~12 compare {\opp}/{\hp} CEL abundances derived from our
optical spectra with those derived by Rubin et al. (1988) and Simpson et al.
(1995) from infrared fine structure line observations of four of the nebulae.
For M\,42 and M\,17, the IR CEL lines yield {\opp}/{\hp} abundances which are
27~per cent and 32~per cent smaller than those obtained from our optical CEL
observations, while for NGC\,3576 and 30~Doradus the IR CEL lines yield
{\opp}/{\hp} abundances which are 21\,per cent and 26\,per cent larger than
those from the optical CEL data. These differences of only 20--30\,per cent are
close to the 20\,per cent uncertainties estimated by Simpson et al. for the
radio \emph{free-free} fluxes (that measure H$+$) falling in the Kuiper
Airborne Observatory (KAO) beam used for the far-infrared line observations.
It should be noted that the positions in the nebulae that were observed
at far-infrared wavelengths with the KAO usually did not coincide with the
positions covered by our optical long-slit observations. However,
spatial abundance variations within these H~{\sc ii} regions have never
been detected, as exemplified by our own spatially-resolved abundance
analysis of 30~Dor (Section 6).

We therefore conclude that the agreement found between the {\opp}/{\hp}
abundances derived from optical and infrared CELs rules out temperature
fluctuations as the cause of the factor of two discrepancy found between ORL
and CEL {\opp}/{\hp} abundances for M\,17, NGC\,3576 and 30~Doradus. On the
other hand, the large $t^2$ derived for LMC~N11B from the ORL/CEL {\opp} ratio
would imply temperature fluctuations that cannot be explained by chemically and
density homogeneous nebular models.

\subsubsection{The case against high-density clumps and in favour of 
metal-rich inclusions}

High density ionized clumps with {\eld}\,$\geq$\,10$^5$\,{\cmt} embedded in a
medium of lower density could lead to high ORL C, N and O abundances, via an
effect of pseudo-temperature fluctuations, whereby the observationally derived
{\foiii} temperature is significantly overestimated due to collisional
quenching of the nebular $\lambda\lambda$4959, 5007 lines (Viegas \& Clegg
1994). Such high density clumps would also quench emission in the low critical
density far-IR lines of [O~{\sc iii}] and [N~{\sc iii}], leading to lower
abundances being deduced from infrared CELs than from ORLs.

\emph{HST} images of the Orion nebula show numerous clumpy microstructures
surrounded by more or less uniform nebulosity. Walsh \& Rosa (1999) reported
\emph{HST} observations of a partially ionized globule and a filament in the
Orion nebula that displays high densities of up to 5\,$\times$\,10$^6$\,{\cmt}.
They did not however publish the density-sensitive diagnostic. They also
commented that in the core of the nebula the filaments contribute half the
{\fnii} emission, but only 10\,per cent of the Balmer line flux. The existence
of high-density, partially ionized condensations with
{\eld}\,$\sim$\,10$^6$\,{\cmt} has also been postulated by Bautista, Pradhan \&
Osterbrock (1994) in order to explain abnormalities in the emission spectrum of
[Fe~{\sc ii}] in the Orion nebula. Independent observations by Esteban et al.
(1998) and Baldwin et al. (2000) however, showed that [Fe~{\sc ii}] lines are
formed in lower density gas along with [O~{\sc i}] lines. Finally, as Lucy
(1995) and Verner et al. (2000) discuss, the formation of [Fe~{\sc ii}]
emission lines is affected by radiative pumping fluorescence processes in ways
that render them unsuitable as straightforward density diagnostics for {\hii}
regions.

\begin{figure} \label{fig:N35bd}
\begin{center} \epsfig{file=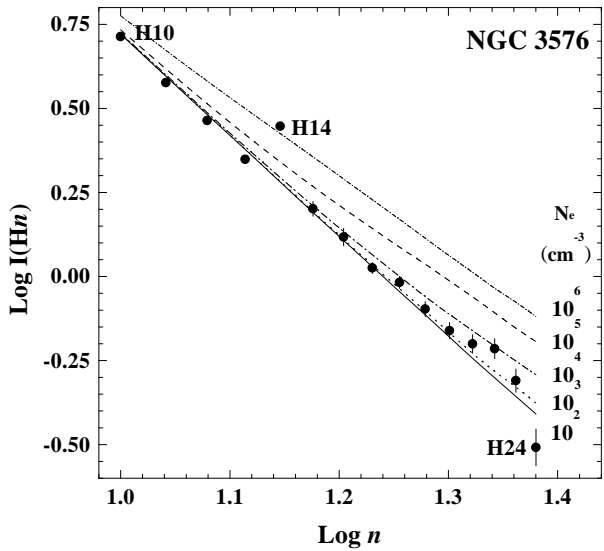, width=8.6 cm, clip=}
\caption[The high-order Balmer lines of NGC 3576 as a density diagnostic.]
{\label{fig:N35bd}Observed log-intensities (in units of {\Hb}~=~100) of
high-order Balmer lines ($n\,\rightarrow\,2$, $n$ = 10,..., 24) as a function
of the principal quantum number $n$; H\,14 at 3721.94\,\AA~is blended with the
{\fsiii} $\lambda$3721.63 line. The various curves show respectively the
predicted Balmer decrements for electron densities from {\eld}~=~10$^2$ to
10$^6$\,{\cmt}. A constant temperature of 8070\,K, derived from the nebular
continuum Balmer discontinuity, has been assumed in all cases.}
\end{center}
\end{figure}

In Fig.\,~9 we plot the reddening-corrected intensities of high-order {\hi}
Balmer lines from our high resolution (1\,\AA) spectrum of NGC\,3576 and
compare them with theoretical predictions for various nebular electron
densities (Storey \& Hummer 1995). Since such lines are sensitive indicators of
ionized, high density regions (assuming those have a normal hydrogen content),
it is interesting to see that our data do not show any evidence for significant
amounts of high density material. In fact, the spectrum can be fitted with a
uniform electron density of 3000\,{\cmt}, consistent with the values derived
from the optical {\fcliii} density diagnostics ({\S 3.2}, Table~3). We 
note that our adopted reddening of c(H$\beta$) = 1.25 for NGC~3576 was 
derived from the observed relative intensities of H$\alpha$, H$\beta$,
H$\gamma$ and H$\delta$. The dereddened intensities of these low 
Balmer lines give an excellent fit to the theoretical Case~B ratios (Table 
2). The high Balmer lines (H10-H24) used for Figure~9 span a
wavelength range of less than 130~\AA , so it would require a 
large change in the adopted reddening to change their decrement slope
to one that was consistent with N$_e = 10^5 - 10^6$~cm$^{-3}$. 
Such a change in reddening would be inconsistent with the 
low Balmer line relative intensities, however. We consider
the high-n Balmer lines to provide a secure upper limit of 
N$_e < 10^4$~cm$^{-3}$ for their emitting region.

The Balmer decrement electron density diagnostic discussed above pertains to
regions dominated by hydrogen, but not necessarily to possible
hydrogen-deficient regions in the nebulae, which could give rise to enhanced
heavy element ORL emission, causing the observed abundance discrepancies.
However, such hydrogen-deficient regions cannot have electron densities
significantly higher than those deduced from the hydrogen Balmer decrement, in
the case of NGC\,3576, or from the standard CEL density diagnostics discussed
in Section~3.2, since, as discussed in Section~5.2.1, the observed relative
intensities of {\oii} V\,1 multiplet components indicate that {\opp}
recombination is occurring in regions with electron densities lower than
3500\,cm$^{-3}$, the critical density of the {\opp} $^3$P$_2$ level.
Collisional quenching of the optical and infrared CELs in high-density clumps
can therefore be ruled out as the cause of the lower abundances derived from
CELs compared to those derived from ORLs. The only remaining way in which CELs
can be collectively suppressed in an ORL-emitting ionized region is for the
temperature of the region to be low enough to suppress not just optical CEL
emission but also infrared fine-structure emission lines such as the 52- and
88-$\mu$m lines of [O~{\sc iii}], implying electron temperatures lower than a
few hundred Kelvin. Is there any evidence for hydrogen-deficient zones inside
{\hii} regions?

Rosa \& Mathis (1987) report the existence of a hydrogen-poor
(He/H~=~0.14) region in the outskirts of the 30~Doradus nebula, enhanced
in most metals, apart from nitrogen. They hypothesize that it might
contain chemically processed ejecta from a massive, late-type Wolf-Rayet
star. It would be interesting to obtain high spatial resolution spectra of
that region to search for regions of enhanced heavy element ORL emission.
Our long-slit spectra sample a slice through the central 1\,arcmin of 30
Doradus, which is known to contain a number of WR stars (Moffat et al.
1987, Parker 1993). Could it be that knots of heavy-element enhanced
material have been ejected from such stars and are the cause of the high
ORL abundances? It is well known from kinematical studies (e.g. Meaburn
1984) that 30~Doradus contains pressure-driven bubbles, shells and sheets
of ionized gas, originating from interactions with the intense winds from
WR, Of stars and supernova remnant ejecta. In fact two such potential
culprits were directly encountered by our long slit, i.e. R\,139
(Of~+~WNL-type) and R\,140 (WC~+~WN-type). The existence of a local
minimum in the ORL {\cpp}/{\opp} ratio (Fig.\,~8) in the nebular region
between R\,139 and R\,140 might be indicative of an outflow of
carbon-depleted material from these evolved stars. On the other hand,
LMC~N11B does not contain any known Wolf-Rayet stars (Parker et al. 1992)
and its ORL/CEL abundance discrepancy factor is more than twice that of
30~Doradus. Moreover, the fact that amongst our sample of {\hii} regions
the ORL {\cpp}/{\opp} ratios of the postulated hydrogen-deficient
component are so similar to the CEL C/O ratios of the `normal' nebular
component seems hard to reconcile with standard nuclear-processing
scenarios for the hydrogen-deficient component.

Concerning the apparent low-density of the ORL-emitting material, it is
interesting to make a comparison with the dual-component nebular models of the
PN NGC\,6153 (Liu et al. 2000), proposed as a solution for the extreme ORL/CEL
abundance discrepancy (a factor of 10) in that nebula. One of those models
(IH2) contained very cold (500\,K), low density ($<$\,1000~{\cmt}) H-deficient
inclusions, immersed in hot gas with normal abundances. This model however was
rejected for NGC\,6153 on account of (a) the required large filling factor of
the metal-rich component, which would significantly increase the total CNO
content of the nebula, and (b) the very large overpressure of the normal
component, which would lead to the rapid collapse of the postulated inclusions.
It is possible that the {\oii} ORLs from our sample of {\hii} regions originate
from cold, low-density, ionized and metal-rich regions located around much
denser neutral cores -- which may also be H-deficient; such material may have
been ejected from evolved, massive Of/WR stars in the vicinity. The existence
of \emph{neutral}, dense inclusions, whose photo-evaporation leads to the
emission of heavy-element ORLs from lower density ionized halos, could help to
alleviate the problem of the survival of fully ionized inclusions subject to
large over-pressures.

\section{Conclusions}

The current dataset supported by published far-IR observations all but rules
out the notion that Peimbert-type temperature fluctuations are the cause of the
high metal abundances derived from optical recombination lines with respect to
the lower abundances yielded by forbidden lines in {\hii} regions. Neither do
our results advance the case for the existence of Viegas \& Clegg-type
\emph{high density, ionized} condensations in this sample of nebulae that could
compromise the observationally derived {\foiii} temperature, since the relative
intensities of high-order {\hi} Balmer lines {\em and} the relative intensities
of {\oii} V\,1 multiplet components both indicate that recombination is
occurring in regions having electron densities similar to those indicated by
the standard CEL nebular density diagnostics. Instead, our analysis points
towards an origin for much of the enhanced heavy-element optical
recombination-line emission (enhanced relative to the optical and infrared CEL
lines from the same ions) in ionized regions that are of \emph{low} density and
which are also cold, so that they do not emit CELs. Thus a resolution of the
ORL/CEL problem appears to require the existence of a hitherto unseen component
in these {\hii} regions, consisting of cold, rarefied and metal-rich, ionized
gas.

\vspace{7mm} \noindent {\bf Acknowledgments}

YGT acknowledges the award of a Perren Studentship. We thank
Daniel P\'{e}quignot for insightful comments and helpful discussions and
the referee, Dr. R. Rubin, for his very careful reading of the manuscript.

\end{document}